\documentclass{aastex631}

\usepackage{newtxtext,newtxmath,float,xspace,threeparttable}
\usepackage{multirow}
\usepackage{csvsimple}
\usepackage[utf8]{inputenc}

\newcommand{\todo}{\ifmmode \text{\color{red}\Huge{\(\bullet\)}} \else {\color{red}{\Huge$\bullet$}}\fi}
\newcommand{\tido}{\ifmmode {{\color{red}\bullet}} \else {\color{red}$\bullet$}\fi}

\newcommand{\Halpha}{\ifmmode {\rm H}\alpha \else H$\alpha$\fi}

\newcommand{\Hbeta}{\ifmmode {\rm H}\beta \else H$\beta$\fi}

\newcommand{\oiii}{\ifmmode \left[{\rm O}\,\textsc{iii}\right] \else [O\,{\sc iii}]\fi}
\newcommand{\OIII}{\ifmmode \left[{\rm O}\,\textsc{iii}\right]\,\lambda5007 \else [O\,{\sc iii}]\,$\lambda5007$\fi}
\newcommand{\nii}{\ifmmode \left[{\rm N}\,\textsc{ii}\right]  \else [N\,\textsc{ii}]\fi}
\newcommand{\NII}{\ifmmode \left[{\rm N}\,\textsc{ii}\right]\,\lambda6584 \else [N\,\textsc{ii}]\,$\lambda6584$\fi}

\newcommand {\nh}{\ifmmode N_{\rm H} \else $N_{\rm H}$\fi}

\newcommand {\Lsoftint} {\ifmmode L^{\rm in}_{\mathrm{2-10\ keV}} \else $L^{\rm in}_{\mathrm{2-10\ keV}}$\fi}

\newcommand {\ergpersec} {\ifmmode {\rm erg~s}^{-1} \else erg~s$^{-1}$ \fi}

\def\micron{{\mbox{$\mu{\rm m}$}}}
\def\arcsec{{\mbox{$^{\prime \prime}$}}}

\def\arcsec{{\mbox{$^{\prime \prime}$}}}

\def\erg{{\rm\thinspace erg}}

\def\g{{\rm\thinspace g}}

\def\K{{\rm\thinspace K}}

\def\km{{\rm\thinspace km}}

\def\m{{\rm\thinspace m}}

\newcommand{\Msun}{\ifmmode M_{\odot} \else $M_{\odot}$\fi}

\def\s{{\rm\thinspace s}}

\def\ergps{\hbox{$\erg\,\s^{-1}\,$}}

\def\kmps{\hbox{$\km\s^{-1}\,$}}

\def\apjl{\hbox{ApJL}}
\def\apjs{\hbox{ApJS}}

\def\aap{\hbox{AAP}}
\def\micron{{\mbox{$\mu{\rm m}$}}}
\def\arcsec{{\mbox{$^{\prime \prime}$}}}

\newcommand{\SiX}{\ifmmode \left[{\rm Si}\,\textsc{x}\right]\,\lambda1.4300 \else [Si\,{\sc x}]\,$\lambda1.4300$\fi}
\newcommand{\SiVI}{\ifmmode \left[{\rm Si}\,\textsc{vi}\right]\,\lambda1.9640 \else [Si\,{\sc vi}]\,$\lambda1.9640$\fi}
\newcommand{\SXI}{\ifmmode \left[{\rm S}\,\textsc{xi}\right]\,\lambda1.9196 \else [S\,{\sc xi}]\,$\lambda19196$\fi}
\newcommand{\SVIII}{\ifmmode \left[{\rm S}\,\textsc{viii}\right]\,\lambda0.9915 \else [S\,{\sc viii}]\,$\lambda0.9915$\fi}
\newcommand{\SIX}{\ifmmode \left[{\rm S}\,\textsc{ix}\right]\,\lambda1.2520 \else [S\,{\sc ix}]\,$\lambda1.2520$\fi}
\newcommand{\FeXIII}{\ifmmode \left[{\rm Fe}\,\textsc{xiii}\right]\,\lambda1.0747 \else [Fe\,{\sc xiii}]\,$\lambda1.0747$\fi}
\newcommand{\SiXI}{\ifmmode \left[{\rm Si}\,\textsc{xi}\right]\,\lambda1.9320 \else [Si\,{\sc xi}]\,$\lambda1.9320$\fi}

\newcommand{\lledd}{\ifmmode L/L_{\rm Edd} \else $L/L_{\rm Edd}$\fi}
\newcommand{\mbh}{\ifmmode M_{\rm BH} \else $M_{\rm BH}$\fi}

\newcommand{\kms}{\ifmmode {\rm km\,s}^{-1} \else ${\rm km\,s}^{-1}$\fi}

\newcommand{\nuvr}{\ifmmode {\rm NUV}-r \else NUV-$r$\fi}
\newcommand{\mh}{\ifmmode M_{\rm H_2} \else $M_{\rm H_2}$\fi}
\newcommand{\mhi}{\ifmmode M_{\rm HI} \else $M_{\rm HI}$\fi}
\newcommand{\mstar}{\ifmmode M_{\ast} \else $M_{\ast}$\fi} 
\newcommand{\must}{\ifmmode \mu_{\ast} \else $\mu_{\ast}$\fi}
\newcommand{\hmol}{\ifmmode H_2 \else $H_2$\fi}
\newcommand{\rmol}{\ifmmode R_{\rm mol} \else $R_{\rm mol}$\fi}
\newcommand{\tdep}{\ifmmode t_{\rm dep}({\rm H_2}) \else $t_{\rm dep}({\rm H_2})$\fi}
\newcommand{\tdepHI}{\ifmmode t_{\rm dep}({\rm HI}) \else $t_{\rm dep}({\rm HI})$\fi}
\newcommand{\fgas}{\ifmmode f_{\rm H_2} \else $f_{\rm H_2}$\fi}
\newcommand{\fhi}{\ifmmode f_{\rm HI} \else $f_{\rm HI}$\fi}
\newcommand{\xco}{\ifmmode \alpha_{\rm CO} \else $\alpha_{\rm CO}$\fi}

\usepackage[T1]{fontenc}
\usepackage{ae,aecompl}

\usepackage{graphicx}	
\usepackage{amsmath}	
\usepackage{amssymb}	
\usepackage{hyperref} 
\turnoffeditone
\turnoffedittwo
\turnoffeditthree

\received{}
\revised{}
\accepted{}

\shorttitle{BASS XXVIII: NIR DR2}
\shortauthors{den Brok et al.}

\begin{document}


\title{BASS XXVIII: Near-infrared Data Release 2, High-Ionization and Broad Lines in Active Galactic Nuclei\footnote{Based on observations collected under programs 086.B-0135(A), 089.B-0951(A), 090.A-0830(A), 091.B-0900(B), 093.A-0766(A), 098.A-0635(B), 099.A-0403(B), 0101.A-0765(A), and 0102.A-0433(A) with X-shooter at the Very Large Telescope of the Paranal Observatory in Chile,  operated by the European Southern Observatory.}}

\correspondingauthor{Jakob S. den Brok}
\email{jdenbrok@astro.uni-bonn.de}

\author[0000-0002-8760-6157]{Jakob S. den Brok}
\affiliation{Institute for Particle Physics and Astrophysics, ETH Z{\"u}rich, Wolfgang-Pauli-Strasse 27, CH-8093 Z{\"u}rich, Switzerland}
\affiliation{Argelander Institute for Astronomy, Auf dem H{\"u}gel 71, D-53231, Bonn, Germany}

\author[0000-0002-7998-9581]{Michael J. Koss}
\affiliation{Eureka Scientific, 2452 Delmer Street, Suite 100, Oakland, CA 94602-3017, USA}
\affiliation{Space Science Institute, 4750 Walnut Street, Suite 205, Boulder, Colorado 80301, USA}

\author[0000-0002-3683-7297]{Benny Trakhtenbrot}
\affiliation{School of Physics and Astronomy, Tel Aviv University, Tel Aviv 69978, Israel}

\author[0000-0003-2686-9241]{Daniel Stern}
\affiliation{Jet Propulsion Laboratory, California Institute of Technology, 4800 Oak Grove Drive, MS 169-224, Pasadena, CA 91109, USA}

\author[0000-0001-5804-1428]{Sebastiano Cantalupo}
\affiliation{Institute for Particle Physics and Astrophysics, ETH Z{\"u}rich, Wolfgang-Pauli-Strasse 27, CH-8093 Z{\"u}rich, Switzerland}
\affiliation{Dipartimento di Fisica G. Occhialini, Università degli Studi di Milano Bicocca, Piazza della Scienza 3, 20126 Milano, Italy}

\author[0000-0003-3336-5498]{Isabella Lamperti}
\affiliation{Department of Physics and Astronomy, University College London, Gower Street, London WC1E 6BT, UK}
\affiliation{European Southern Observatory, Karl-Schwarzschild-Strasse 2, D-85748 Garching bei M{\"u}nchen, Germany}
\affiliation{Centro de Astrobiología(CAB, CSIC–INTA), Departamento de Astrofísica, Cra. de Ajalvir Km. 4, 28850 – Torrejón de Ardoz, Madrid, Spain}

\author[0000-0001-5742-5980]{Federica Ricci}
\affiliation{Instituto de Astrofísica, Facultad de Física, Pontificia Universidad Católica de Chile, 306, Santiago 22, Chile}

\author[0000-0001-5231-2645]{Claudio Ricci}
\affiliation{N\'ucleo de Astronom\'ia de la Facultad de Ingenier\'ia, Universidad Diego Portales, Avenida Ej\'ercito Libertador 441, Santiago, Chile}
\affiliation{Kavli Institute for Astronomy and Astrophysics, Peking University, Beijing 100871, People's Republic of China}
 \affiliation{George Mason University, Department of Physics \& Astronomy, MS 3F3, 4400 University Drive, Fairfax, VA 22030, USA}

\author[0000-0002-5037-951X]{Kyuseok Oh}
\affiliation{Korea Astronomy \& Space Science institute, 776, Daedeokdae-ro, Yuseong-gu, Daejeon 34055, Republic of Korea}
\affiliation{Department of Astronomy, Kyoto University, Kitashirakawa-Oiwake-cho, Sakyo-ku, Kyoto 606-8502, Japan}
\affiliation{Japan Society for the Promotion of Science Fellow}

\author[0000-0002-8686-8737]{Franz E. Bauer}
\affiliation{Instituto de Astrof\'{\i}sica  and Centro de Astroingenier{\'{\i}}a, Facultad de F\'{i}sica, Pontificia Universidad Cat\'{o}lica de Chile, Casilla 306, Santiago 22, Chile}
\affiliation{Millennium Institute of Astrophysics (MAS), Nuncio Monse{\~{n}}or S{\'{o}}tero Sanz 100, Providencia, Santiago, Chile}
\affiliation{Space Science Institute, 4750 Walnut Street, Suite 205, Boulder, Colorado 80301, USA}

\author[0000-0002-1321-1320]{Rogerio Riffel}
\affiliation{Departamento de Astronomia, Universidade Federal do Rio Grande do Sul Porto Alegre, Brazil}

\author[0000-0002-7608-6109]{Alberto Rodr{\'\i}guez--Ardila}
\affiliation{Laborat{\'o}rio Nacional de Astrof{\'\i}sica, Itajubá, MG, Brazil}

\author[0000-0001-5481-8607]{Rudolf B\"{a}r}
\affiliation{Institute for Particle Physics and Astrophysics, ETH Z{\"u}rich, Wolfgang-Pauli-Strasse 27, CH-8093 Z{\"u}rich, Switzerland}

\author{Fiona Harrison}
\affiliation{Cahill Center for Astronomy and Astrophysics, California Institute of Technology, Pasadena, CA 91125, USA}

\author[0000-0002-4377-903X]{Kohei Ichikawa}
\affiliation{Frontier Research Institute for Interdisciplinary Sciences, Tohoku University, Sendai 980-8578, Japan}

\author[0000-0001-8450-7463]{Julian E. Mej\'ia-Restrepo}
\affiliation{European Southern Observatory, Casilla 19001, Santiago 19, Chile}

\author[0000-0002-7962-5446]{Richard Mushotzky}
\affiliation{Department of Astronomy and Joint Space-Science Institute, University of Maryland, College Park, MD 20742, USA}

\author[0000-0003-2284-8603]{Meredith C. Powell}
\affiliation{Kavli Institute of Particle Astrophysics and Cosmology, Stanford University, 452 Lomita Mall, Stanford, CA 94305, USA}
\author[0000-0003-2704-599X]{Rozenn Boissay-Malaquin}
\affiliation{Center for Space Science and Technology, University of Maryland, Baltimore County, Baltimore, MD 21250, USA}
\affiliation{X-ray Astrophysics Laboratory, NASA / Goddard Space Flight Center, Greenbelt, MD 20771, USA}
\affiliation{Center for Research and Exploration in Space Science and Technology, NASA / Goddard Space Flight Center, Greenbelt, MD 20771, USA}

\author[0000-0001-5146-8330]{Marko Stalevski}
\affiliation{Astronomical Observatory, Volgina 7, 11060 Belgrade, Serbia}
\affiliation{Sterrenkundig Observatorium, Universiteit Gent, Krijgslaan 281-S9, Gent, B-9000, Belgium}

\author[0000-0001-7568-6412]{Ezequiel Treister}
\affiliation{Instituto de Astrof\'isica and Centro de Astroingenier\'ia, Facultad de F\'isica, Pontificia Universidad Cat\'olica de Chile, Casilla 306, Santiago 22, Chile}

\author[0000-0002-0745-9792]{C. Megan Urry}
\affiliation{Yale Center for Astronomy \& Astrophysics, Physics Department, PO Box 208120, New Haven, CT 06520-8120, USA}

\author[0000-0002-3158-6820]{Sylvain Veilleux}
\affiliation{Department of Astronomy and Joint Space-Science Institute, University of Maryland, College Park, MD 20742, USA}


\begin{abstract}

We present the BAT AGN Spectroscopic Survey (BASS) Near-infrared Data Release 2 (DR2), a study of 168 nearby ($\bar z = 0.04$, $z<0.6$) active galactic nuclei (AGN) from the all-sky \textit{Swift Burst Array Telescope} X-ray survey observed with Very Large Telescope (VLT)/X-shooter in the near-infrared (NIR; 0.8 -- 2.4~$\mu$m). We find that 49/109 (45\%) Seyfert 2 and 35/58 (60\%) Seyfert 1 galaxies observed with VLT/X-shooter show at least one NIR high-ionization coronal line (CL, ionization potential $\chi>100$ eV). Comparing the emission of the \SiVI\ CL with the X-ray emission {for the DR2 AGN},  we find a significantly tighter correlation, with a lower scatter (0.37\,dex) than for the optical \OIII\ line {(0.71\,dex)}. We do not find any correlation between CL emission and the X-ray photon index $\Gamma$. We find a clear trend of line blueshifts with increasing ionization potential in several CLs, such as \SiVI, \SiX, \SVIII, and \SIX, indicating the radial structure of the CL region. Finally, we find a strong underestimation bias in black hole mass measurements of Sy 1.9 using broad H$\alpha$ due to the presence of significant dust obscuration. In contrast, the broad Pa$\alpha$ and Pa$\beta$ emission lines are in agreement with the $M$--$\sigma$ relation.
{Based on the combined DR1 and DR2 X-shooter sample, the NIR BASS sample now comprises 266 AGN } {with rest-frame NIR spectroscopic observations, the largest set assembled to date.}
\end{abstract}




\section{Introduction} \label{sec:intro}


{Active galactic nuclei (AGN) are accreting, supermassive black holes (SMBHs) located in the center of certain galaxies. They can be among the most luminous, nontransient objects in the known universe \citep{Banados2018}.} While AGN spectra have been extensively analyzed in many wavelength regimes from radio to gamma rays, the rest-frame near-infrared (NIR) wavelength regime ($0.8 - 2.4~\mu$m) has, to date, only been sparsely studied.
{Early works include studies of large samples (27 sources, \citealt{Glikman2006}; 47 sources, \citealt{Riffel:2006}; 23 sources, \citealt{Landt2008}). Over the past few years, studies have increased the number of sources investigated (50 sources, \citealt{Mason2015}; 41 sources, \citealt{Onori2017}; 102 sources, \citealt{Lamperti2017}; 40 sources, \citealt{Muller2018}).} Spectroscopic NIR observations are advantageous because the NIR wavelengths are less susceptible to interstellar dust extinction by up to a factor 10 as compared to the optical regime (\citealt{Goodrich1994}; \citealt{Veilleux1997}; \citealt{2002Veilleux}), allowing more obscured AGN to be studied (e.g. \citealt{Lamperti2017}).
The NIR band also contains a wealth of emission lines that can help to characterize the ionization structure of the material that may eventually feed the accreting SMBH. 

{Hydrogen Pa$\alpha$ ($\lambda = 1.8751$$\mu$m) and Pa$\beta$ ($\lambda = 1.2818$$\mu$m) are prominent emission lines that are regularly found in the NIR regime.}
Previous studies have used these lines to derive black hole mass estimates ($M_{\rm BH}$) based on their width and strength (\citealt{Kim2010}; \citealt{Landt2013}; \citealt{Kim2015}; \citealt{LaFranca2015}; \citealt{Ricci_F2017}). In certain sources, broad NIR line components have been detected in galaxies that lack broad H$\alpha$ or H$\beta$ (e.g., \citealt{Goodrich1994}; \citealt{Veilleux1997}; \citealt{Smith2014}; \citealt{Lamperti2017}). This is explained by dust obscuration within the host galaxy. Consequently, the Paschen lines provide an additional way to derive black hole masses for obscured AGN \citep[e.g.][]{Ricci_F2017}. Furthermore, NIR [Fe \textsc{ii}] emission lines can be used to study physical characteristics, as they give important clues on the detailed structure of the emitting gas and they constitute important cooling lines \citep{Riffel2013,Marinello2016}.
In addition, several high-ionization coronal lines (CLs; ionization potential $\chi > 100$ eV) can be found in the NIR spectral region, such as \SiVI, \SiX, \SIX, \SVIII, and \FeXIII. {But CLs are not just unique to the NIR regime. They can also be found in the optical spectral region (e.g. [Ne \textsc{v}] $\lambda3425$ and [Fe \textsc{vii}] $\lambda6087$; see \citealt{Mazzalay2010}) or the mid-IR region (e.g. [Ne \textsc{v}] $14.3\,\mu$m; see \citealt{Sturm2002})}. {Because of their high-ionization potential (IP), CLs are hard to produce in starburst regions \citep{Marco2005}. While type II supernovae can also cause CL emission \citep{Komossa2009}, the lines are generally weak and short-lived \citep{Izotov2009}. Since CLs mostly survive only very close to a hard ionization source, they are generally unique tracers of AGN.} {A} proposed mechanism for producing these lines is a strong, central source of intense ionizing continuum in the energetic ultraviolet (EUV) and soft X-ray bands that photoionizes the species (\citealt{Shields1975}, \citealt{Rodriguez2011}). Another proposed mechanism is shocks of high-velocity gas clouds that interact with the narrow-line region (NLR) gas (\citealt{Osterbrock1964}, \citealt{Oke1968}). These shocks heat the gas to high temperatures $T\ge10^6$ K \citep{Olivia1997}.  {With greater sensitivity of observations, however, emission mechanisms such as shocks} { could produce detectable CL emission in the absence of AGN, though only rarely in some of the highest star formation} { mergers in nearby luminous infrared galaxies \citep{Rich2011}.}
Finally,  both mechanisms may occur simultaneously to explain the observed line ratios \citep{Rodriguez2006,Geballe2009,Rodriguez2011}.
If photoionization is the main generator of these emitting species, a hard radiation field is needed in order to consistently match up the levels of ionizing photons required to produce CL emission \citep{Olivia1997}. This is consistent with the EUV and soft X-rays seen in many AGN, suggesting that CL emission scales with the AGN X-ray emission.

The main interest in NIR CLs mainly derives from the fact that they may be used to detect AGN in dusty environments because of the lowered effect of extinction in the NIR. {In the UV to IR regime, the dominant source of obscuration is dust, while high columns of gas are the most important cause of extinction in the X-ray \citep[see the review by ][]{Hickox2018}.} Theoretical arguments indicate that the accretion rate onto SMBHs peaks during the period when the AGN is obscured  by dust and gas (e.g. \citealt{Hopkins2009}). Furthermore, hard X-ray observations show that a large fraction of SMBHs are located in gas-rich \citep[e.g.,][]{Koss:2013,Koss:2021:29}, dusty nuclei of galaxies \citep[e.g.][]{Koss:2011:57}, and a large fraction are obscured by high columns of gas \citep[e.g.][]{Brandt2015,Kocevski2015,Koss2016,Ricci2017MNRAS}. {This is further highlighted by the fact that recent \textit{NuSTAR} observations have found an increasing number of nearby, low-luminosity, Compton-thick AGNs \citep[e.g.][]{Annuar2015, Ricci2016, Annuar2017}. Finding and correctly identifying obscured AGN has implications for observational cosmology. As a majority of the AGN population is obscured, a complete census of all sources, obscured and unobscured, is needed to correctly constrain the evolution of SMBH growth over cosmic times. With the advent of the \textit{James Webb Space Telescope} (\textit{JWST}), it will be possible to perform infrared spectroscopic observations with an unprecedented sensitivity \citep{Gardner2006}. NIR CLs thus provide several advantages for the identification of AGN activity. }

In this work, we investigate CL emission from  AGN selected above 10 keV from the Burst Array Telescope (BAT) on the \textit{Neil Gehrels Swift Observatory}. 
We examine the properties of CLs in the largest sample of AGN with NIR spectra to date with the goal of learning about the physical mechanisms behind their production. An additional goal is to determine the rate of appearance of such lines in the NIR to determine their variability as a robust tracer of AGN activity. 
For the distance calculations in this work, we use the concordance cosmological model with $\Omega_M = 0.3$, $\Omega_\Lambda = 0.7$ and $H_0$ = 70 km~s$^{-1}$Mpc$^{-1}$. 


\section{Survey Description, Samples, and Data Reduction}
    \subsection{Sample}
          {The BAT AGN Spectroscopic Survey (BASS) project\footnote{\href{https://www.bass-survey.com/}{https://www.bass-survey.com/}} is a collaborative effort to characterize a complete survey of local hard X-ray selected AGN {\citep{Koss_2017, Ricci_2017}}, based on the \textit{Swift}--BAT all-sky survey.}
          This 105 month \textit{Swift}--BAT all-sky survey has identified 1632 objects, of which 1105 (68\%) are AGN \citep{Oh_2018}. {Due to the hard X-ray (14$-$195 keV) AGN selection, the sample is nearly unbiased with respect to obscuration up to Compton-thick AGN {\citep{Ricci2015,Koss2016}} { and very faint AGN due to X-ray flux limits {\citep[e.g][]{Ichikawa2017}}}.}  {For the second data release(DR2),} Very Large Telescope (VLT) X-shooter observations in queue mode were obtained for {269} 
          AGN over several semesters (098.A-0635, 099.A-0403, 0101.A-0765, 0102.A-0433, and 0103.A-0521; these were  carried as filler programs), focusing on Type 1.9 or Type 2 AGN or newly identified AGN.  
          {A key goal of the high spectral resolution was to measure black hole masses from velocity dispersions in Type 1.9 or Type 2 AGN \citep{Restrepo_DR2_mbh, Koss_DR2_overview}, but the NIR arm also provides access to less obscured features.}
          The median seeing was $1\farcs0$ based on the  {Differential Image Motion Monitor in the $V$ band with a standard deviation of 0\farcs7}. A summary and information on the individual observations can be found in \autoref{tb:summary_obs} in appendix \ref{app:obsdat}.
          
\begin{figure}
            \centering
            \includegraphics[width =\textwidth]{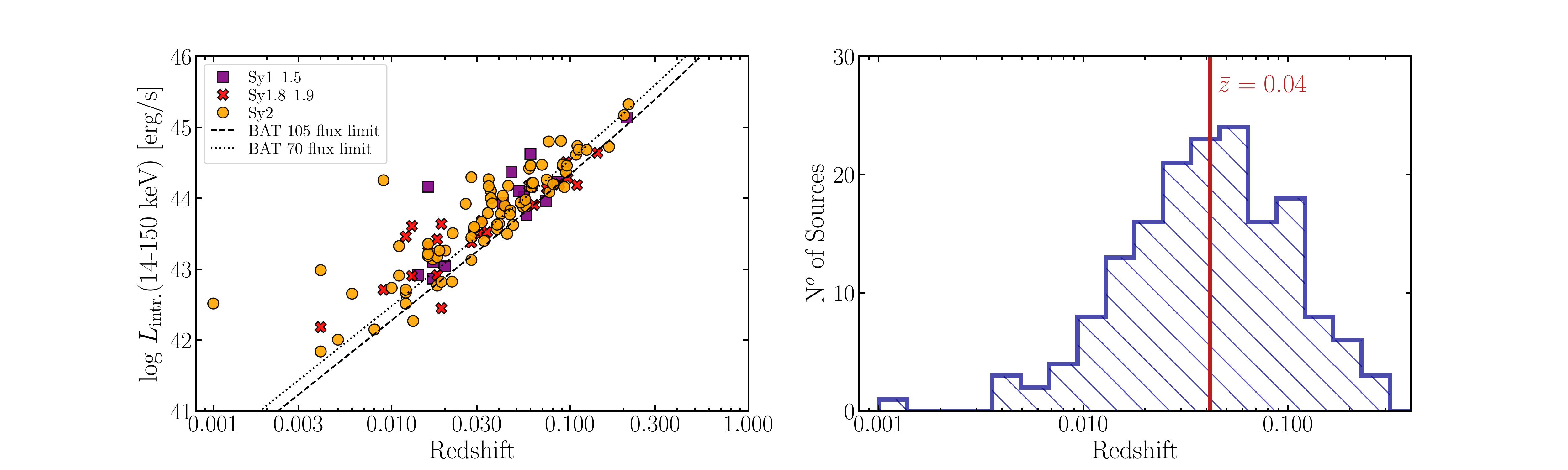}
            \caption[Distribution of the X-ray luminosity of the BAT-detected AGN]{(Left) Distribution of the X-ray luminosity of the BAT-detected AGN used in this work as a function of redshift. The dashed line shows the flux limit of the 105 month BAT all-sky survey for 90\% of the sky ($8.4\times10^{-12}$~erg~cm$^2$~s$^{-1}$) and the dotted line shows that for the 70 month survey ($1.34\times10^{-11}$~erg~cm$^2$~s$^{-1}$). (Right) Redshift distribution of our sample. The median redshift is $\bar z = 0.04$, {which is consistent with the median of the parent BASS sample \citep[i.e., $\bar z = 0.037$,][]{Koss_DR2_overview}.}}
            \label{fig:redsh_distrbution}
        \end{figure}

          {From the X-shooter DR2, we only selected nearby AGN ($z<0.5$) and excluded beamed AGN \citep{Vaidehi2019}} { to avoid sources with differential beaming of the X-ray emission.}
          Additionally, we have included 10 archival observations of BAT AGN in our sample (e.g. from 086.B-0135) fulfilling the conditions mentioned above (low redshift and no beamed AGN). 
          Our final sample totals 
          168 unbeamed AGN (\autoref{tb:summary_obs}), of which 110/168 (66~\%) are Seyfert 2, 28/168 (17~\%) are Seyfert 1.9 and 30/168 (18~\%) are Seyfert 1 -- 1.8 type AGN with broad \Hbeta. The final sample is biased toward Seyfert 1.9 and Seyfert 2 AGN compared to BAT-detected AGN, which show equal fractions of Type 1 and Type 2 AGN \citep{Koss_2017}.  
           {Depending on the spectral setup of the instrument, this includes sources with either $z>0.3$ (if the full NIR range of 9940 -- 24\,790 was covered) or $z>0.1$ (for which we only have limited NIR coverage of {9940} -- 21\,010)}.
          \autoref{fig:redsh_distrbution} (left) shows the hard X-ray versus redshift plane of the sample of AGN for which NIR spectra were obtained. \autoref{fig:redsh_distrbution} (right) shows the redshift distribution of our sample.

          {For completeness, we include all sources from BASS NIR data release 1 (DR1) \citep{Lamperti2017} in our analysis. This sample consists of 102 NIR spectra of nearby AGN from several observation programs. Most of the sources (55/102) were observed from the 2.2 m NASA Infrared Telescope Facility telescope, with  resolution of R = 800 -- 1000. Seven of  the 102 sources were taken with the Florida Multi-object Imaging Near-IR Grism Observational Spectrometer (FLAMINGOS) at the Kitt Peak 4m telescope. Additional sources were taken from archival data from Gemini. The DR1 sample shows a bias toward Seyfert 1 galaxies (${\sim}${68}\% are Seyfert 1$-$1.9), due to the setup of the archival surveys. We refer the reader to \cite{Lamperti2017} for a full description of the sample.
          The total NIR BASS DR1+DR2 sample consists of 266 BAT-detected AGN (four AGN overlap between the samples).}

          The sample of additional DR2 data of reduced spectra will be made public on the BASS survey website.  We note that  the BASS follow-up with X-shooter is ongoing; in this study we use X-shooter observations taken through 2019 October 13.  {The additional X-shooter observations taken since 2019 October 13 will be presented in later BASS releases and are part of ongoing European Southern Observatory (ESO) programs.  The additional data will include other NIR spectroscopy efforts within BASS that are currently ongoing, including follow-up of 65 BASS AGN with Magellan/FIRE \citep{Ricci_DR2_NIR_Mbh} and with Palomar/Triplespec (M. Balokovic, in preparation).}
        
    \subsection{Observational Setup}
     \label{sec:obs_set}
        The observations were all carried out with X-shooter, a multiwavelength  (0.3~--~2.5~$\mu$m) echelle spectrograph with medium spectral resolution $R = 4000-18\,000$  \citep{Dodorico2006, Vernet_2011}. 
        It has three spectroscopic arms,  each equipped with optimized optics, dispersive elements, and detectors. Two dichroics are used to split the incoming light into the three arms for efficient observation of all three arms simultaneously. The NIR arm has a wavelength coverage of 1 -- 2.5~$\mu$m\ and includes the traditional atmospheric bands \textit{J}, \textit{H}, and \textit{K}.
    
        The bulk of the observations were carried out between 2017 and 2018. {A summary of all observations is listed in \autoref{tb:summary_obs}}. Two spectral setups were chosen for the NIR arm: 
        for 84/168 (50\%) we obtained full coverage between 0.994 and 2.479 $\mu$m, while the other 84/168 (50\%) had more limited coverage between {0.994} and 2.101 $\mu$m.
        The slit width was set to 0\farcs9, giving a spectral resolution of {$R \sim 5400$} (note that in one archival observation, a slit width of 0\farcs4 was used; see \autoref{tb:summary_obs}). For most observations the typical total integration time was set to 480 s (for 60/168) or 960 s (for 71/168). {To remove the thermal background, sky emission lines, and detector artifacts,} the science targets were observed in two positions on the slit in an ABBA nodding sequence\footnote{ \url{https://www.aanda.org/component/article?access=doi&doi=10.1051/0004-6361/201936825}} with a 5\arcsec\ nod throw. In one archival source a STARE observation was used.
        
        {We also obtained an independent estimate of the spectral resolution with the penalized pixel fitting method \citep[ppxf;][]{Cappellari2004, Cappellari2017} by fitting stellar absorption lines to individual stars that were observed with the 0.9\arcsec\ slit} {with a default pipeline extraction of 4\arcsec along the slit.}  { To measure the X-shooter spectral resolution, we followed the approach of \cite{Gonneau2020}, which was also used for measuring} { resolutions in the other BASS DR2 optical spectra \citep[][]{Koss_DR2_overview}}. {We use the $\mathtt{PHOENIX}$ theoretical spectral library \citep{Husser2013} as templates, which have much higher resolution ($R\sim500,000$) than the observations.  We fit the 1.45--1.78 $\mu$m and 2.285--2.38 $\mu$m regions, respectively, to target stellar absorption features in the CO bandheads in the $H$ and $K$ bands.  In five different stars, we measured $\sigma$=$20\pm1$ \kmps.  This corresponds to $R=6150$, or an FWHM of 0.00026 $\mu$m at 1.6 $\mu$m, slightly better than the nominal instrumental resolution listed in the manual.}

    \subsection{Data Reduction}
        The spectra were first reduced using the standard pipeline in the ESO reflex software \citep{Freudling2013}. Pipeline v2.9.3 was used in all of the sources presented in this paper. We used the default parameters for the creation of the calibration frames.  We used the \texttt{xsh\_scired\_slit\_nod} recipe\footnote{\url{https://www.eso.org/sci/facilities/paranal/instruments/X-shooter/doc/VLT-MAN-ESO-14650-4942_P105v1.pdf}} to transform the science and flux-standard frames into flat-fielded, rectified, and wavelength-calibrated 2D-order spectra. {The standard 4$\arcsec$ extraction region along the slit was used for each spectrum.} One of the X-shooter spectrophotometric standard stars was selected\footnote{List of standard stars given here: \url{https://www.eso.org/sci/facilities/paranal/instruments/X-shooter/tools/specphot_list.html}} for the flux calibration. We corrected atmospheric absorption features that contaminated the spectra (H$_2$O, CO$_2$, CH$_4$ and O$_2$) using the software tool \texttt{molecfit} \citep[v1.5.9;][]{Kausch2015,Smette2015}.   \texttt{Molecfit} uses a radiative transfer code to simulate the atmosphere adopting the observed atmospheric parameters including the ambient temperature, pressure, mirror temperature, and outside humidity.
        
        For the software to work properly the observed spectrum needs to have distinctive, but not saturated telluric features for correction and should avoid intrinsic emission or absorption features from the AGN.
        With \texttt{molecfit}, no observation time needs to be allocated to telluric standard stars, and because \texttt{molecfit} simulates the atmosphere, small atmospheric changes over a night are better accounted for \citep{Ulmer2019}.


\section{Spectroscopic Measurements}
    \subsection{NIR Emission Line Measurements}
    
        For the emission line fitting, the software tool \texttt{PySpecKit} { \citep[v0.1.20;][]{Pyspeckit}} was used following the procedure of \cite{Lamperti2017}. The software is an extensible, spectroscopic toolkit. The fitting procedure relies on the Levenberg--Marquardt algorithm. For the modeled emission lines, a single Gaussian profile is used, or the combination of two Gaussian profiles if the second is detected above 2$\sigma$ {above the standard deviation in amplitude above the noise}. 
        Before we fit the spectra, we first correct for Galactic extinction, using the built-in \texttt{deredden} function, which takes the $E_{B-V}$ value into consideration { \citep[values from][]{Schlegel1998}}.
        The following physical quantities are fitted: the width $\sigma_{\rm line}$ and  height/amplitude $A_{\rm line} $, as well as the wavelength position $\lambda_{\rm line}^{\text{obs}}$ of the Gaussian profile. {
        For the final line FWHM measurement, we subtract the instrumental dispersion (i.e., ${\sim}56\,{\rm km\,s^{-1}}$) in quadrature.}
        
        In order to facilitate the fitting procedure, the NIR spectrum is split into smaller wavelength regions to best fit the varying continuum. 
        The separately fitted regions are (see \autoref{tab:spec_reg}) Pa$\epsilon$ (0.94$-$0.98 $\mu$m), $[$S \textsc{viii}$]$ (0.97$-$1.0 $\mu$m), Pa$\gamma$ (1.0$-$1.15 $\mu$m), Pa$\beta$ (1.15$-$1.35 $\mu$m), $[$Si \textsc{x}$]$ (1.4$-$1.5 $\mu$m) and Pa$\alpha$ (1.8$-$2.02 $\mu$m).
        An example where all lines are fitted successfully is presented in appendix \ref{sec:fits} in \autoref{fig:fitting_plts}.
        The reason why the $[$S \textsc{viii}$]$ spectral region is fitted separately and not included in the Pa$\epsilon$ region is that the spectra are cut at 1 $\mu$m, meaning that depending on the redshift, part of the region 9400 -- 10\,000 might be in the NIR arm and part of it in the VIS arm. By separating the $[$S \textsc{viii}$]$ region, issues from the separation of  spectra and flux calibrations are minimized. {The emission lines we fit in the NIR regime are described in appendix \ref{sec:em_lines}.}
        
         The first step in fitting the emission lines is determining the continuum of the spectrum.  To allow for more flexibility, especially in telluric-corrected regions with possible residuals, a fourth-order polynomial is fitted to the spectrum. {Fitting the AGN continuum using a fourth-order polynomial has been done in several previous studies  \citep[e.g][]{Krajnovic2007,Raimundo2013,Zeimann2015,Husemann2020}}. The continuum level is estimated individually for each of the specific spectral regions (described in \autoref{tab:spec_reg}). Emission lines and heavily affected telluric regions are masked.

        \begin{table}
        \centering
        \caption{Overview of the Different spectral regions.}
        \label{tab:spec_reg}
        \begin{tabular}{c c}
             {Spectral Region} &  {Wavelength Range [$\mu$m]}\\ \hline
             Pa$\epsilon$ & 0.94 -- 0.98 \\
             $[$S \textsc{viii}$]$ & 0.97 -- 1.0 \\
             Pa$\gamma$ & 1.0 -- 1.15 \\
             Pa$\beta$ & 1.1 -- 1.35 \\
             $[$Si \textsc{x}$]$ & 1.4 -- 1.5 \\
             Pa$\alpha$ & 1.8 -- 2.02
        \end{tabular}
        
    \end{table}
    
       In certain cases, the continuum shape is irregular. Either the intrinsic continuum or the telluric correction residuals cause an irregular continuum shape and a spline fit is used to estimate the continuum level. {In 123 spectral regions for 88/168 (52\%) AGN mainly due to strong telluric residuals a spline fit is applied to correct for the continuum. }  An example where lines are heavily affected by tellurics and the spline fit is applied is presented in \autoref{fig:fitting_plts_2} (see Pa$\alpha$ region in bottom panel). In Appendix \ref{app:spline}, we provide more details on the spline fit.
        
        The emission lines are fit using Gaussian profiles. We distinguish between narrow lines (FWHM < 1200 km s$^{-1}$) and broad lines (FWHM > 1200 km s$^{-1}$). A broad component is only allowed for the hydrogen recombination lines (Pa$\alpha$, Pa$\beta$, Pa$\gamma$, and Br $\delta$), the strong He \textsc{i} lines, and the [S \textsc{iii}]$\lambda$9531 emission line, which is the strongest narrow line in the NIR wavelength range. For [S \textsc{iii}]$\lambda$9531 we also use a third, blueshifted component, which is empirically motivated. {Such a blueshifted component is for example also seen in the bright \oiii\ emission lines \citep[e.g.][]{Rojas2020}.} {The other NIR lines we fit do not show evidence of significant blue shifted narrow-line components.}

        As a first step, the Pa$\beta$ region is used to set constraints on the width and offset of the other emission lines. The relative velocity centers of the narrow lines are tied together and the width of the strongest narrow line is used to constrain the width of the other narrow lines in velocity space (with an allowed difference of 200 km s$^{-1}$ {for narrow and 500 km s$^{-1}$ for broad lines}). If no line is found in the Pa$\beta$ range, the Pa$\gamma$ range is used instead to constrain parameters. 
        The broad lines are similarly tied together, if detected in the Pa$\beta$ or Pa$\gamma$ region. The broad component's centroid wavelength can be shifted by a larger amount. This is empirically motivated by a study of a large sample of AGN looking at shifts of H$\beta$ with respect to the systemic redshift \citep{Shen2017} finding shifts up to 1000 km s$^{-1}$ with a mean velocity shift of 109 km s$^{-1}$.
        For high-ionization lines, the allowed offset (of the line's position and width) is set to 400 km s$^{-1}$, motivated by observations that these lines tend to be blueshifted.
        
        For a detection, the amplitude $A_{\rm line}$ has to be above a certain threshold $n \cdot \sigma_{n}$, where $\sigma_n$ is the noise level of the surrounding continuum and $n$ is the targeted threshold limit. For the determination of the noise level, a window of 0.015\,{$\mu$}m toward the blue and red of where the line is expected to be, while masking the line itself, is used to calculate the rms value. The threshold is set to $n=3$ with a width set to the FWHM of other more prominent emission lines.
        Thus the sample is equivalent width rather than flux limited. For nondetected \SiVI\ emission lines, we determine upper limits using $F_\text{UL} = 3\sigma$.
        
        All fits are inspected visually to see whether the lines were fit well. In 14 cases manual intervention is needed for a high-ionization line because a residual is fitted instead of an actual emission line. In 16 cases, the emission line needs to be fitted manually because of complications with the surrounding noise of the telluric correction. 
        
        Errors in the fitted parameters are estimated by performing 20 Monte Carlo simulations drawn from a normal distribution with a standard deviation equal to the noise level in the spectrum. {The full table with the measurements is described in Appendix \ref{sec:measdata}.} \\
        
    \begin{figure*}
        \centering
        \includegraphics[width = 0.45\textwidth]{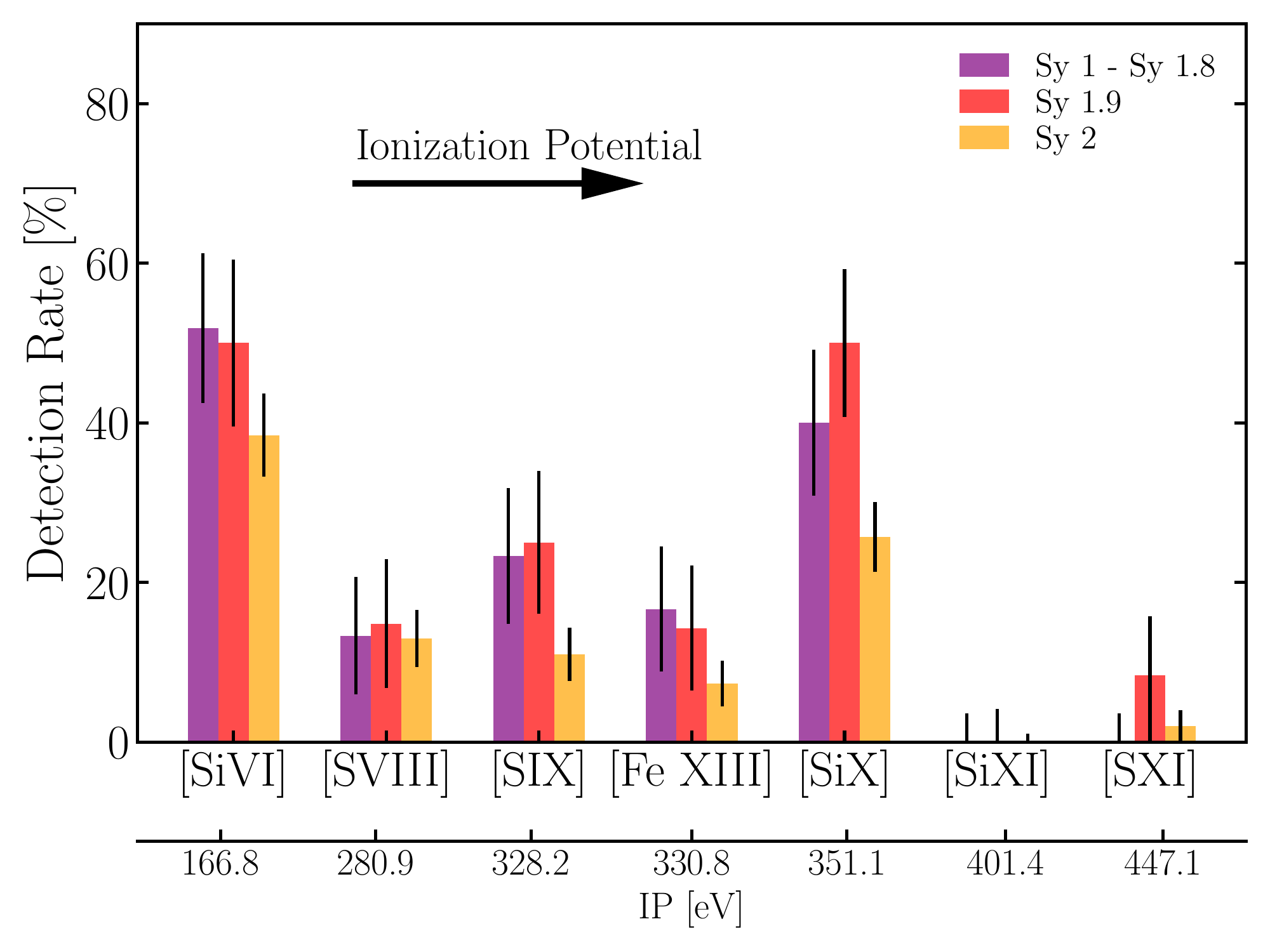}
        \includegraphics[width = 0.45\textwidth]{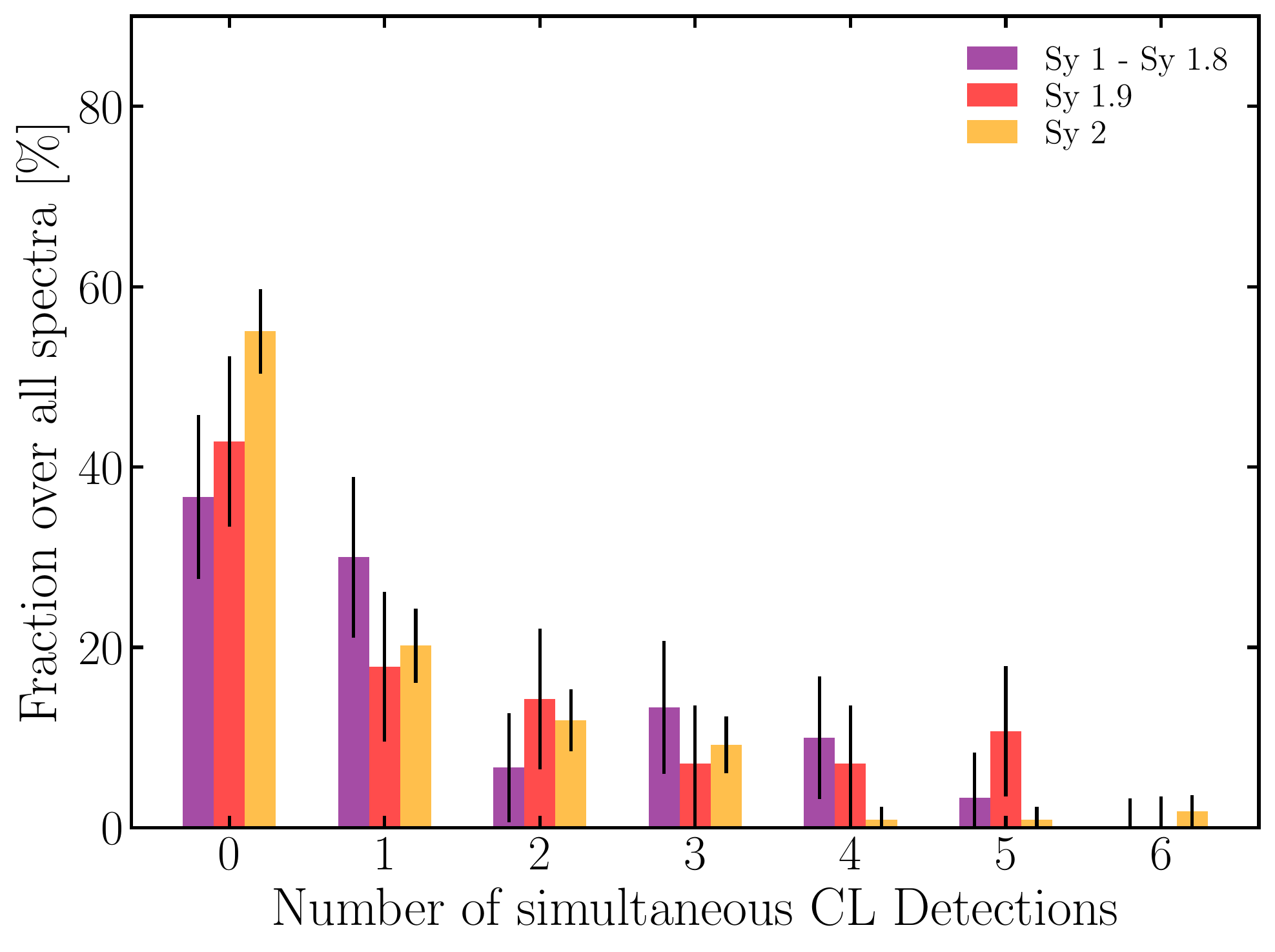}
        \caption{(Left) The percentage of detection of high-ionization lines by AGN type (Seyfert 1--1.8, Seyfert 1.9, and Seyfert 2) sorted according to their IP. (Right) Number of CL detections in a single spectrum separated by AGN type (i.e., Seyfert 1 and 2). In 54/110~(49\%) Seyfert 2 galaxies one or more high-ionization emission lines are detected. {The error bars are estimated using a 1$\sigma$ binomial proportion confidence interval}. }
        \label{fig:detec_rate}
    \end{figure*}
    \begin{figure}
        \centering
        \includegraphics[width = 0.4\textwidth]{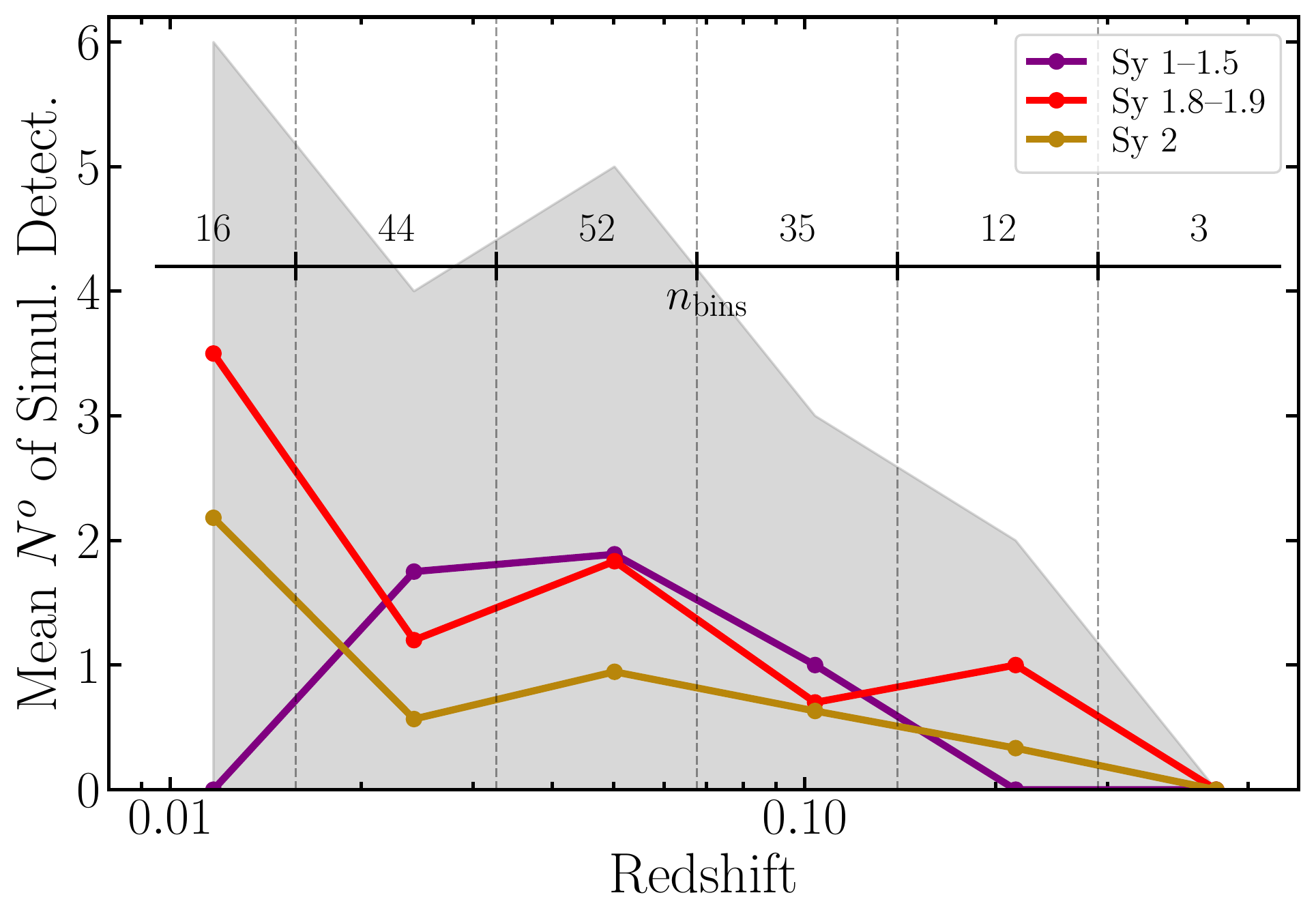}
        \caption{Average number of simultaneously detected CLs binned by redshift. {The gray shaded area shows the range in number of detections for each of the different sources in a given bin}, while the colored lines indicate the mean of detections per redshift bin. {The number of sources per bin is shown.} The dashed lines indicate the boundaries of the redshift bins.}
        \label{fig:red_num_dist}
    \end{figure}
 
        \subsubsection{Black Hole Mass Estimation}
        For narrow-line sources the black hole mass measurements used in this paper are obtained from  velocity dispersion measurements using the Ca \textit{H} + \textit{K}, Mg \textsc{i}, or Ca \textsc{ii} triplet (around 0.845 -- 0.87 $\mu$m) absorption lines using the $M$--$\sigma$ relation \citep{Kormendy2013}. {The method is described in detail in \cite{Koss_DR2_sigs}, which is part of this special ApJ series.}   For broad-line sources, black hole masses are obtained from Balmer lines (mostly H$\beta$; see \citet{Restrepo_DR2_mbh} for a description of methods).
        In a future paper, the CO bandheads in the NIR $H+K$ bands will be used to estimate the mass. For 138/168 (82\%) of the sample, black hole mass estimations are available from either Paschen lines or optical velocity dispersions or both.
        
        In certain cases, broad NIR emission line components are detected while optical broad Balmer lines are not. For these cases, we use the width and strength of so-called NIR hidden broad lines to estimate the black hole mass and compare the result with the values from other methods described above.  
        Specifically, we use the Pa$\alpha$ and Pa$\beta$-based prescriptions from \cite{Kim2010}. 
        {{We scale down these mass prescriptions by $-0.13$ dex, to bring them into agreement with the virial factor of $f=1$ used throughout the BASS/DR2 analyses.}}\footnote{{
        Throughout the BASS/DR2 analyses, a virial factor of $f=1$ is used for virial $M_{\rm BH}$ estimates that rely on the FWHM of broad emission lines. If one uses the respective line velocity dispersion ($\sigma$) instead, this choice would correspond to $f_\sigma = 5.5$, assuming a Gaussian line profile. 
        The Paschen line prescriptions in \cite{Kim2010} are calibrated against H$\alpha$-based $M_{\rm BH}$ estimates from \cite{Greene2005}, which in turn assume $f=0.75$. \cite{Kim2010} corrected scaled these up by a factor 1.8 (0.26 dex), while the BASS/DR2-wide choice of $f=1$ reflects a correction by a factor of only $1/0.75=4/3$ (0.125 dex). 
        To bring the \cite{Kim2010} prescriptions into agreement with the BASS/DR2-wide mass prescriptions, we scale them down by a factor of $1.8 / (4/3) = 1.35$ (0.13 dex).
        } 
        } 
        {{Although there is a range of relevant virial factors discussed in the literature, generally in the range $f\approx0.7{-}1.1$ \cite[e.g.,][and references therein]{Greene2005,LaFranca2015,Woo2015,Yong2016,Restrepo2018}, we stress that the differences between them are much smaller than the scatter that dominates the resulting black hole mass estimates in our present analysis (see below).
        }} 
        The resulting \mbh\ prescriptions are therefore
        \begin{align*}
            \text{Pa~}\alpha: \frac{M_{\rm  BH}}{M_{\odot}} &= 10^{7.16}\left(\frac{L_{\rm  Pa\alpha}}{10^{42}{\text{erg s}^{-1}}}\right)^{0.43}\times~~~ \left(\frac{\text{FWHM}_{\rm Pa\alpha}}{10^3 \text{km s}^{-1}}\right)^{1.92}\\
            \text{Pa~}\beta:\frac{M_{\rm BH}}{M_{\odot}} &= 10^{7.20}\left(\frac{L_{\rm Pa\beta}}{10^{42}{\text{erg s}^{-1}}}\right)^{0.45}\times~~~ \left(\frac{\text{FWHM}_{\rm  Pa\beta}}{10^3 \text{km s}^{-1}}\right)^{1.69} \, .
        \end{align*}
        
        \noindent  
        In cases where both broad H$\alpha$ and broad Paschen line measurements are available, we can compare the mass estimates from the Paschen lines with the H$\alpha$ emission line. 
        {{The \mbh\ estimates based on the broad H$\alpha$ emission line are taken from \citet{Restrepo_DR2_mbh}.
        They used the prescription from \cite{Greene2005}, but scaled up by  $4/3$ (0.125 dex) so it corresponded to the virial factor $f=1$.}}
        
        \subsubsection{Ancillary Measurements}
        \label{sec:anc_data}
        In addition to the NIR line measurements, we use X-ray data as well as the \OIII\ emission line, which is located in the optical rest-frame regime. {The  \OIII\ observations are from the same X-shooter} { spectrum; hence instrumental offsets and differences from the NIR in emitting regions are minimized.}  {We note that we did not account for aperture effects in the different slit sizes of the \OIII emission (1.6\arcsec\ in the UVB} {arm) and CL emission in the NIR (0.9\arcsec)}. {However, as shown by \cite{Berney2015}, such aperture effects are negligible} { even in the more extended \OIII emission, since the emitting region is} {very concentrated. Likewise for the more } { compact NIR CL emission, adaptive optics integral-field units (IFU) studies have found the emission to extend to as much as 150 pc \citep{Mueller2011},} { which would correspond to missing extended emission only in $z<0.007$ AGN with a 0.9\arcsec\ NIR slit, which represents only 6/168 AGN of} { our sample suggesting aperture effects are very minimal for the very nearest of our AGN.}          The \OIII\ emission line measurements are presented in a companion paper \citep[][]{Oh_DR2_NLR}. They are detected in the optical data of the X-shooter observations used in our study and have been corrected for Galactic extinction in the same manner.
        The intrinsic X-ray luminosity and column density $N_{\rm H}$ are determined using X-ray observations from \textit{Swift}--BAT in combination with soft X-ray telescopes such as \textit{XMM-Newton}, \textit{Suzaku}, \textit{Chandra}, and \textit{Swift}--XRT (see \citealt{Ricci_2017} for a description of the models). The \textit{Swift}--BAT telescope provides the observed 14 -- 195 keV flux. Additionally, spectral fitting of AGN-specific models to the combined X-ray spectra provides intrinsic luminosities and column density estimates for 116/168 (69\%) AGN. {X-ray spectral fitting of all 105 month sources will be included in a future release (C. Ricci et al. in preparation). As shown in \cite{Koss:2016:85} and \cite{Ricci_2017}, for BAT observations, the observed flux significantly underestimates the intrinsic flux for $N_{\rm H}>10^{24}{\rm \, cm^{-2}}$ which only affects a small number of sources (only 7.6\% of the full BASS sample are Compton-thick AGN; \citealt{Ricci2015}}). {Because we do not yet have intrinsic flux measurements for the complete sample, we will use the intrinsic 14 -- 195 keV flux for X-ray luminosity measurements for sources of the 70\,month sample (116/168), and use the observed 14 -- 195 keV flux for the remaining (52/168) sources. In practice the observed BAT 14 -- 195 keV flux is significantly different (i.e., $>$20\%) for Compton-thick AGN which are rare in the Swift sample \citep[i.e., 7.6\%; see][]{Ricci2017}, which would only correspond to $\sim$3 sources in our sample of 52 observed 14 -- 195 keV fluxes.} {For the derived intrinsic X-ray luminosity the error is $<0.1$ dex \citep{Lanz2019}, unless the AGN are} { Compton-thick, for which the typical errors are $0.4$ dex \citep{Ricci2015}. The typical uncertainty for the observed X-ray luminosity is} { ${\sim}0.25$ dex \citep{Ricci_2017}.}  {In this study, when talking about the \textit{``hard X-ray''} flux we are refering to the 14 -- 195 keV X-ray flux.}


\begin{figure*}
            \centering
            \includegraphics[width =     0.44\textwidth]{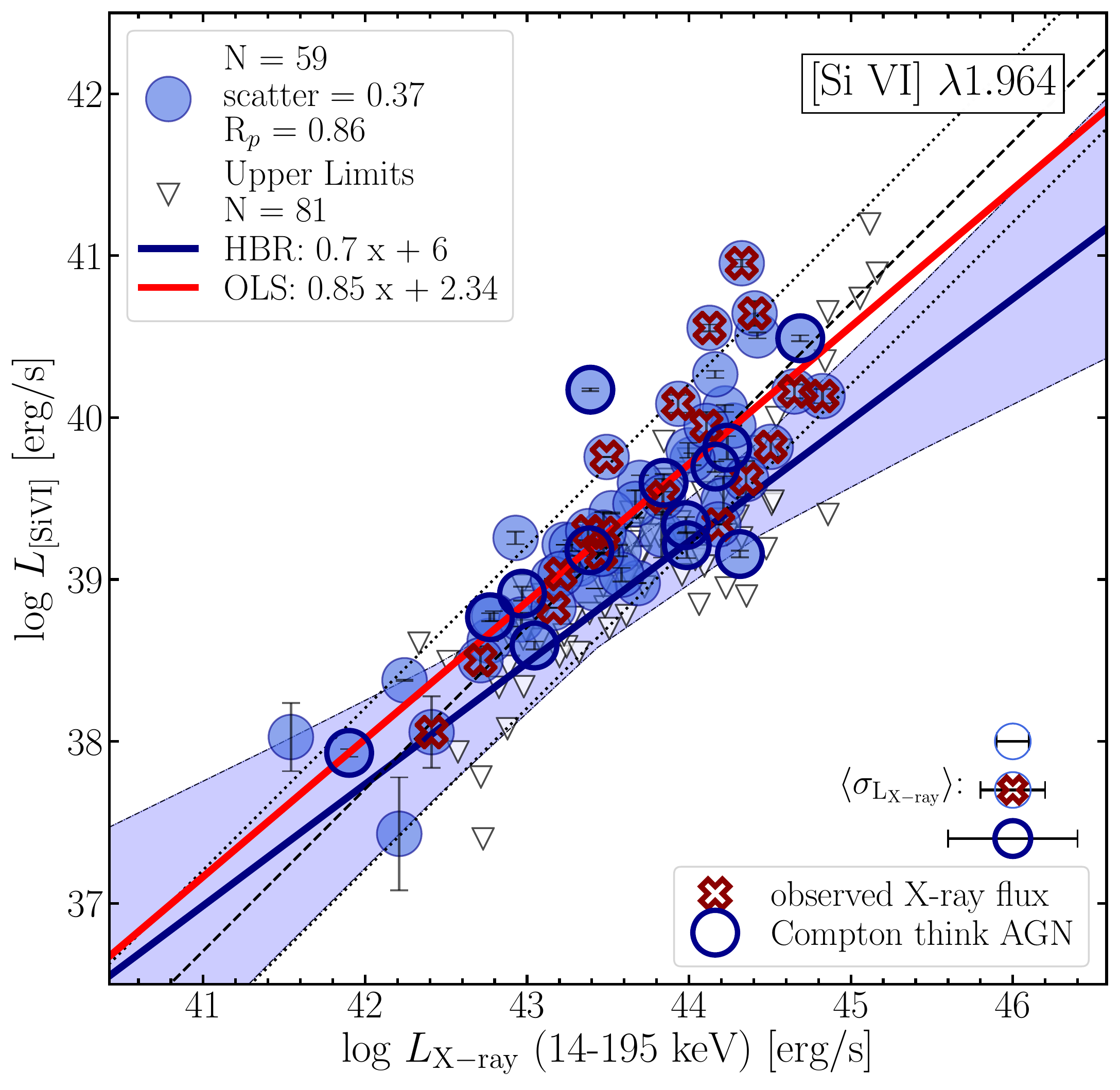}
            \includegraphics[width = 0.44\textwidth]{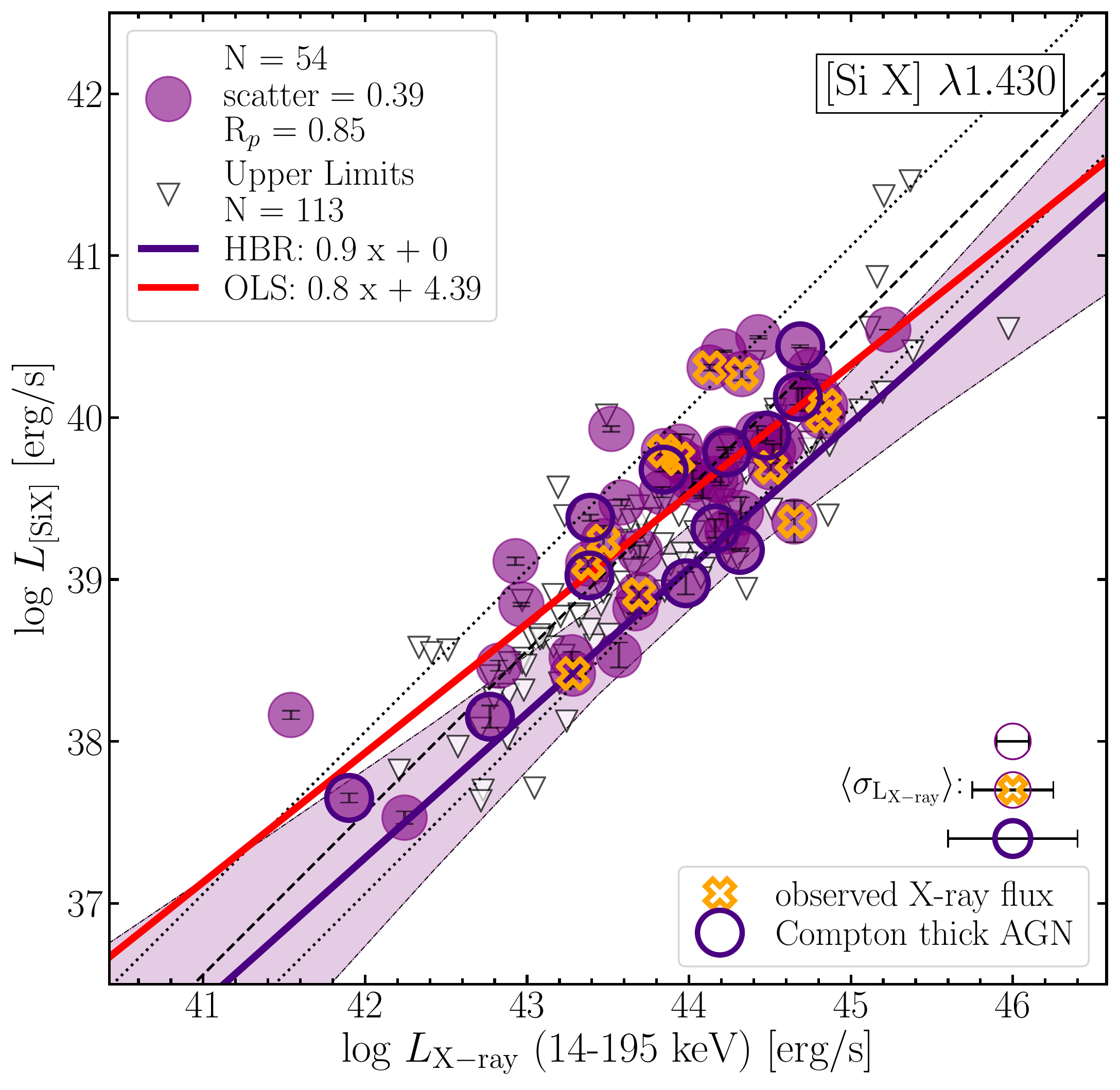}
            \caption{The blue/purple fit is the HBR line from \texttt{Linmix}, which can take upper limits into account. {The shaded region indicates the pointwise  3$\sigma$ confidence interval of the regression line.} The dashed line is scaled to the median flux ratio and has slope 1. The dotted lines are the 0.5 dex offset from the dashed line. The empty points are the 3$\sigma$ upper limits. The red line shows the OLS bisector. (Left) Comparison of \SiVI~ emission vs  X-ray emission. (Right) Comparison of \SiX~ emission vs. X-ray emission.  {We use the intrinsic 14-195\,keV X-ray flux for sources from the Swift-BAT 70\,month survey and the observed 14-195\,keV X-ray flux for the remaining sources } {(indicated by the points marked with a cross). Compton-thick AGN ($N_{\rm H}>10^{23.5}$\,cm$^{-2}$) are indicated by circles with increased edgewidth.} { The typical uncertainty of the X-ray luminosity is indicated by the points in the lower right corner (for a description, see Section \ref{sec:anc_data}).}}
                \label{fig:SiVI_general}
\end{figure*}
\begin{figure*}
            \centering
            \includegraphics[width = 0.9\textwidth]{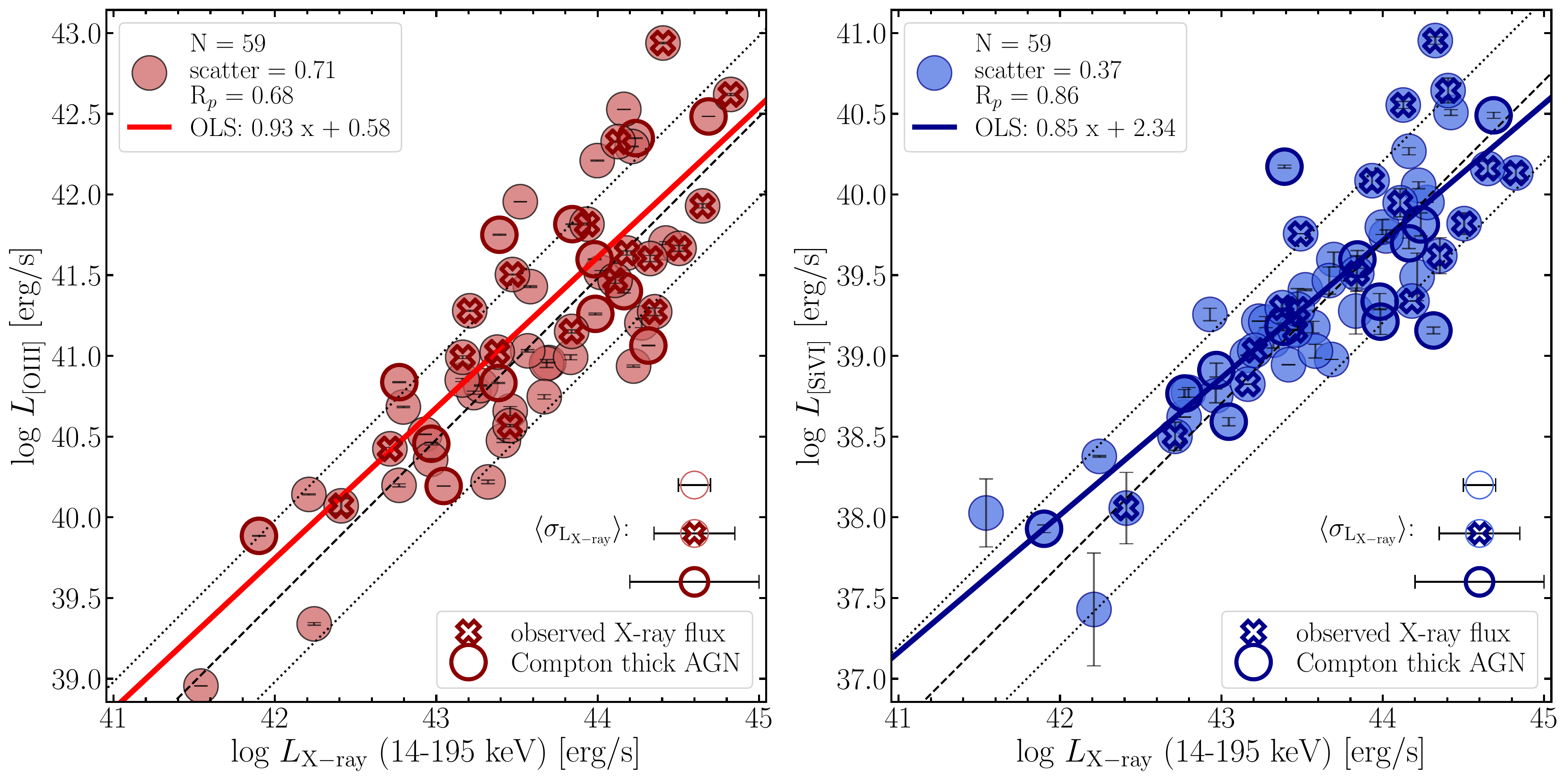}
            \caption{Emission of \OIII~(left) and \SiVI~(right) vs. X-ray emission for sources where both lines are detected simultaneously. The scatter of \OIII\ emission with respect to  X-ray emission is slightly above the value found in \cite{Berney2015} ($\sigma=$0.62 dex). Lines and points follow the description of \autoref{fig:SiVI_general}. 
            }
            \label{fig:OIII_SiVI_comp}
            \end{figure*}
\begin{figure*}
            \centering
            \includegraphics[width = 0.4\textwidth]{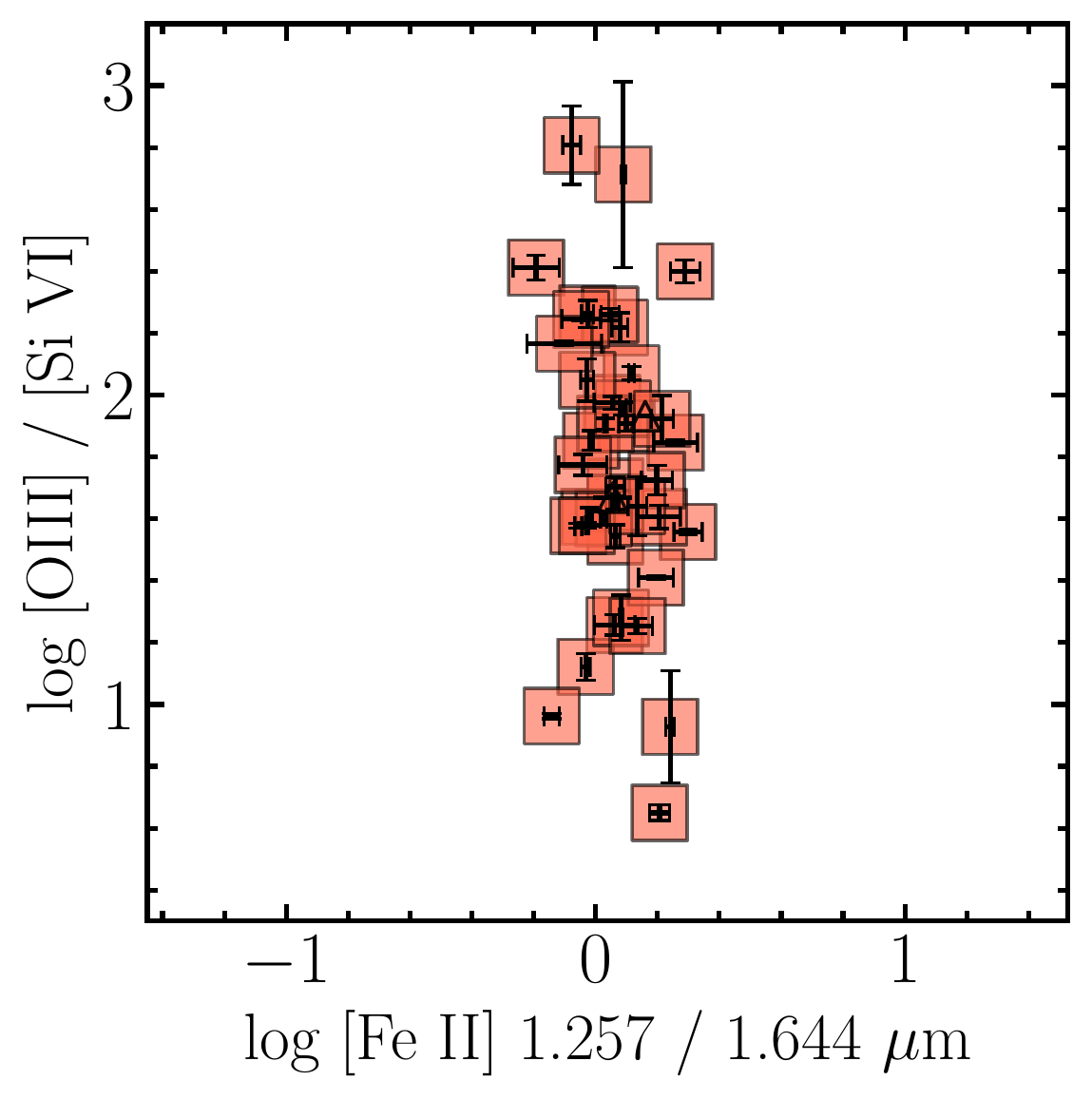}
            \caption{Ratio of  \OIII\ to \SiVI\ emission as a function of the $[$Fe \textsc{ii}$]\lambda1.257/1.644\,\mu$m ratio. The iron line ratio traces to some degree the obscuration \citep{Riffel:2006}, while the line ratio on the $y$-axis traces the IP. 
            }
            \label{fig:OIII_SiVI_obsc}
\end{figure*}

\section{Results}
\subsection{CLs}

            If CLs are an efficient tracer of AGN activity they should be detectable in all bright nearby AGN detected in \textit{Swift}--BAT. \autoref{fig:detec_rate} summarizes the detections of CLs for the sample.
            With the exception of the \SiX\ emission line, a trend can be observed of the number of detections going down with increasing IP. 
            The line with the highest absolute number of detections and the highest detection rate is the \SiVI~ CL (59/140 Seyfert 1 and Seyfert 2 galaxies,\footnote{The sample size is smaller than 168 because we exclude here spectra that do not cover the \SiVI\ emission line due to a combination of the galaxy's redshift and wavelength coverage.} 42~\%). The \SiX\ CL is detected in {54}/167 (32\%) Seyfert 1 and 2 galaxies. 

            \autoref{fig:detec_rate} (right) shows the distribution of the number of detections per spectrum. In 49/109~(45\%) Seyfert 2 spectra, at least one CL is detected. For 3/109~(4\%) Seyfert 2 spectra, five or more CLs are detected in a single spectrum. In 19/30~(63\%) Seyfert 1 -- 1.8 and in 16/28 (57\%) Seyfert  1.9 spectra at least one CL is detected. 
            The uncertainties of the detection rates are estimated using binomial proportion confidence intervals. The probability confidence interval is set to 1 $\sigma$.

            \autoref{fig:red_num_dist} shows the average number of simultaneously detected CLs binned by redshift. 
            {The gray area indicates the range in number of detections for each of the different sources in a given bin}. A trend can be seen such that we have fewer simultaneous CL detections with increasing redshift, for Seyfert 1--1.8, Seyfert 1.9, and Seyfert 2 galaxies. This decrease is due to a number of factors: less spectral coverage for higher redshifts, the shift of CLs into heavy telluric absorption regions, and generally weaker line fluxes due to increased distance {\citep{Rodriguez2011, Lamperti2017}}.

    \subsection{Comparison of CLs and X-Ray Luminosity}
        
        Naively speaking, CL emission is expected to be driven by soft X-ray and far-UV high-energy photons (>{100} eV), which ionize the CL species \citep{Done2012}. So as a first step, we check the correlation between high-ionization and X-ray emission. We use the model-independent \textit{Swift}-BAT-observed 14--195~keV X-ray emission, and for the CL emission we focus on the \SiVI\ and \SiX \ luminosities, which have the highest detection rates and intensities.
        
        \begin{table*}
            \centering
            \caption{ Details of Regression Fits between CLs and \textit{Swift}--BAT X-Ray (14-195 keV).}
        \label{tab:summary_SiVI_SiX}
        \begin{tabular}{c c c c c c c c c}
             \hline
            Line & $N_\text{det}$ & $N_\text{un}$ & Line Ratio & Slope & Intercept & Scatter (dex) &$R_\text{pear}$ & $p_\text{pear}$ \\
            & (1) & (2) & (3) & (4) & (5) & (6) & (7)&(8)\\ \hline \hline
            \SiVI & 59 & 81 & 27000 & OLS: $0.85\pm0.06$ & $-7\pm4$ & 0.37 & 0.86&$2.0\times10^{-18}$ \\
            & & & & HBR: $0.74\pm0.09$ & 2$\pm$3 & $0.30\pm0.1$ & & \\
            & & & & &  &>0.42  & &\\\hline
            \SiX & 54 & 113 & 36000 & OLS: $0.80\pm0.08$ & 4$\pm$3 & 0.39 & 0.85&$3\times10^{-17}$ \\
            &  &  &  & HBR: $0.89\pm0.08$ & $0\pm3$ &  $0.3\pm0.2$& &\\ 
            & & & & &  & >0.4 && \\\hline
            \SVIII & 22 & 143 & 52000 & 0.9 $\pm$ 0.1& 0$\pm$4 & 0.37 & 0.89 &$4\times10^{-8}$\\ \hline
            \SIX & 29 & 139 & 39000 & $0.7\pm0.1$ & 10$\pm$4 & 0.52 & 0.78&$5\times10^{-7}$\\ \hline
            \FeXIII & 17 & 151 & 48000 & $1.0\pm0.1$ & $-$8$\pm$6 & 0.46 & 0.88&$2\times10^{-6}$
        \end{tabular}
        \vspace{0.3 cm}
    
        { \footnotesize Notes. 1) Number of sources with line detection. (2) Number of sources without line detection. (3) Ratio of mean X-ray luminosity to mean line luminosity. (4) Slope of the OLS bisector (only detections are considered). For \SiVI~and \SiX, also the slope of the HBR is also given. (5) Intercept of the OLS bisector (only detections are considered). For \SiVI~and \SiX, also the intercept of the HBR is also given. (6) Scatter of the data points in dexes. For \SiVI~and \SiX, the intrinsic scatter is an estimate from the \texttt{Linmix} module, which takes nondetections into account (second value). The third value for \SiVI~and \SiX\ is a conservative estimate of the lower limit by treating the nondetections as detections. (7) Pearson correlation coefficient. (8) Pearson $p$-value with null hypothesis of slope zero.}
    
        \end{table*}

        The result can be seen in \autoref{fig:SiVI_general}. 
        We fit the data using an ordinary least squares (OLS) bisector. In addition, the Python module \texttt{Linmix}\footnote{Software module created by Joshua E. Meyers (\href{https://linmix.readthedocs.io/en/latest/index.html}{https://linmix.readthedocs.io}) based on model described in \citealt{Kelly2007}.} is used for regression analysis. The package uses hierarchical Bayesian regression (HBR), which can take the upper flux limits into account. \autoref{tab:summary_SiVI_SiX} lists the regression fit parameters.
        {We note that a positive correlation will be induced due to the correlated axes in a luminosity--luminosity plot. However, in the subsequent analysis, we will mainly focus on and describe the quality of the regression using the scatter around the regression because it is the same in a luminosity--luminosity or a flux--flux plot.}

        For the relation of \SiVI\ and $L_{\rm X{-}ray}$ (14--195 keV), the scatter is $\sigma = 0.37$ dex (\autoref{fig:SiVI_general}, left). This scatter takes only detections into account. Consequently, the actual intrinsic scatter is most likely higher. We can get a conservative estimate of the lower limit of the scatter assuming the nondetections are 3$\sigma$ detections. This shows that the scatter is actually $\sigma > 0.42$ dex. Using the Python \texttt{Linmix} package we estimate the intrinsic scatter taking nondetections into account. Because the module runs a Markov Chain Monte Carlo (MCMC), we further estimate the uncertainties in the intrinsic scatter. For comparison of \SiVI\ and X-ray emission, we get $\sigma_{\rm intr.} = 0.30\pm0.10$ dex. The Pearson correlation coefficient is $R_\text{pear} = 0.9$ ($p_\text{pear} = 3\times10^{-21}$) showing a strong correlation. As expected when considering flux values rather than luminosities the correlation is more moderate with $R_\text{pear} = 0.74$ ($p_\text{pear} = 6\times10^{-11}$). 

        \begin{figure}
            \centering
            \includegraphics[width = 0.4\textwidth]{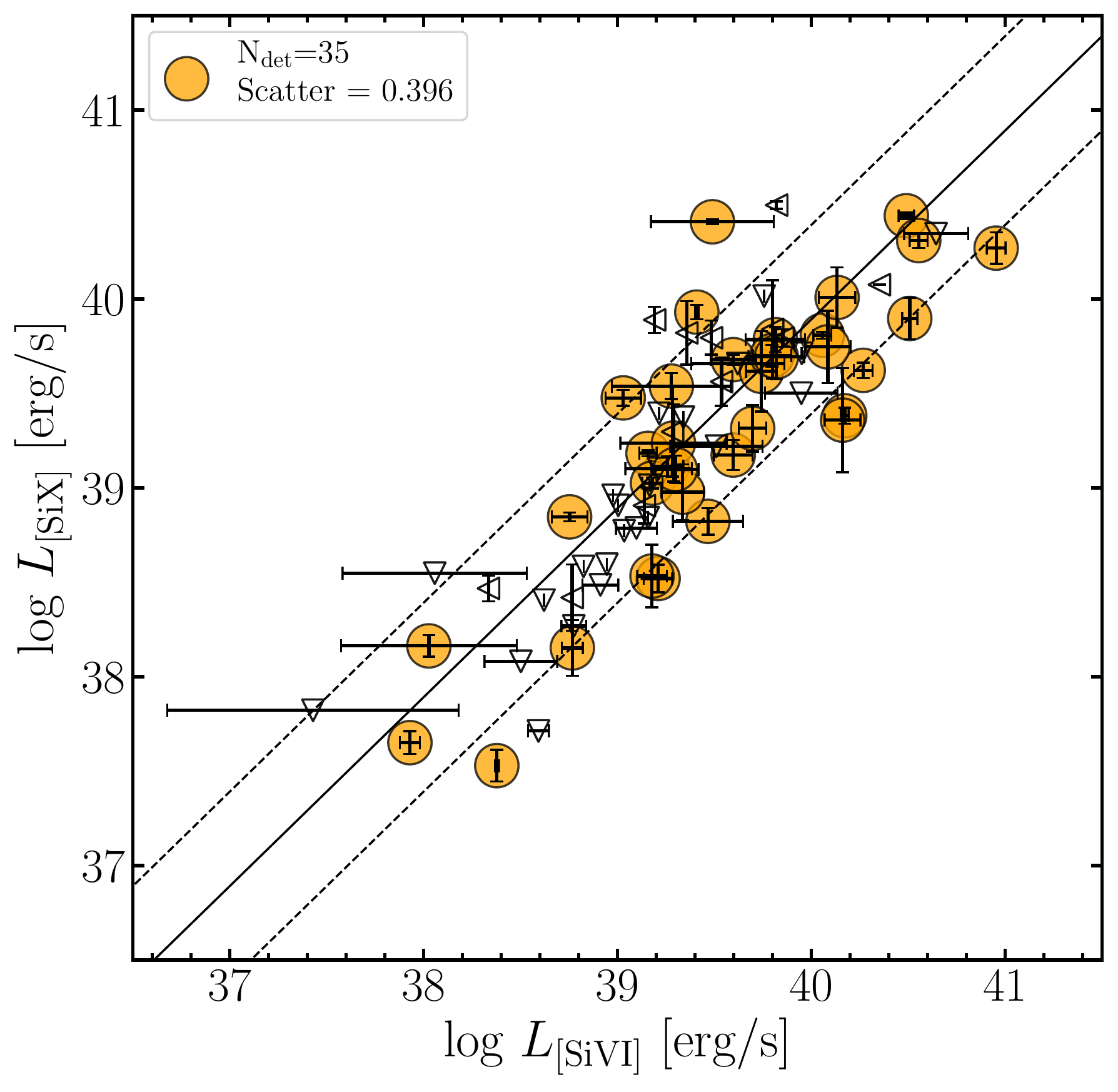}
            \caption{Luminosity of \SiVI~vs. \SiX. Empty triangles indicate upper limits. The lines follow the description in \autoref{fig:SiVI_general}.}
            \label{fig:comp_SiVI_SiX}
        \end{figure}
        \begin{figure}
            \centering
            \includegraphics[width = 0.44\textwidth]{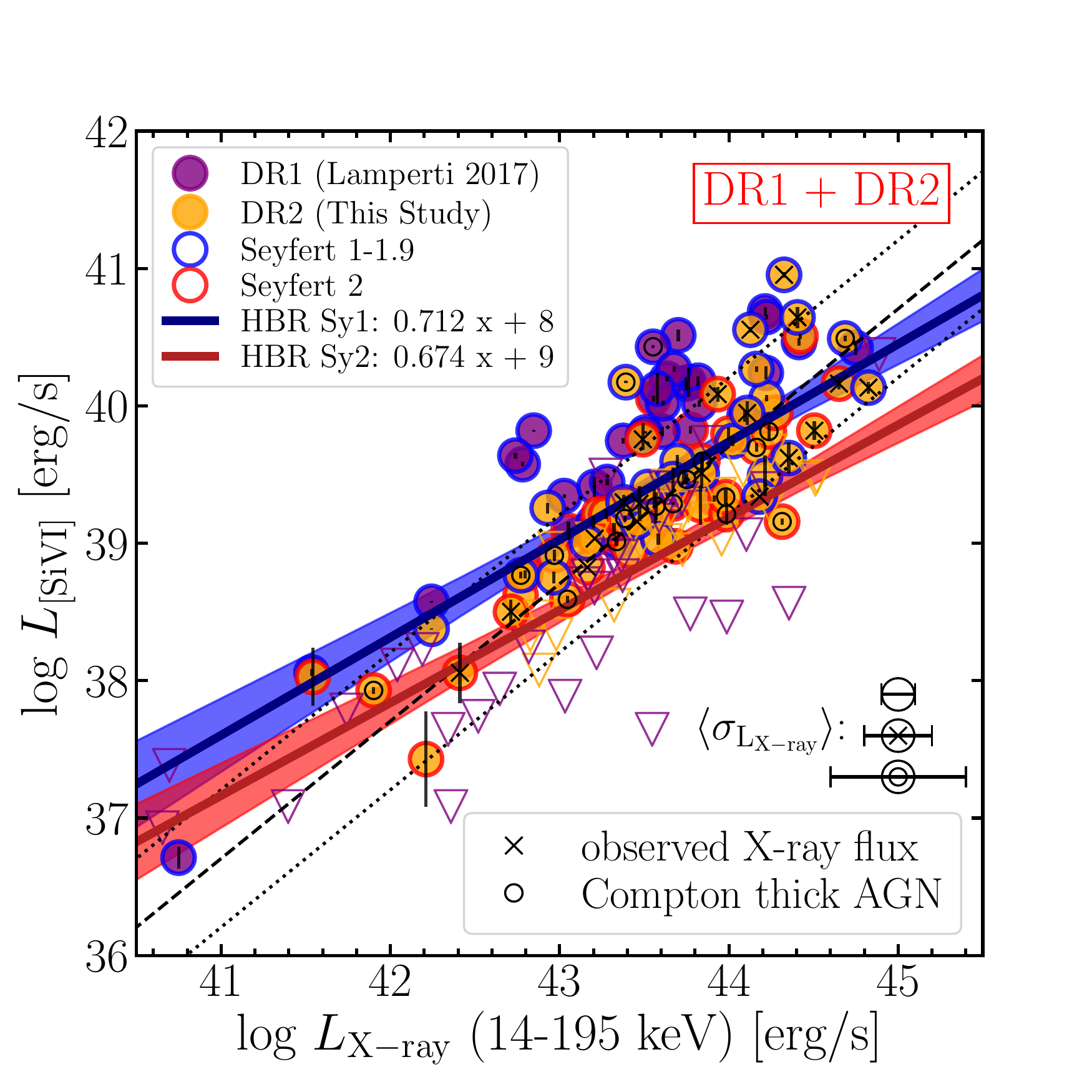}
            \caption{Luminosity of \SiVI~vs. observed $14{-}195$ keV X-ray luminosity for sources with detections in BASS DR1 and this work, BASS DR2. Purple markers indicate sources from DR1 ($N = 44$) and orange markers indicate sources from  DR2 ($N = 57$). Edges are color-coded by AGN type (Type 1 vs. Type 2). {The shaded region indicates the 1$\sigma$ confidence interval of the respective regression line. Downward triangles indicate upper limits (only separated by DR1 and DR2).} The lines and points follow the description in \autoref{fig:SiVI_general}.}
            \label{fig:Dr1andDr2}
        \end{figure}
        In \autoref{fig:SiVI_general} (right), the correlation of \SiX\  emission with X-ray emission is shown, again using an OLS bisector and an HBR to fit the data. For the relation of \SiX\ and $L_{\rm X{-}ray}$ (14--195 keV), the scatter of the detections is $\sigma = 0.39$ dex. Again assuming the nondetections to be detections, the lower limit of the scatter is estimated to be $\sigma>0.4$ dex. The intrinsic scatter estimate from the MCMC method is $\sigma_{\rm intr.} = 0.30\pm0.20$~dex. The Pearson correlation with the hard X-ray luminosity is $R_\text{pear} = 0.85$ ($p_\text{pear} = 3\times10^{-17}$). For the flux, the correlation is more moderate with  $R_\text{pear} = 0.61$ ($p_\text{pear} = 2\times10^{-6}$). 

        In order to quantitatively investigate whether the correlations of \SiVI\ and \SiX\ with X-ray luminosity differ significantly, the Fisher $z$-test is used, based on the two Pearson correlation coefficients of the luminosity correlation. The two-tailed $p$-value is 0.3, indicating the two distributions are not significantly different.
        
        As a further step concerning the comparison of the correlation of \SiVI\ and \SiX\ with the X-ray emission, we only include sources that show emission from both CLs simultaneously in their spectrum. 
        \autoref{fig:OIII_SiVI_comp} shows the correlation of \OIII\ (left) and \SiVI\ (right) versus X-ray luminosity, but only in those sources in which both lines are detected simultaneously.
         The \OIII\ species has an IP of 35.1 eV; \SiVI\ has an IP of 166 eV. The \OIII\ emission line is measured as a part of BASS DR2 \citep{Oh_DR2_NLR}.
         {We detect \oiii\ in all of our sources. The lines are measured using the same  spectra we use in this study.}
        
        In 57 sources, \SiVI~ is observed simultaneously with \OIII. Compared with the hard X-ray luminosity, the scatter of \OIII~is $\sigma = 0.71$ dex and the Pearson correlation coefficient is $R_{\rm pear} = 0.68$. This result of the scatter is consistent with \cite{Berney2015} ($\sigma=$0.62 dex). For \SiVI, the scatter is $\sigma = 0.37$ dex and $R_{\rm pear} = 0.86$.
        Applying the Fisher $z$-test, the $p$-value is $p<0.001$, meaning that the two correlations are different.
        
        The scatter of the \SiVI-$L_{\rm X{-}ray}$ relation is lower  than that of the \OIII-$L_{\rm X-ray}$ relation. However, this scatter is a lower limit as it only takes detections into account and the actual intrinsic scatter might be higher. 
        In light of what we find it would seem that CLs are a better proxy for AGN power. In \autoref{fig:OIII_SiVI_obsc} we show the ratio of  \OIII\ to \SiVI\  emission as a function of the iron line ratio $[$Fe \textsc{ii}$]\lambda1.257/1.644\,\mu$m. While the \OIII\ to \SiVI\ ratio traces ionization, the iron line ratio traces to some degree the obscuration {as it is independent of temperature and density \citep{Rodriguez2004,Riffel:2006,Deb2010}}. We find a larger scatter in the \OIII\ to \SiVI\ ratio ($\sim$1.5 dex) than in the iron line ratio ($\sim$0.2 dex).

        In \autoref{fig:comp_SiVI_SiX}, \SiVI\ luminosity is compared to \SiX\ luminosity. The correlation coefficient is $R_{\rm pear} = 0.82$ and $p_{\rm pear} = 3\times10^{-9}$  (scatter $\sigma$ = 0.4 dex). 
        
        \cite{Lamperti2017} studied the NIR emission for a subset of AGN as part of  the first data release of the BASS project. The DR1 analysis has a sufficient number of \SiVI~detections ($N_{\rm det} = 42$). In their study, \cite{Lamperti2017}  noted that Seyfert 1s showed a higher \SiVI\ luminosity than Seyfert 2s. In regard to the \SiX~emission, \cite{Lamperti2017}  had only 17 detections.

        \begin{figure}
                \centering
                \includegraphics[width = 0.5\textwidth]{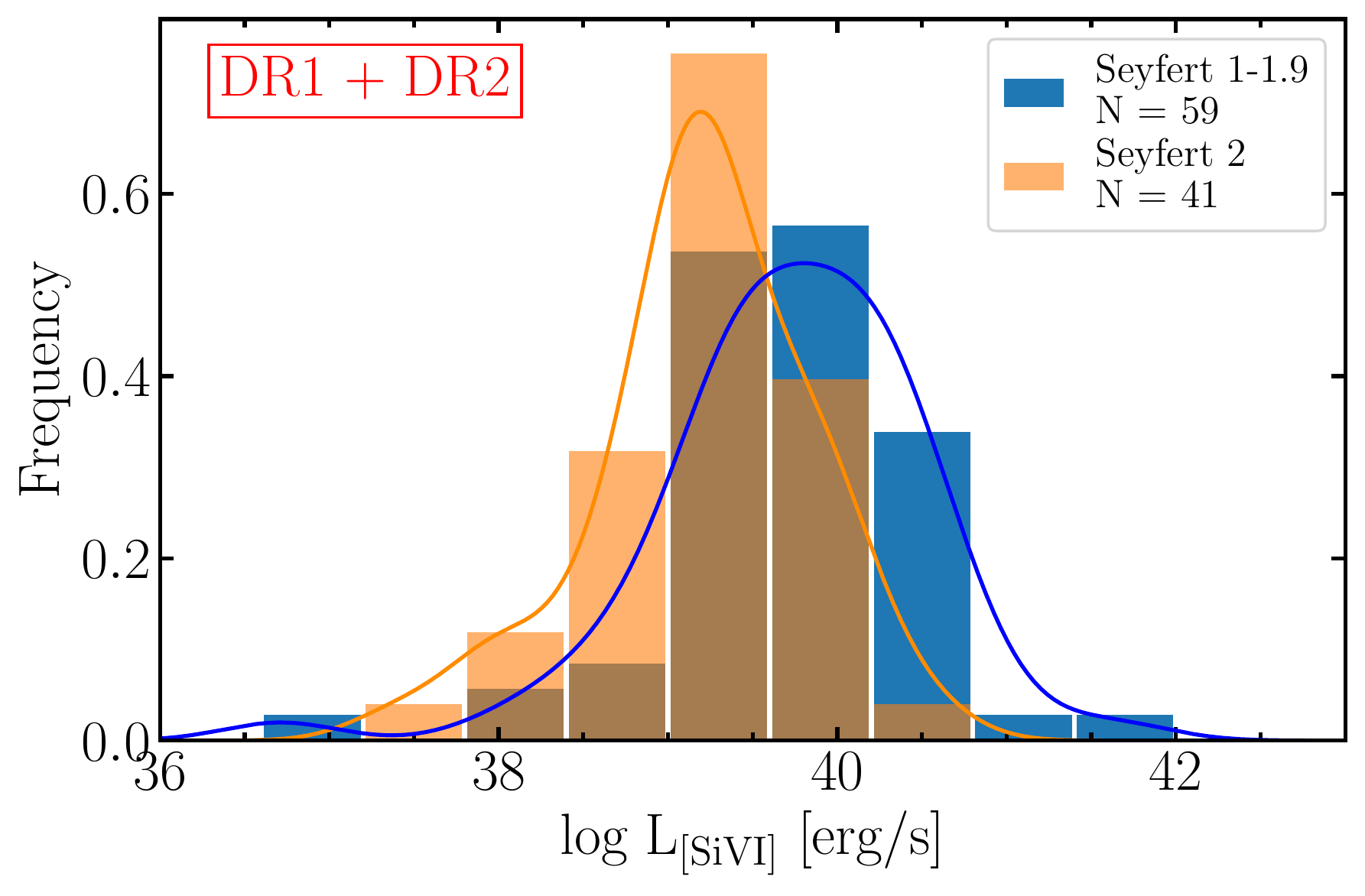}
                \caption{Histogram of \SiVI~luminosity, separated by Seyfert 1 and 2. The lines indicate the kernel density estimation of the two distributions.}
                \label{fig:hist_SiVI}
            \end{figure}
        In \autoref{fig:Dr1andDr2}, the emission of \SiVI~ and the X-ray is shown for both DR1 and DR2. Seyfert 1 galaxies show a statistically higher luminosity. 
        The median luminosity {and the 16$^{\rm th}$ and 84$^{\rm th}$ percentile ranges} of the Seyfert 1 sample are $\langle\log L_{[\text{Si VI}]}/{\rm erg~s^{-1}}\rangle_{\rm Sy 1} = 39.8^{+0.7}_{-0.7}$ and those for the Seyfert~2 galaxies are $\langle\log L_{[\text{Si VI}]}/{\rm erg~s}^{-1}\rangle_{\rm Sy 2} = 39.2^{+0.6}_{-0.5}$. {Furthermore, we find that the scatter is smaller for the \SiVI\ emission with the hard X-ray for Seyfert 2 galaxies (0.37 dex) than for Seyfert 1--1.9 galaxies (0.45 dex). The scatter across the full sample (Seyfert 1--2 galaxies) is 0.47 dex. \autoref{tab:DR1_DR2} provides a summary of the scatter and regression parameters for the combined DR1 and DR2 set.}

        Applying the $t$-test to investigate whether indeed Seyfert 1 and Seyfert 2 luminosity values of the \SiVI\ emission line differ, we get a $p$-value of $p = 3\times 10^{-5}$. We therefore can reject the null hypothesis that the distributions are equal.
        \autoref{fig:hist_SiVI} shows a histogram of the luminosity distributions of Seyfert 1--1.9 and Seyfert 2 galaxies. {We see that the \SiVI\ emission from Seyfert 1--1.9 galaxies is shifted toward brighter values.}
        \begin{table}
        \caption{Summary of DR1 and DR2 Combined.}
        \centering
        \begin{tabular}{c l c c c c}
            \hline
             Line & AGN Type & $N_{\rm det}$& Scatter & $R_\text{pear}$ & $p_\text{pear}$\\
             &  & (1)& (2) & (3) &(4) \\ \hline \hline
             \multirow{3}{*}{\SiVI} & Sy 1--1.9& 59 & 0.45 & 0.84 & 1$\cdot 10^{-16}$\\
             & Sy 2& 41& 0.37 & 0.88&3$\cdot 10^{-13}$\\
             & all& 97& 0.47 & 0.83 &1$\cdot 10^{-25}$
        \end{tabular}
        {\\ Note. (1) Number of detected emission lines. (2) The scatter of the data points in dexes (3) Pearson correlation coefficient.(4) Pearson $p$-value with null hypothesis being a slope of zero.}
      \label{tab:DR1_DR2}
    \end{table}

\begin{figure}
            \centering
            \includegraphics[width =0.65 \textwidth]{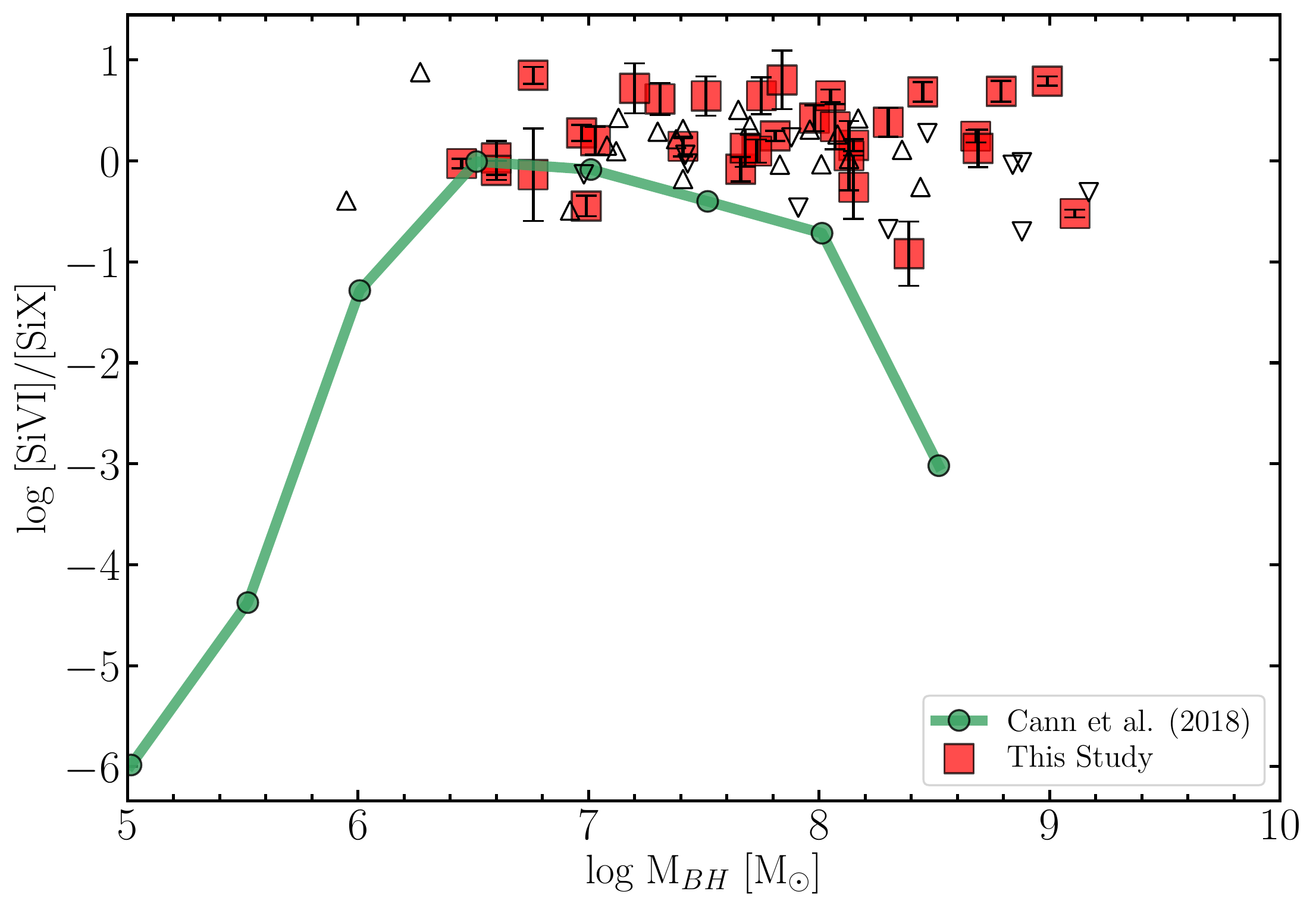}
            \caption{The dependence of the ratio of CL emission \SiVI/\SiX. The green line is the theoretical prediction from Figure 8 in \citeauthor{Cann_2018} (2018). The authors assumed a fixed $L/L_{\text{Edd}} = 0.1$, $n_{\rm H} = 300$ cm$^{-2}$ and $\log U = -2$. The red squares indicate the observed ratios. Upper and lower limits are indicated by upward- and downward-facing triangles. Based on the lack of a downturn in the observed data, we conclude that the conditions of the model in \cite{Cann_2018} need to be broadened. }
            \label{fig:M_bh_ratio}
        \end{figure}
\begin{figure*}
            \centering
            \includegraphics[width = 0.9\textwidth]{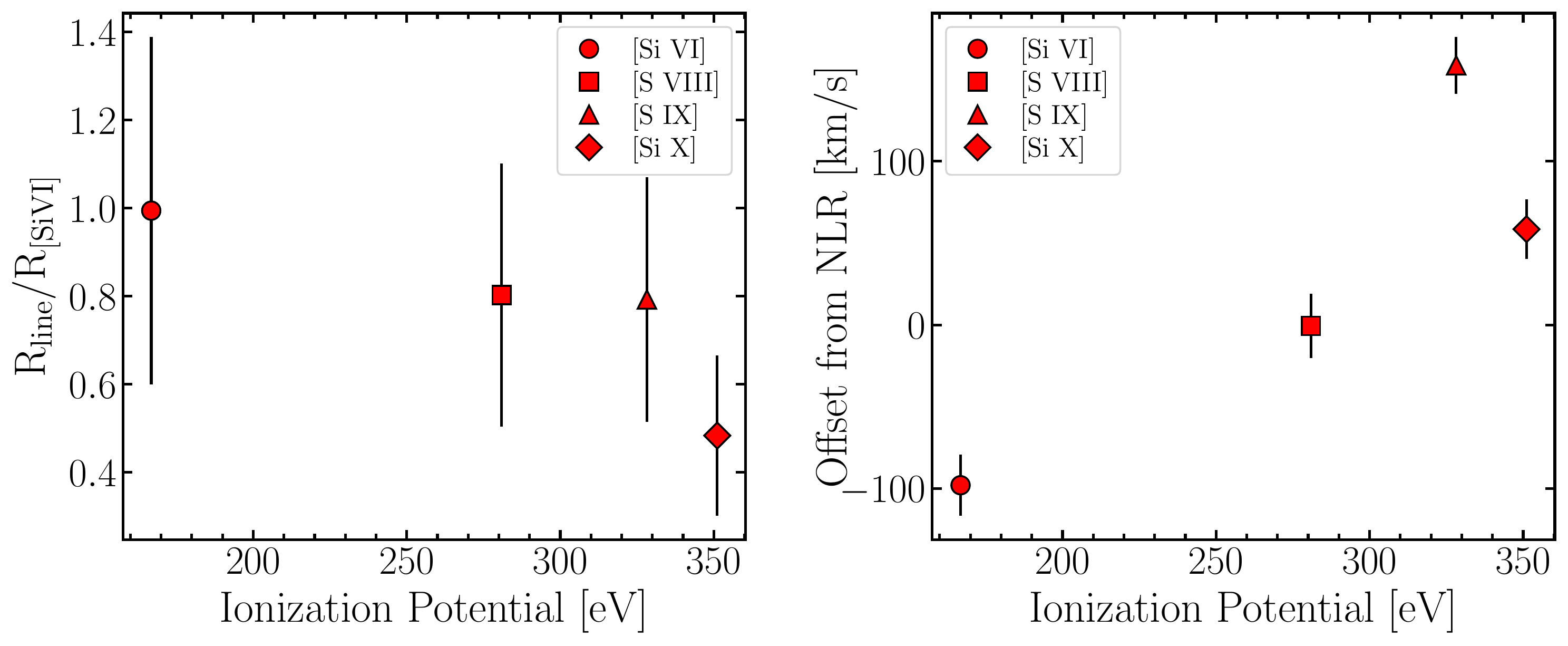}
            \caption{(Left) The averaged $R_{\rm line}$/$R_{[\text{Si\ \textsc{vi}}]}$ of the CLs compared with their IP. (Right) The averaged offset of the CLs with respect to the NLR compared with their IP. For both cases only sources where all CLs are present are selected.  }
            \label{fig:FWHM_analysis}
        \end{figure*}
        
    \subsection{Scaling Relations with Black Hole Mass}
        Theoretical calculations predict a tight dependence between CL emission and the mass of the central black hole, \mbh\ \citep{Cann_2018}, since AGN spectral energy distributions strongly depend on the black hole mass. 
        { {Appendix \ref{App:bhmass}}} presents the correlation of the emission of the two CLs with the most detections (\SiVI~and \SiX) with black hole mass. A moderate correlation with $R = 0.56$ for \SiVI\ and $R=0.44$ for \SiX\ is found.
        
        \autoref{fig:M_bh_ratio} shows the theoretical mass dependence of the ratio of \SiVI/\SiX\ emission. For the calculation,  \cite{Cann_2018} assumed a fixed $L/L_{\text{Edd}}$ ratio $ = 0.1$, a gas density of $n_{\rm H} = 300$ cm$^{-3}$ and a dimensionless ionization parameter 
        of $\log U = -2$ (see Figure 8 in \citealt{Cann_2018}). The red squares indicate the observed values. The ratio is normalized such that the maximum ratio has a value of 1.0. The predicted drop at high masses is not observed, hinting that the boundary conditions chosen in \cite{Cann_2018} are too narrow.

        \subsection{CL FWHM and Offset}

        For the FWHM analysis, we take into account how the velocity of the CL-emitting gas clouds depends on the black hole mass (see \citealt{Netzer2013}). If we assume virial motion, we have
        
        $$
         \mbh \approx f(R)\frac{\Delta v_{\rm line}^2 R}{G},
        $$
        where $G$ is the gravitational constant, $R$ is the distance from the black hole, $\Delta v_{\rm line}$ is the velocity measure from the line profile, and $f(R)$ is the virial factor, which takes into account the unknown geometry and orbital structure of the CL region (CLR). Therefore, 
        $$
            R\propto \frac{M_\text{BH}}{\Delta v_\text{line}^{2}}.
        $$

         We calculate the ratio $R_{\rm line}$/$R_{[\text{Si\ \textsc{vi}}]}$ for  sources for which we detect \SiVI, \SVIII, \SIX, and \SiX\ simultaneously in the NIR spectrum. {Because we look at the ratio, we use for the velocity measure the FWHM determined from our line fitting with \texttt{PySpeckit}. Furthermore, the virial factor cancels out, as we assume similar inclinations between orbits of the CLR}. There are seven sources for which this is the case. \autoref{fig:FWHM_analysis} (left) shows the median of $R_{\rm line}$/$R_{ [\text{Si\ \textsc{iv}}]}$ for the seven sources and errors based on the standard deviation. We see that CLs with higher IP tend to be closer in. 

       The CL {velocity} offset is another interesting parameter to analyze, as it can give further information about the kinematics of the CLR. {Line} offsets are calculated relative to the NLR{'s velocity offset}, as determined from {looking at} the Pa$\beta$ or He \textsc{i} emission line.
        \autoref{fig:FWHM_analysis} (right) shows the {average} mean {velocity offset} for seven spectra that show all CLs simultaneously. There is a trend toward increasing blueshift with decreasing IP.   
        \begin{figure}
            \centering
            \includegraphics[width =0.75 \textwidth]{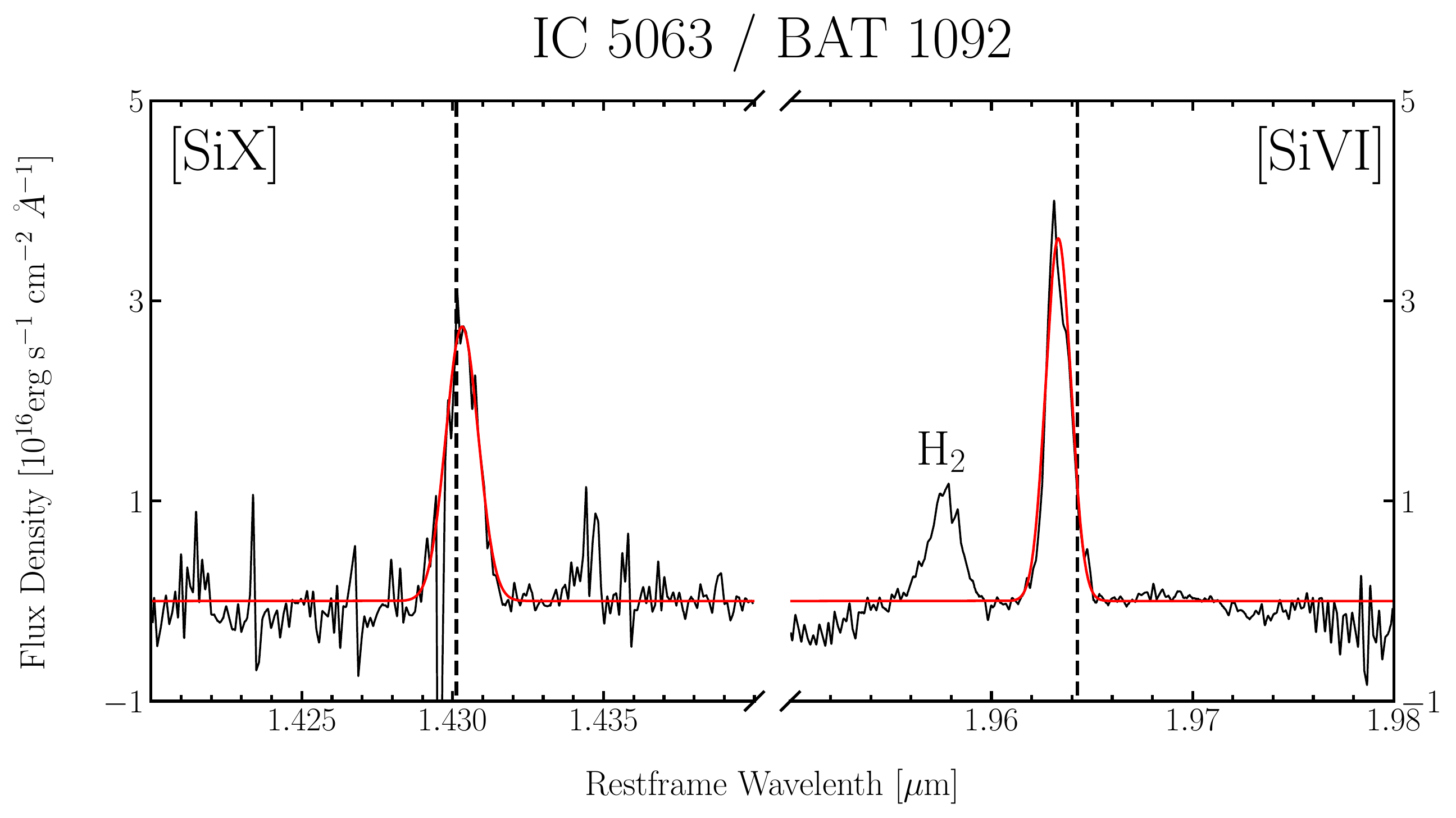}
            \caption{Example of offset of \SiX~and \SiVI. The \SiX~ line is clearly not blueshifted like the $[$Si VI$]$ line. The line blueward of $\SiVI$\ is H$_2$.  The dashed lines mark the systemic redshift, based on lower-ionization lines (Pa$\beta$ or He \textsc{i}).}
            \label{fig:offset_example}
        \end{figure}

\begin{figure}
            \centering
         \includegraphics[width = 0.45\textwidth]{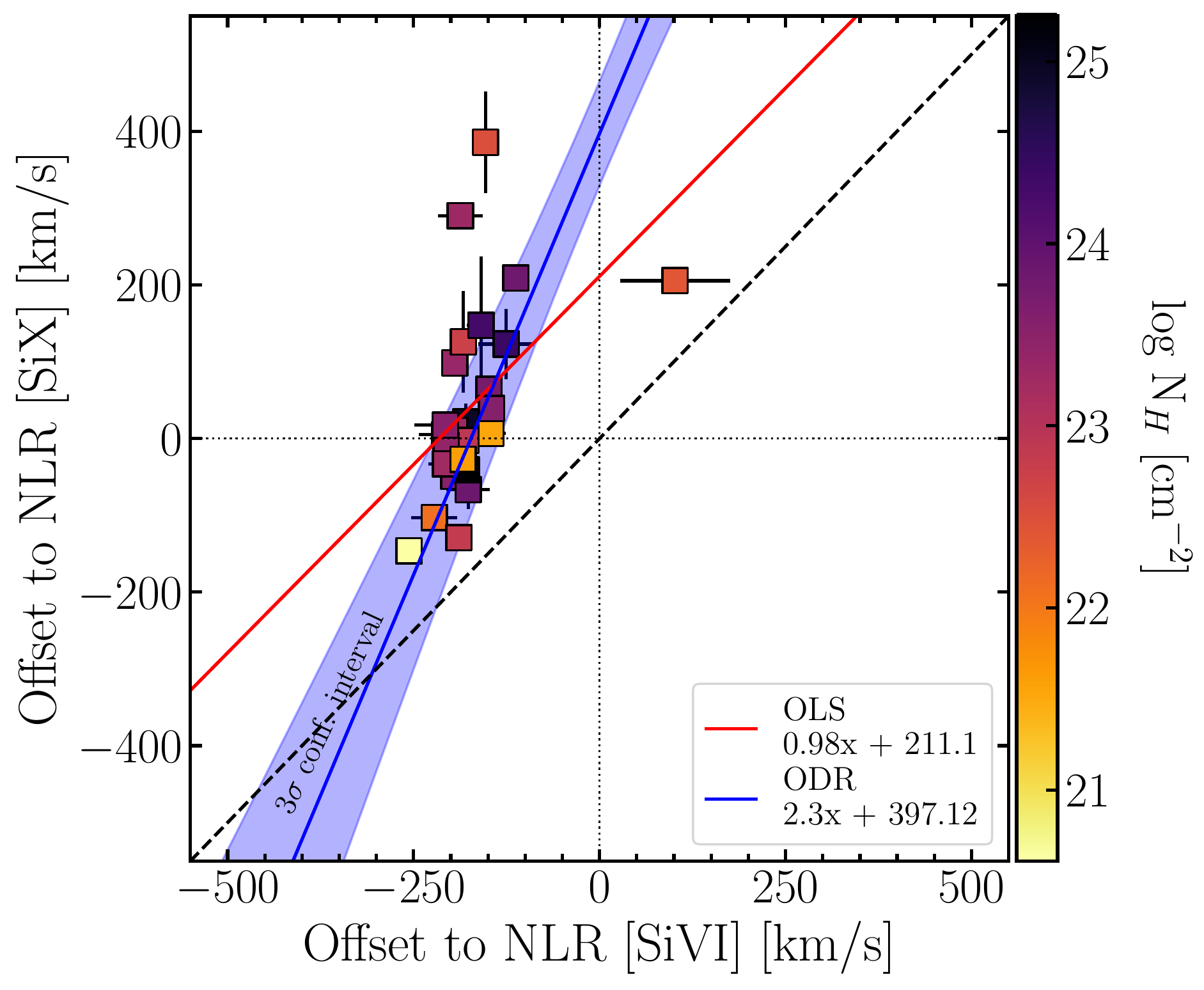}
            \caption{Velocity offset of \SiVI~and \SiX. The OLS regression and the orthogonal distance regression (ODR) are computed. The ODR takes into account the uncertainties of the measurements. The blue shaded region marks the 3$\sigma$ (pointwise) confidence interval of the ODR fit. The points of individual sources are color-coded according to the hydrogen column density. The dashed line marks the 1:1 locus.}
            \label{fig:offset_analysis_SiX_SiVI}
        \end{figure}

        \autoref{fig:offset_example} illustrates that significant shifts are robustly seen, even by eye, in our data, and cannot be explained by poor statistics or a low signal-to-noise ratio (S/N).
        We focus {as an example} on the \SiX\ and \SiVI\ CLs because they have the highest detection rates of all the high-ionization lines. {For example, }\autoref{fig:offset_example} shows {the spectrum of BAT 1092. It can be clearly seen } that the \SiVI\ line has a systematic blueshift, while {such a shift is less clear for} \SiX\ {, which shows a similar offset to the NLR region (as indicated by the dashed line).} In \autoref{fig:offset_analysis_SiX_SiVI}, we see that such a blueshift is systematically observed for \SiVI throughout our sample. In the figure, we color-coded the targets by their respective hydrogen column density. However, we do not find any clear trend with the column density and the magnitude of the velocity offset. 

        \begin{figure}
            \centering
            \includegraphics[width = 0.7\textwidth]{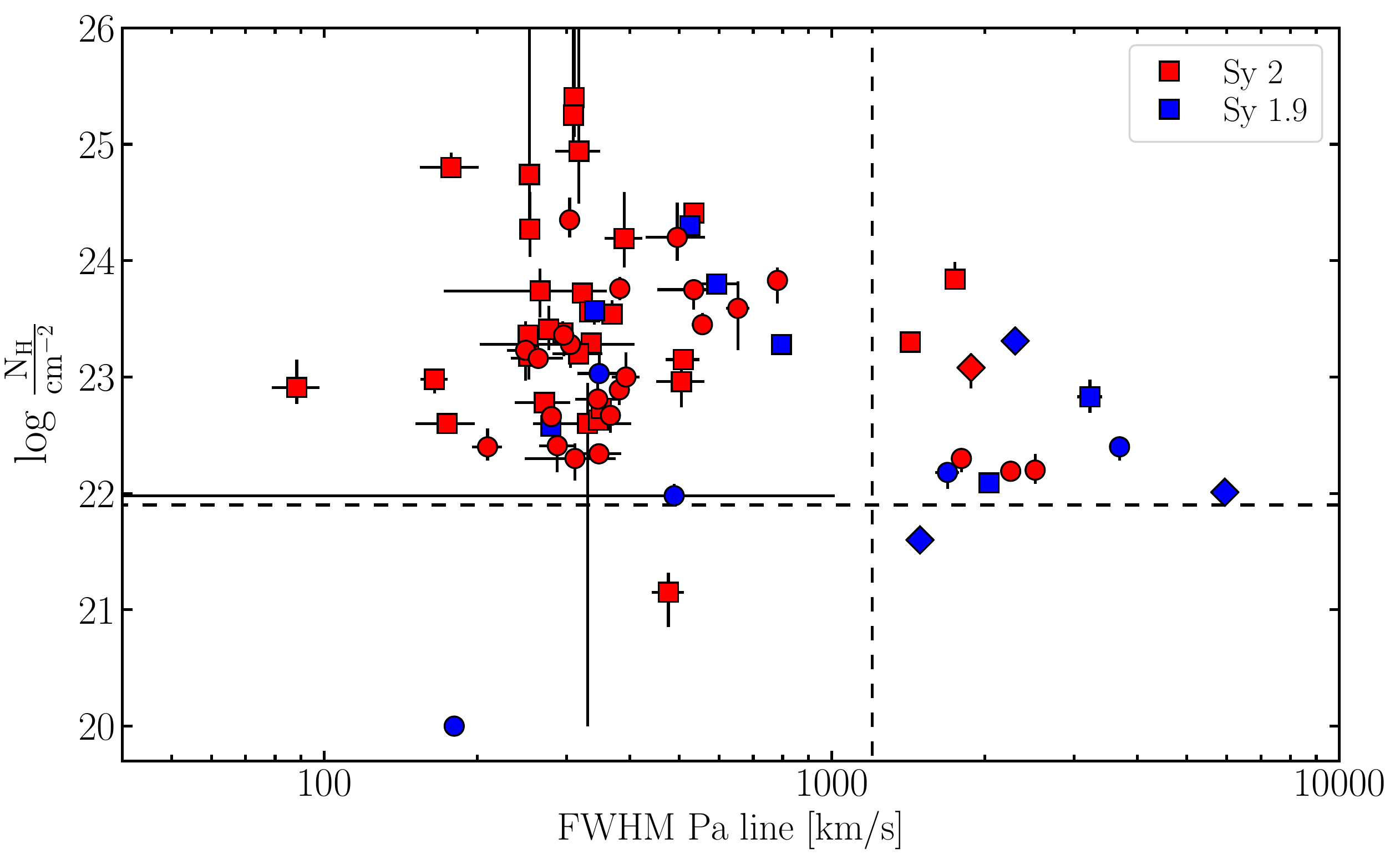}
            \caption{Scatter plot of Paschen FWHM and hydrogen column density $N_{\rm H}$. Squares indicate Pa$\beta$, circles indicate Pa$\alpha$, and diamonds indicate that the FWHMs of Pa$\alpha$ and Pa$\beta$ are taken. The instrumental FWHM of 56\,km\,s$^{-1}$ is removed. The horizontal dashed line is the nominal column density threshold separating X-ray obscured and unobscured AGN. The vertical dashed line is the separation into Seyfert 1 and 2 based on the FWHM, where Seyfert 2 galaxies lack broad Balmer lines. A small fraction of optical Seyfert 2 galaxies however show broad NIR Pa$\beta$ or Pa$\alpha$ emission lines, indicating hidden broad lines.}
            \label{fig:HBL}
        \end{figure}
\begin{figure*}
            \centering
            \includegraphics[width = 0.8\textwidth]{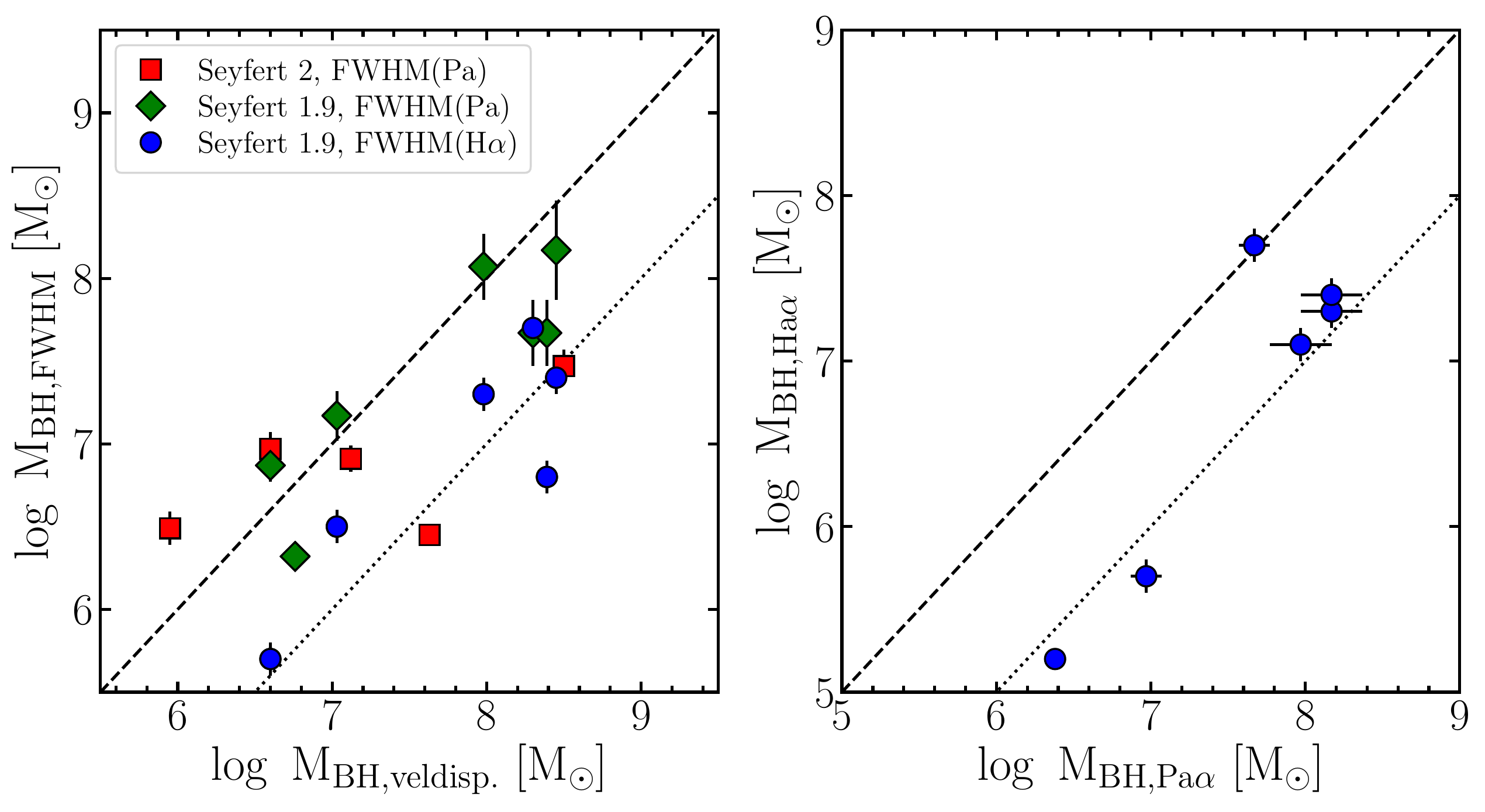}
            \caption{(Left) Comparison of black hole mass estimates from velocity dispersion measurements ($M_\text{BH, veldisp.}$) and from Paschen or H$\alpha$ line widths ($M_\text{BH, FWHM}$) for Seyfert 2 and 1.9 galaxies with hidden broad lines. For the Seyfert 1.9 galaxies, no offset is seen when using Paschen lines, while H$\alpha$ gives lower mass by almost 1 dex. The dashed line indicates a 1:1 relation. The dotted line is a linear correlation shifted by 1 dex. (Right) Comparison of the estimated mass of the central black hole using the FWHM of the broad component of H$\alpha$ and Pa$\alpha$ for Seyfert 1.9 galaxies. }
        \label{fig:Dust_atten}
    \end{figure*}
    \begin{figure*}
            \centering

            \includegraphics[width = 0.4\textwidth]{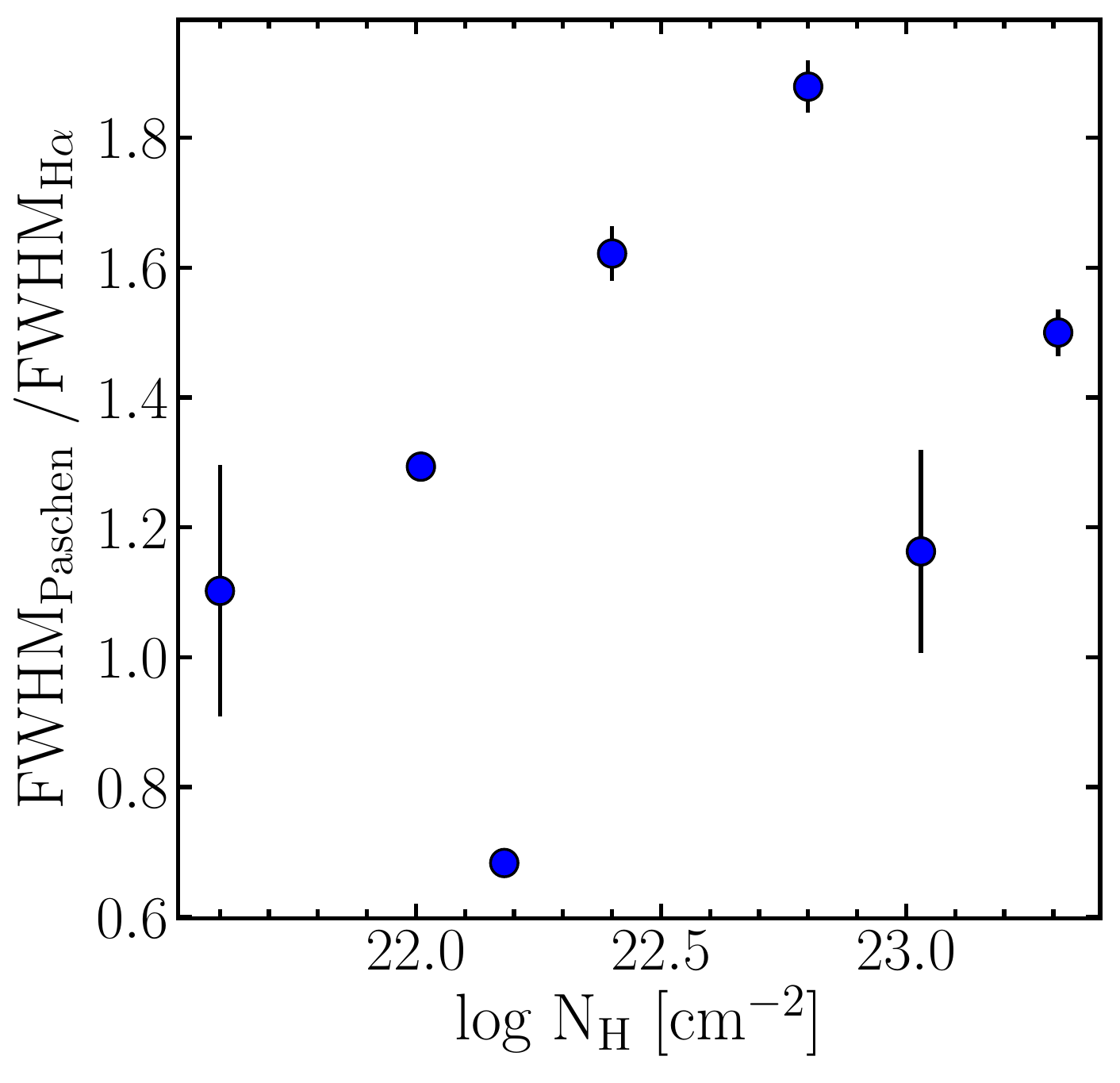}
            \caption{Ratio of FWHMs of Paschen and H$\alpha$ lines as a function of the line-of-sight X-ray hydrogen column density. 
            }
        \label{fig:Dust_atten_2}
    \end{figure*}

    \subsection{Hidden Broad Lines}
        Our sample consists of 110 sources that are classified as Seyfert 2 galaxies based on the lack of a broad Balmer line component (FWHM > 1200 km s$^{-1}$) in the optical spectrum.  For 59 cases, we have Pa$\beta$ or Pa$\alpha$ measurements together with line-of-sight X-ray column density measurements. \autoref{fig:HBL} shows the distribution of the hydrogen column density as a function of the NIR Paschen emission line FWHM. If both the Pa$\alpha$ and Pa$\beta$ emission lines have broad components, the average is taken. The vertical dashed line indicates the separation of AGN into Seyfert 1 and Seyfert 2 based on the emission line width and the horizontal dashed line that based on the hydrogen column density (AGN with $\log (N_H/{\rm cm^{-2}})>$21.9 are considered to be Seyfert 2 {galaxies \citep{Koss_2017}}. This is consistent with the fact that the bulk of Seyfert 2 galaxies have a narrow FWHM ($<1200$ km s$^{-1}$) and a high column density $\log (N_{\rm H}/{\rm cm^{-2}}) = 21.1-25.4$ (mean column density: $\log (N_{\rm H}/{\rm cm^{-2}}) = 23.4 \pm 0.9$). 
        In the following, we only take into account those sources that have velocity dispersion \mbh\ estimates and/or broad H$\alpha$ FWHM measurements, as we are interested to see whether the Paschen lines can be used for mass estimations. In \autoref{tab:HBL_summary} the Seyfert 1.9 and 2 sources that match these criteria are summarized.
        In 59 Seyfert 2 sources a Paschen line is found and in 6/59 (10\%) a broad Paschen  component is detected, but there is no detection of broad H$\beta$ components. These are so-called hidden broad lines. Theses sources show a column density in the range of $\log (N_{\rm H}/{\rm cm^{-2}}) = 22.3 - 23.8$  (mean: $\log (N_{\rm H}/{\rm cm^{-2}}) =23.1 $.  
        As an example, the spectrum of LEDA 157443 (BAT ID 597) is shown in \autoref{fig:HBL_example} in appendix \ref{sec:hbl}, showing its broad Pa$\alpha$ line, while the Balmer lines do not have a clear broad component.
        
        Additionally for the 15 Seyfert 1.9 sources considered, 7/15 (47\%) show a broad Paschen component despite a column density above $\log (N_{\rm H}/{\rm cm^{-2}}) = 21.9$ . These cases have column densities in the range of $\log (N_{\rm H}/{\rm cm^{-2}}) = 22.0 - 23.0 $ (mean: $\log ({N_{\rm H}}/{\rm cm^{-2}}) =22.5 $).
        
        \autoref{fig:Dust_atten} (left) compares the black hole mass estimates using the FWHMs of Paschen and Balmer lines and the stellar velocity dispersion. \autoref{fig:Dust_atten} (right) compares the mass estimates when using the H$\alpha$ and the Pa$\alpha$ emission line, demonstrating an offset from the 1:1 locus. In \autoref{fig:Dust_atten_2}, the ratio between the FWHMs of H$\alpha$ and Pa$\alpha$ is compared with the column density. The ratio FWHM(Paschen)/FWHM(H$\alpha$) might increase with column density. However, this is based on the few Sy 1.9 objects having both H$\alpha$ and Paschen broad lines, and more data are needed to definitely understand whether this ratio changes with $N_{\rm H}$ (\citep[see, e.g.,][]{Ricci_DR2_NIR_Mbh}).

        \begin{table*}
            \caption{Summary of Sources with Hidden Broad Lines.} 
            \label{tab:HBL_summary}
            \scriptsize
    
            \centering
         \begin{tabular}{l l c l l l l l c c}
            \hline
         BAT ID & Counterpart & AGN Type & $\log N_{\rm H}$  & FWHM bPa$\alpha$ & FWHM bPa$\beta$ & FWHM bH$\alpha$ & $\log M_{\text{BH, vel}}$&$\log M_{\text{BH, Pa.}}$   \\
          &  &  & $[$cm$^{-2}]$ & $[$km s$^{-1}]$&$[$km s$^{-1}]$& $[$km s$^{-1}]$&$[M_\odot]$&$[M_\odot]$ \\ 
         (1) & (2) & (3) & (4)& & &&(5) &(6) \\\hline \hline
         63 & NGC 454E &  Sy 2 & $23.3\pm 0.04$ --&-- &$1430\pm30 $&--&7.63&  $6.45\pm0.05$\\
         218 & LEDA 15023 &  Sy 2 & $23.84\pm0.10$ &--& $1710\pm 30$&--&6.6&$7.0\pm0.1$\\
         493 & LEDA 1063109 &  Sy 2 & $22.3\pm0.1$ &$1650\pm160$&--&--&7.12 & $6.91\pm0.08$ \\
         511 &  SDSS J104208.36+004206.1& Sy 2  & $22.2\pm0.1$ &$2500\pm1$&-- &-- &-- &$7.8\pm0.2$ \\
         597 & LEDA 157443 & Sy 2  &$22.19\pm0.05$ &$2150\pm150$&-- &-- &8.5&$7.5\pm0.1$ \\
         1085 & ESO 234-G 050 & Sy 2 & $23.1\pm0.1$ & $1570\pm90$&$1950\pm170$&--&5.95&$6.5\pm$0.1\\
         72 & NGC 526A & Sy 1.9& $22.01\pm0.01$ & $5200\pm50$&$6620\pm130$&$4600\pm30$&7.98&$8.1\pm0.2$\\
         246 & LEDA 146662 & Sy 1.9& $22.18\pm0.10$ & $3450\pm60$&--&$5050\pm140$&8.3&$7.7\pm$0.2\\
         457 & LEDA 97526 & Sy 1.9 & $22.4\pm 0.1$ & $3600\pm80$&--&$2220\pm30$&--&$8.0\pm0.2$\\
         677 & ESO 383-18 & Sy 1.9 & $23.31\pm 0.02$& $2200\pm 30$& $2300\pm 40$&$1500\pm 30$&6.6&$6.9\pm$0.1 \\
         1138 & 2MASX J22+03 & Sy 1.9 & $22.8\pm0.1$& --&$3250\pm200$&$1730\pm30$&8.39&$7.7\pm0.2$\\
         1157 & NGC7314 & Sy 1.9 & $21.60\pm0.03$& $1510\pm20$&$1370\pm40$&$1210\pm40$&6.76&$6.32\pm0.05$\\
         1604 & 2MASX J21480531-5359413 & Sy 1.9 & $23.03\pm0.1$& --& $3140\pm240$&$2700\pm300$&7.03&$7.17\pm0.15$\\
         1625 & 2MASX J23061656-5147462 & Sy 1.9 & $21.08\pm0.2$& $4000\pm1000$&$10000\pm3000$&$2460\pm10$&8.45&$8.2\pm0.3$
         
        \end{tabular}
    
        \vspace{0.4 cm}
    
        \footnotesize 
    
        {Note. These are Seyfert 2 and Seyfert 1.9 galaxies that show broad hydrogen recombination lines and column densities $N_{\rm H}>10^{21.9}\,\rm cm^{-2}$. Only sources that have velocity dispersion $M_{\rm BH}$ estimates and/or broad H$\alpha$ FWHM measurements are considered. (1) \textit{Swift}--BAT 105 month survey identification number. (2) Name of host galaxy. (3) Optical AGN classification according to \citeauthor{Osterbrock1981} (1981). (4) Line-of-sight column densities measured by \cite{Ricci_2017}. (5) Black hole mass estimates using optical velocity dispersion measurements. (6) Black hole mass estimate from NIR Pa$\alpha$ and/or Pa$\beta$ emission line; if both lines are detected, the average mass is used.}

    \end{table*}

    \begin{table*}
        \centering
        \caption{Comparing the Mass Estimates Using Different Emission Lines: H$\alpha$, Pa$\beta$, and Pa$\alpha$. }
        \label{tab:mass_estimate_comp}
        \begin{tabular}{c c c c c c}
        \hline
         BAT ID & $\log M_{\text{BH, H}\alpha}$&$\log M_{\text{BH, Pa}\beta}$ &$\log M_{\text{BH, Pa}\alpha}$ &$\log M_{\text{BH, veldisp}}$ & bH$\alpha$/bH$\beta$   \\
         & [$M_{\odot}$]&[$M_{\odot}$] &[$M_{\odot}$] &[$M_{\odot}$] \\ \hline \hline
         72 & $7.3\pm0.1$&$8.1\pm0.2$&$8.2\pm0.2$ & 7.98 & 7\\
         246 & $7.7\pm0.1$&--&$7.7\pm0.2$& 8.3 & 23 \\
         457 & $7.1\pm0.1$&--&$8.0\pm0.2$ & -- & >300 \\
         677 & $5.7\pm0.1$&$6.8\pm0.1$&$7.0\pm0.1$ & 6.6 & 7\\
         1138 & $6.8\pm0.1$&$7.7\pm0.2$&-- &8.39 & 2\\
         1157 & $5.20\pm0.05$&$6.14\pm0.05$&$6.38\pm0.05$ &6.76 & 15\\
         1604 & $6.5\pm0.1$&$7.2\pm0.2$&-- &7.03 & 24\\
         1625 & $7.4\pm0.1$&$8.7\pm0.2$&$8.2\pm0.2$ &8.45 & >3000
        \end{tabular}
        {Note. The last column shows the Balmer decrement, a measure of the dust content, based on the broad components of H$\alpha$ and H$\beta$. }
    \end{table*}
    

\newpage
\section{Discussion}
        \subsection{Detection of Coronal Lines}

           A necessary condition for a line to be an efficient tracer of AGN activity is that it should be detected in a large number of targets.        
           We see a trend that with increasing IP, the fraction of detected lines decreases (see \autoref{fig:detec_rate} left). The most interesting CL in terms of detection and strength is the \SiVI\ emission line. A challenging factor for the detection however is that most of the X-shooter spectra are cut at around 2.1 $\mu$m. 
           {For 27/168 sources that are observed with a spectral range of 0.994-2.1\,$\mu$m, the \SiVI\ line is no longer covered for objects with $z>0.03$. In total, 65\% of our sample are observed with this limited spectral coverage setup. This partially explains why, while the detection percentage for \SiVI\  is the highest, in terms of absolute numbers, there are more detections of \SiX, which does not have a similar redshift limitation. } 
           
            Another challenge for the detection of \SiVI\ is the strong telluric CO$_2$ absorption band at similar wavelengths (1.95--2.05\,$\mu$m). Even with a good telluric correction, the S/N might not be sufficient to detect the line. However, because the \SiVI~emission line tends to be stronger (on average $F_{\rm [Si~VI]} \approx 2 F_{\rm [Si~X]}$), the main analysis is focused on the \SiVI~emission line.

            \cite{Lamperti2017} found a higher detection rate of CLs for Seyfert 1 -- 1.9 galaxies {than for Seyfert 2 galaxies (they find a rate of 53\% for Seyfert 1--1.9 and 20\% for Seyfert 2 galaxies with at least one CL)}. 
            {Despite the large bias toward Seyfert 2 galaxies in the DR2 sample}, we detect a larger absolute number of CLs compared to \cite{Lamperti2017} because we have a larger sample, and our VLT observations are of higher quality in terms of spectral resolution and sensitivity, which is essential for deblending the lines.
            While \SiX~ has a fairly high detection rate (30--40\%), \SiXI\ is {not detected in our sample}.

            A possible explanation for the nondetection could be a loss of spatial resolution, because of the increased distance to the sources \citep{Rodriguez2011} {or a generally poor resolution of the instrument}. CLs are thought to be produced in the nuclear region, between the broad-line region (BLR) and the NLR \citep{Rodriguez2006} and so they lose contrast as more nearby continuum stellar light from the host galaxy is included in the spectral aperture (\citealt{Mazzalay2007},\citealt{Mazzalay2013}).
            Additionally, the nondetection of \SiXI\ can be explained by the difference in IP and critical density. \SiXI\ has a higher IP than \SiVI~and is thus likely produced closer to the black hole (as also inferred from our discussion of CLR constraints; see section \ref{Sec:Const_CLR}). \cite{Rodriguez2011} suggest a density gradient toward the center of the AGN and the critical density of \SiXI\ is lower than the local density of where the emission is produced. As a consequence, the emission of \SiXI\ might be suppressed due to collisional deexcitation. 
            
            Concerning the strength of the \SiVI~emission line, looking at the distributions of observed luminosities (see \autoref{fig:hist_SiVI}), we see that most \SiVI~ lines are located in the luminosity range of $L_{\rm X-ray} =  10^{39} - 10^{41}$ erg s$^{-1}$ for Seyfert 1 galaxies and $L_{\rm X-ray} =  10^{38.5} - 10^{40}$ erg s$^{-1}$ for Seyfert 2 galaxies.
            Based on the \cite{Lamperti2017} NIR data in DR1, which consists mainly of Seyfert 1 galaxies, and the DR2 sample here, we find that the average flux of Seyfert 1 CLs is higher than that for Seyfert 2 galaxies, indicating that torus obscuration might play a role. 
            We do not find that Seyfert 1 sources show a higher CL detection rate than Seyfert 2 sources.
            Of the NIR high-ionization lines, the \SiVI~emission line is the strongest. 
            In terms of line flux, the median and 16$^{\rm th}$ and 84$^{\rm th}$  percentile ranges of the detected \SiVI~emission line flux are $\langle\log F_{[\text{Si VI}]}/{\rm erg~s^{-1}\,cm^{-2}}\rangle = -14.9^{+0.5}_{-0.6}$ whereas for \SiX\ $\langle\log F_{[\text{Si X}]}/{\rm erg~s^{-1}\,cm^{-2}}\rangle = -15.2^{+0.6}_{-0.3}$.

\begin{figure*}[h]
                \centering
                \includegraphics[width = 0.8\textwidth]{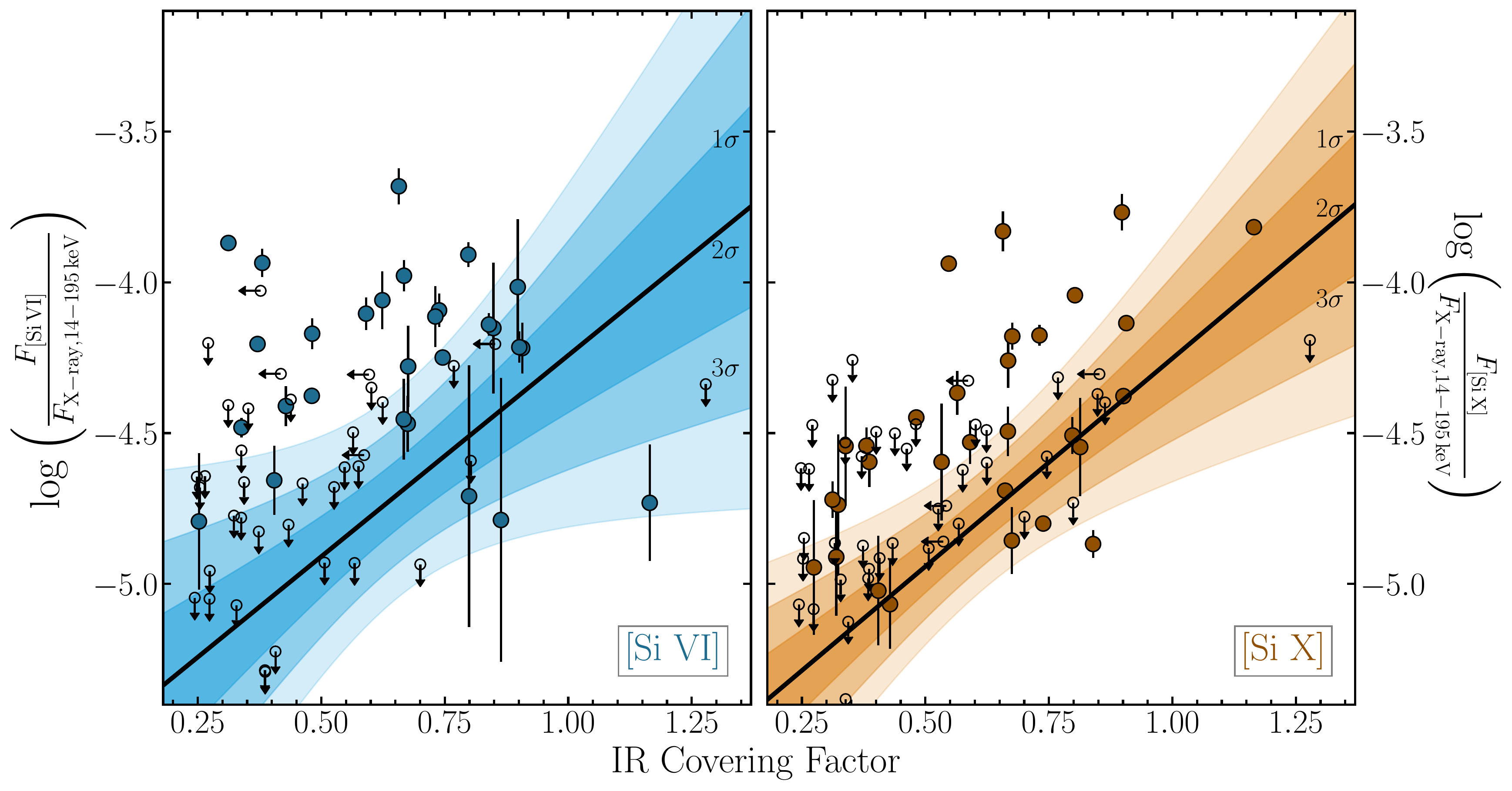}
                \caption{ { Correlation of the dust covering factor with the high-ionization line vs X-ray 14--195\,keV flux. The IR (dust) covering factor values are taken from \cite{Ichikawa2019}. We use the \texttt{Linmix} module to fit a linear regression to the data. The package can take upper limits along one axis into account when performing a linear regression using MCMC. For the fit, we exclude the upper limits in the covering factor ($x$-axis), amounting to five dismissed points for \SiVI\ (left) and four for \SiX\ (left). We include the flux ratio ($y$-axis) upper limits. The shaded regions show the pointwise 1$\sigma$, 2$\sigma$, and 3$\sigma$ credibility intervals of the linear regression fit from \texttt{Linmix}.}}
                \label{fig:cov_fitting}
            \end{figure*}
            
            {To further expand on torus obscuration, we investigate in \autoref{fig:cov_fitting} whether there is a correlation between the dusty torus covering factor and the high-ionization lines vs. the X-ray (14-195\,keV) flux. \cite{Ichikawa2019} measured the dusty (IR) covering factor in several sources in our sample.} {They computed the geometrical covering factor by assuming the two-phase torus model (described in \citealt{Stalevski2016}).} {\cite{Ricci2017Natur} found a trend of the covering factor with the Eddington ratio. Upon including the upper limit in our regression analysis with the \texttt{Linmix} module, we find evidence of a positive correlation between the covering factor and the high-ionization line flux.} {Including the upper limits in the regression analysis is important because they place strong constraints on emission at lower covering factors.} {This trend coincides with the assumption that CL emission originates from a layer close to the torus. In X-ray heated wind, the CL emission becomes more efficient \citep{Pier1995}, driving the correlation. }

            For tracing potentially obscured AGN, \SiVI\ is the most promising CL in terms of detection rate and line detection. \cite{Lamperti2017}  also found \SiVI\ to have the highest detection rate. But for the detection of intermediate-mass black holes (IMBHs; $M_{\rm BH}<10^5M_\odot$), \cite{Cann_2018} postulate that \SiVI~emission might be suppressed, making line detection more difficult. That is why we also include the \SiX~emission in our analysis, {which \cite{Cann_2018} propose} might be more prominent in the case of IMBHs.

        \subsection{Comparison of CL and X-Ray Luminosities}
        \label{Sec:Const_CLR}

             Looking at the scatter between the \SiVI~and \OIII~line and hard X-ray luminosities, the relation is tighter for \SiVI~($\sigma_{\rm [Si\ \textsc{vi}]} = 0.37$ dex compared to $\sigma_{\rm [O\ \textsc{iii}]} = 0.71$ dex) and more linear ($R_{\rm pear, [Si\ \textsc{vi}]} = 0.89$ as opposed to $R_{\rm pear, [O\ \textsc{iii}]} = 0.68$).
            A similar trend is observed when looking at the \SiX~emission line.

            There are several explanations for the scatter (besides measurement uncertainties). \cite{Rodriguez2011} suggest obscuration, as small differences in obscuration can have a significant effect on the  ratio of the luminosity of the optical [O\ \textsc{iii}] line to that of the IR [Si\ \textsc{vi}] line. Their study found a linear correlation between X-ray emission and CL emission. They found a tighter correlation for Seyfert 1 galaxies and claimed that the scatter is mainly introduced by Seyfert 2 galaxies. \cite{Lamperti2017}, however, found that there is no tighter correlation between \SiVI~emission and CL hard X-ray flux when looking at Seyfert 1 galaxies compared to Seyfert 2 galaxies. 
            We find, using the full DR1+DR2 sample, that the scatter for Seyfert 2 is tighter ($\sigma_{\rm Sy1} = 0.45$ dex and $\sigma_{\rm Sy2} = 0.37$ dex). Therefore the scatter in the CL emission compared to X-ray emission is not primarily caused by obscuration. 
            This is also evident from \autoref{fig:OIII_SiVI_obsc}, as we see that the scatter is larger for the $[$O \textsc{iii}$]$/$[$Si \textsc{vi}$]$ line ratio, which traces the IP, than for the $[$Fe \textsc{ii}$]$ line ratio, which traces obscuration \citep{Riffel:2006}. We note that aperture effects, while playing a role, do not fully explain the scatter in the figure, as even after excluding the most redshifted sources {(sources with $z>0.1$ or $z>0.3$ depending on the spectral coverage of the X-shooter setup used)}, the scatter is around 1 dex. Furthermore, also metallicity cannot be the primary cause of the large scatter of the $y$-axis, as the majority of BAT AGN host galaxies have a stellar mass $>10^{10}$ \citep{Koss2011} and have a constant metallicity gradient.

            Another factor is the physical state of the gas in the emitting media. For example, the electron gas density ($N_e$) influences the strength of the CL emission. \cite{Rodriguez2011} estimated the CL-emitting region to have a density straddling typical values for the NLR and BLR ($10^8-10^9$ cm$^{-3}$). Using detailed IFU and spectrograph studies of Seyfert 2 galaxies, \cite{Rodriguez2017a,Rodriguez2017b} also found that high values ($N_e > 10^5$ cm$^{-3}$) are very likely required. However, \cite{Landt2015} contradicted this, finding the CL gas density is low with $N_e \approx 10^3$ cm$^{-3}$.
            To fully understand the influence of the CL gas density on  high-ionization emission, more detailed studies of the gas conditions are necessary.

            A further potential explanation for the scatter is AGN variability. The NIR and X-ray observations are not made contemporaneously, leading to increased scatter. 
            Looking at \autoref{fig:comp_SiVI_SiX}, where the luminosity of \SiVI~is compared with the \SiX~luminosity (thus avoiding scatter caused by differences in observation time), the scatter is not found to be significantly smaller than when looking at the comparison of \SiVI~or \SiX~emission with the X-ray emission. So AGN variability is unlikely to account for the scatter. For the comparison of the two CLs, the scatter is $\sigma = 0.40$, while for the comparison of \SiVI\ with $L_{\rm X-ray}$ the scatter is $\sigma = 0.36$ and for \SiX\ with $L_{\rm X-ray}$  the scatter is $\sigma = 0.4$

            Another aspect is radius-dependent variability based on varying distances of the emission regions to the ionizing source. If CLs indeed originate from a region between the BLR and the NLR, then we expect them to be more correlated with the X-ray emission than, for example, with\OIII, which is simply due to the light traveling time; regions that are located further away than the typical X-ray variability timescales will show less variability.
            
            Since the detection frequency and the flux of CLs in Seyfert 2 galaxies are (on average) lower than those in Seyfert 1 galaxies (see \autoref{fig:hist_SiVI}), obscuration could play a role in CL detection.
            This strengthens the argument that the CLs are produced in the region between the BLR and NLR, and are thus affected by obscuration by the torus. Thus the CLRs seem to be an extended region in accordance with previous studies (e.g. \citealt{Rodriguez2006}, \citealt{Landt2015}). 
            In fact, the \OIII\ emission shows indeed a larger scatter and lower correlation with the X-ray luminosity, when compared with the \SiVI~emission, though the difference is not significant.
            
\begin{figure*}
                \centering
                \includegraphics[width = \textwidth]{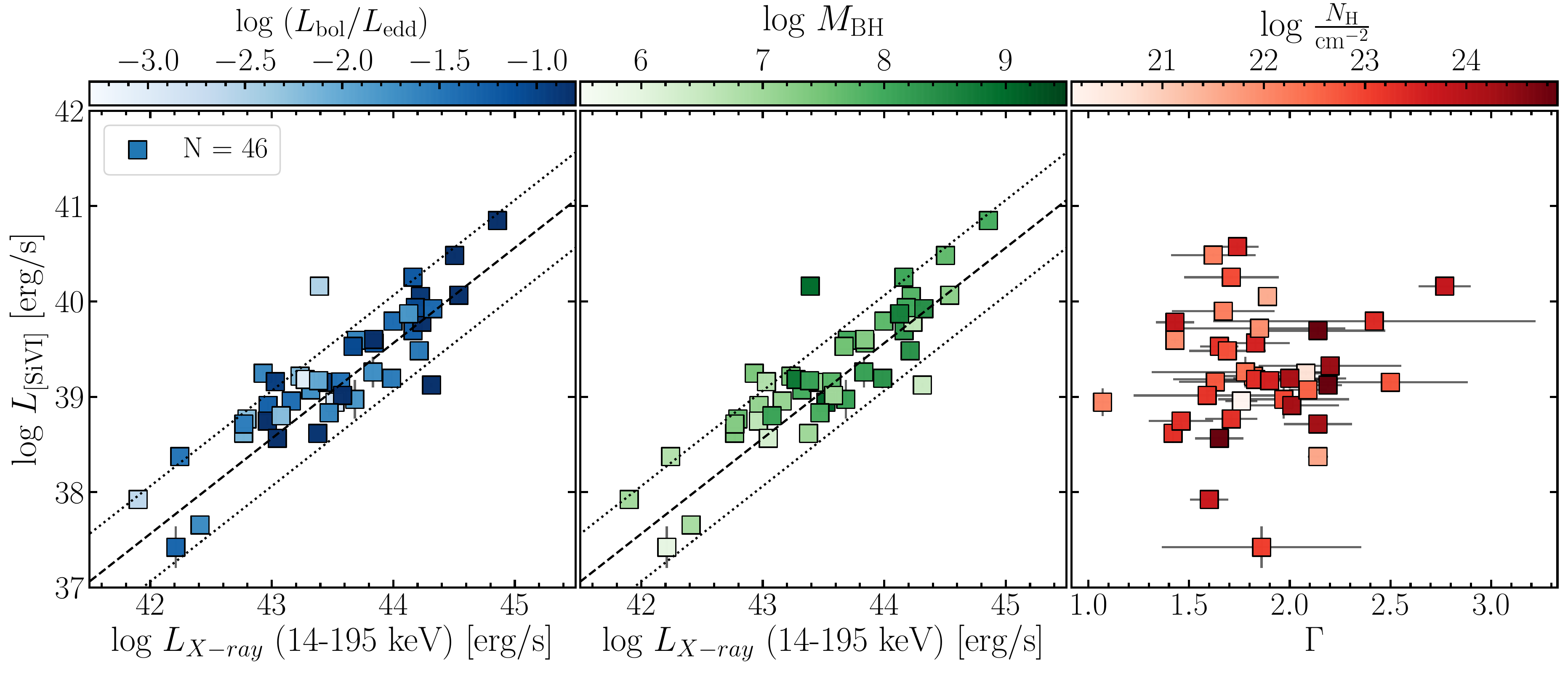}
                \caption{(Left) Correlation of \SiVI~emission with intrinsic hard X-ray luminosity ($14-195$\,keV), colored by Eddington ratio. (Center) Comparison of \SiVI\ and observed hard X-ray 14-195 keV emission, color-coded by SMBH mass. (Right) Comparison of the X-ray photon index $\Gamma$ as a function of the \SiVI\ line emission colored by hydrogen column density. {Many of our sources are from the \textit{Swift}--BAT 105 month sample, which do not yet have intrinsic X-ray measurements. Consequently a lower number of sources are presented here in this plot.} }
                \label{fig:SiVI_Ledd}
            \end{figure*}

            To further investigate the correlation of \SiVI~with X-ray emission, we consider the Eddington ratio ($L_{\rm bol}/L_{\rm Edd}$) dependence (see \autoref{fig:SiVI_Ledd} top left). Previous studies by \cite{Oh2017_2,Oh2019} found a correlation between the Eddington ratio and narrow-line emission. The correlation is most likely caused by X-ray heating processes or removal of material by an energetic outflow.

            Looking at \autoref{fig:SiVI_Ledd}, there is no clear dependence visible for the correlation with the Eddington ratio ($L_{\rm bol}/L_{\rm Edd}$), at least for the high-ionization CL considered here.

           In \autoref{fig:SiVI_Ledd} (right), we compare the \SiVI~emission with the power-law photon index $\Gamma$. X-ray spectra can be described to first order by a power law, parameterized by the photon index $\Gamma$ (for a more detailed description see \citealt{Ricci_2017}). In a previous study, \cite{Rodriguez2011}  claimed to have found a linear correlation between $\Gamma$ and CL emission, while we find none. They postulated that CL emission is predominantly present in sources with a soft excess (when $\Gamma\ge2.5$). 
           We do not see any correlation when looking at the data. Furthermore, we observe hardly any sources with $\Gamma\ge2.5$. This could result from bias in our sample, which are predominantly Seyfert 2. Another reason why we do not find any correlation could be that we have a larger sample than they had (we have $N = 36$, \cite{Rodriguez2011} had $N=13$), we cover a larger energy-range ($0.5{-}150$ keV against $0.1{-}2.4$ keV) and the photon index is estimated using a more sophisticated model that includes higher-energy photons.  It is also possible the soft excess at low energies may be important in the correlation. The column density is also shown in \autoref{fig:SiVI_Ledd} (right). We also do not observe any column density dependence, again indicating that obscuration seems not to cause the large scatter, nor any obvious bias.

           Another influencing factor in CL emission could be the central mass of the black hole. Comparing the mass of the black hole ($M_{\rm BH}$) with CL emission, only a weak correlation ($R_{\rm pear} =0.48$ for \SiVI) is found (see \autoref{fig:M_bh_CL}). This is also seen in \autoref{fig:SiVI_Ledd}, where higher-mass sources are located more frequently in the range of higher \SiVI~ luminosity values. The mass range covered by our measurements reaches $\log M_{\rm BH}/M_\odot = 6.5-9 $. A reason for the weak correlation is also the fact that only a small luminosity range is covered. 
           Based on our data, the luminosity of CLs is not a good indicator of black hole mass.
            However, CLs might be a good tracer of IMBHs according to \cite{Cann_2018}, who found a correlation between \mbh\ and CL ratios. According to their theoretical models, for masses $\log M_{\rm BH}/M_\odot<6$, the ratio between the fluxes of \SiVI\ and \SiX~changes by over seven orders of magnitude. Unfortunately however, our data points do not cover this mass range and go only to $\log M_{\rm BH}/M_\odot>6.5$.
            The drop of over seven orders of magnitude is explained by the interplay of black hole mass, ionization parameters, and physical properties of the gas, in which, at low masses, the effective number of ionizing photons is a strong function of black hole mass for a fixed Eddington ratio based on standard disk theory. This drop, however, also has implications for the search for IMBHs using, for example, the \SiVI\ emission line. If the calculations are correct, the ratio peaks precisely in the range $\log M_{\rm BH}/M_\odot = 6-8$. This would mean that the high detection fraction of \SiVI\ may potentially be a selection effect.
            However, the drop for sources with $\log M_{\rm BH}/M_\odot>8$ is not seen in our data. The fact that we do not see the predicted drop of the CL emission ratio at $\log\mbh/M_\odot>8$ suggests that the sources have a strong UV emission even at high masses (that are capable of ionizing the species). It must be noted that the theoretical calculations are based on fixed parameters, such as $L_{\rm bol}/L_{\text{Edd}} = 0.1$. The BASS sample covers a broad range of $L_{\rm bol}/L_{\text{Edd}}$ at every \mbh\ (see also \citealt{Koss_2017}), providing  a broader range in parameter space than the \cite{Cann_2018} models.  
            {We also note that in the BAT sample, high-mass sources have lower Eddington ratios \citep{Ricci2017Natur}. Such lower Eddington ratios may change the UV ionizing spectrum \citep{Lusso2010} altering the relation used in \cite{Cann_2018}.}
        
            {We also investigate the connection between the S/N and the scatter of the CL emission with X-ray luminosity. We separate the data into high and low S/N and compare the scatter. We find that the scatter stays constant irrespective of the S/N cut applied and we see that the points follow the same distribution according to the 2D Kolmogorov--Smirnov test described in \cite{Fasano1987}.}
        
            So to answer the question about CL correlation with X-ray emission, we indeed see the trend that with increasing X-ray emission, the emission of the CLs \SiVI~and \SiX~ increases. The scatter is smaller, but comparable with that of \OIII. Will this suffice for CLs to be considered efficient tracers of AGN activity? This is not evident from our analysis. What we have shown is that high-ionization lines are detectable, even with the challenges of telluric absorption. Furthermore, they show a relation to other properties of the accreting system, such as the mass $M_{\rm BH}$ or the X-ray luminosity. However, the scatter
            is still quite large ($\sigma \sim 0.4$ dex). An additional advantage is that in obscured AGN (Seyfert 2 galaxies) the CLs are also detected. To fully answer the question about the efficiency of production, {we need a wider range of luminosities and Edington ratios,} including, e.g., galaxies that are likely to host IMBHs (i.e. not the typical BASS sources).

    \begin{figure*}
        \centering
        \includegraphics[width =0.65\textwidth]{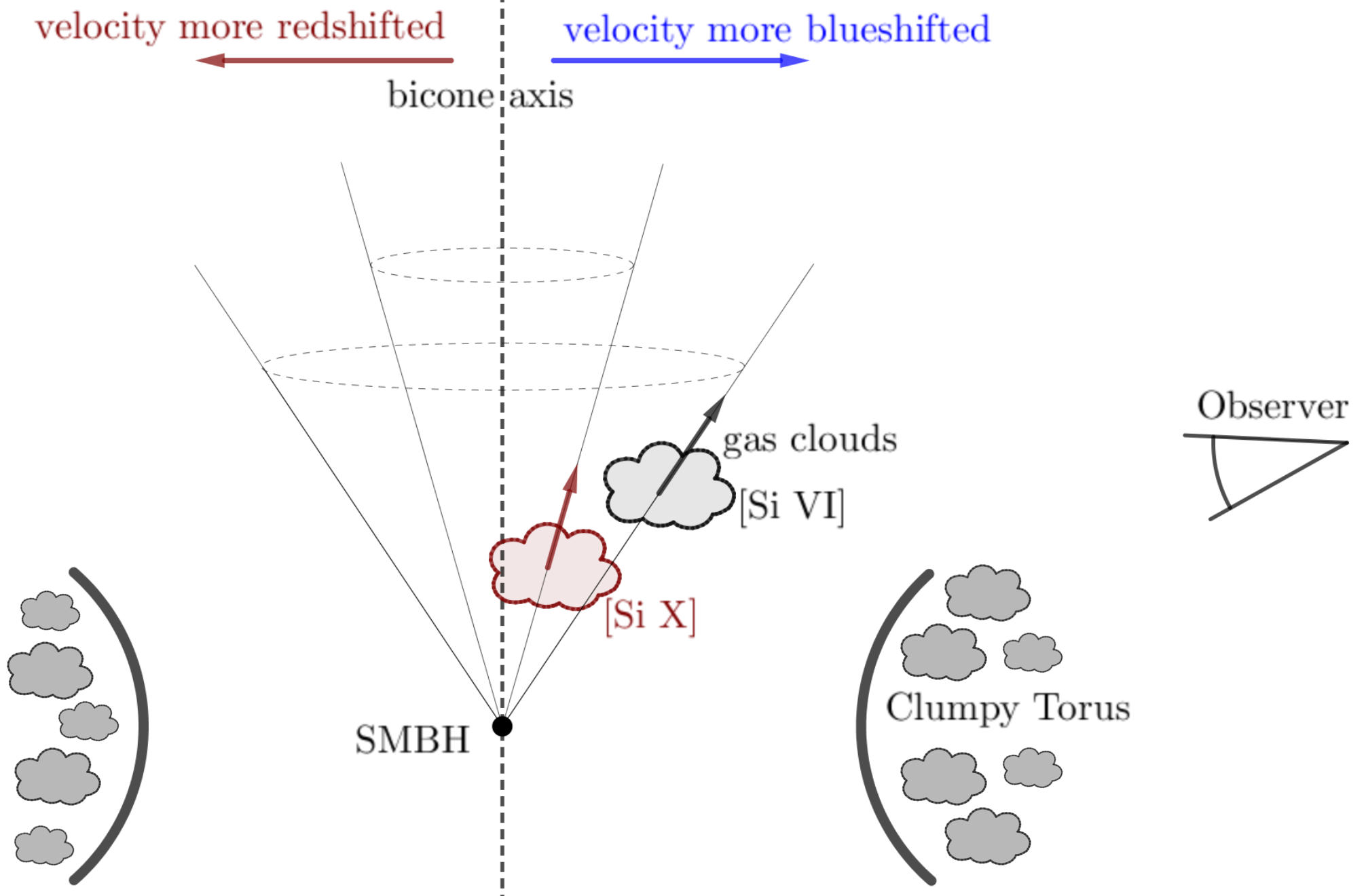}
        \caption[Outflow of Gas Clumps]{Outflow movement of gas clumps. Clumps closer to the bicone axis are more highly ionized. In a Seyfert 2 configuration, the projected velocity (i.e. the velocity observed by us) is indeed more blue shifted for species further away from the axis, because the velocity component toward the observer is larger.}
        \label{fig:CLR_outflow}
    \end{figure*}
    
    \subsection{Constraining the Geometry of the CLR}

    The link between the FWHM and the IP of CLs has already been intensively studied in previous works. \citeauthor{Giannuzzo1995} (1995) postulated that the CLR might occupy different regions in different galaxies based on the wide range of CL FWHMs. Later studies (\citealt{Reunanen2002}, \citealt{Rodriguez2011}) have found a correlation between the FWHM and IP up to some IP for certain cases.
    Up to energies of 200 -- 300 eV, we also observe an increase of the FWHM with increasing IP. If we take further high-ionization lines into account, we see a drop in the FWHM again. 
    However, because the highest ionizing species (>400 eV) are relatively weak, it is difficult to make conclusions.
    \cite{Rodriguez2011} attribute the finding that the increase of the FWHM with IP is only seen up to 300 eV to the combined effect of the electron density gradient toward the center and a spatial extension of the emitting material.
    The critical density can give insights into possibly why the highest-ionizing species are not observable. One of the common lines detected, \SiVI, has a critical density of $n_c = 2.5\times10^9$ cm$^{-3}$. This sets an upper limit, because certain higher-ionization lines, such as \SXI, have an order of magnitude lower critical density ($n_c =3.2\times10^{8} $ cm$^{-3}$). If they are produced closer to the center and there is an increase in density toward the center, \SXI~emission might be suppressed due to collisional deexcitation. The lower critical densities for lines with IPs higher than 350eV are a possible explanation for why the detection frequency drops sharply at high IP (see \autoref{fig:detec_rate}).
    The FWHM measurements tell us approximately how far from the central ionizing source a certain line is produced  within the CLR for a given black hole. Different ionization species will be dominant in different regions, with a mild dependence on \mbh.
    
    \cite{Murayama1998} proposed that high-density clumps that are radially moving outward produce the CLs. The high density clumps are separated into various segments of different dominating ionizing species. This model cannot explain the difference in FWHM with increasing IP, as the FWHM should be similar, because the high-density clump moves as a whole.
    The model by \cite{Fischer2017} can already better explain the finding, because it allows for different velocity dispersions in the infalling dust spirals.
    
    Another finding is that the offset of the line peak, i.e. the bulk motion of the emitting material with respect to the observer, is systematically blueshifted for the \SiVI~emission line (see \autoref{fig:offset_analysis_SiX_SiVI}), while for higher-IP species, the offset is actually redshifted with respect to \SiVI\ (i.e., they always have a smaller bulk offset).
    This can be understood in terms of the general geometry of the CLR: Incoming or accreting gas is mainly ionized as it enters the bicone axis (according to \citealt{Fischer2017}). Due to the hard radiation field or outflows, the material can be accelerated outward. If we observe an AGN in a Type 2 orientation, and the more highly ionized gas is closer to the bicone axis, the higher-ionization lines could be observed at lower apparent radial velocities (the flow being more directed along the plane of the sky). This could explain why all \SiVI~emission lines are blueshifted with respect to the NLR{, while \SiX~emission is both red- and blueshifted (with respect to the NLR)}. Outflows have also been found in previous IFU studies looking at the NLR and CLR \citep{Mueller2011}. {The simplified concept is illustrated in \autoref{fig:CLR_outflow}, similar to the illustration in \cite{Murayama1998}. The central AGN is shielded by a dusty torus \citep{Marinucci2016, Ramos_Almeida2017} from the observer in the case of galaxies with large covering factors, which is generally the case in Seyfert 2 galaxies. The clouds move outward along different ionization cones, which leads to different observed (relative) velocity shifts/offsets of the CL species.} {The highest-ionization lines move along a narrower cone closer to the bicone axis, while CLs with a lower ionization move in a wider cone. In a Type 2 configuration, clumps moving in wider cones have a larger velocity component along the line of sight. This could explain why they are more blueshifted (e.g., \SiVI\ as opposed to \SiX; see \autoref{fig:offset_example}). Especially in the case of a single cone (or because the second cone is more heavily obscured), if the axis points slightly away from the observer, it might explain why some of the highest-ionization lines are even slightly redshifted as compared to the narrow-line emission. 
    }
    A further indication that this phenomenon is an orientation effect can be seen in \autoref{fig:offset_analysis_SiX_SiVI}. A trend can be seen where the sources with lower column density show that both lines are blueshifted, meaning a more face-on look into the center of the bicone.
    Also, there is a component of gas close to the AGN, essentially at the apex of the bicone. This usually has higher ionization (see \citealt{Fischer2017}) than the
    rest of the NLR and does not seem to fit the flow pattern of the more extended (in situ) gas. This could explain the large scatter in offset of the \SiX~emission line.

    So from the offset measurements of the CLs, we get constraints of the overall geometry of the CLR. From the FWHM measurements, we have indications that the emission comes from different regions within the CLR and from the offset analysis, we have indications that most of the ionization takes place along the bicone axis, and as a result of orientation effects, this causes the more highly ionized CL species to show a different offset than the other emission lines.
    
    The structure of the CLR has been addressed in past IFU studies \citep{Mueller2011, Mazzalay2013, Rodriguez2017b, May2018, May2020, Rodriguez2020} and extended CL emission with $\chi>400$\,eV could be observed. However, resolving the innermost parsec region is not yet possible with current instrumentation.
    
    {We would like to emphasize that the analysis presented has some limitations. Outflows in Seyfert galaxies are likely complex and more complicated than the conical outflow depicted in the simple sketch in \autoref{fig:CLR_outflow}. Besides linear outflow kinematics, rotational kinematics are possible for the CLR and NLR \citep{Mueller2011}. In addition, a previous study focusing on the NLR has found that some ionized outflows are hollow \citep{Fischer2013}. In the case of a hollow structure, the geometry of the different ionization cones would be more similar, and the picture presented in \autoref{fig:CLR_outflow} would not sufficiently explain our findings. However, it is not clear that the CLR follows the same geometry as the NLR, as the CLR is generally situated closer to the center of the AGN \citep{Olivia1997, Mazzalay2010}.}
        
    \subsection{Hidden Broad Lines}
                
        We detect broad emission lines in the NIR in a handful of sources that are optically classified as Seyfert 2 galaxies. These sources presumably consist of AGN where the line of sight is impacted by a moderate column density, and hence by extinction, such that the BLR is completely obscured yet they have a column density above $\log (N_{\rm H}/{\rm cm^{-2}}) = 21.9$, to place them in context. 
        Previous studies (e.g. \citealt{Garcet2007}; \citealt{Oh2015}; \citealt{Kamraj2019}) have found sources that have high column density yet show optical broad lines.   

        As can be seen in \autoref{fig:HBL}, for 6/59 (10\%) of Seyfert 2 galaxies, broad Pa$\beta$ or Pa$\alpha$ components are detected. This is consistent with the 9\% fraction found by \cite{Lamperti2017}. Furthermore, if we include Seyfert 1.9 galaxies, we detect broad components in 12/75 (16\%)  sources. This is lower than the 31\% fraction found by \cite{Lamperti2017} and the 32\% fraction found by \cite{Onori2017}. The reason why we have such a low fraction is most likely the low number statistics, as only 12/75 (15\%) in our sample\footnote{Our complete X-shooter sample includes 168 sources. However, for some sources, we do not have column density or Pa$\beta$ measurements, which reduces our sample size from 168 total to 68 Seyfert 1--1.9 galaxies.} are classified as Seyfert 1.8 or 1.9 galaxies. Are these sources challenging the unified model?
        
        \cite{Lamperti2017} found that the Seyfert 2 cases with broad NIR components occupy the bottom 11$^{\rm th}$ percentile of column densities ( $\log(N_{\rm H}/{\rm cm^{-2}})$ = 22.4). Focusing on Seyfert 1.9 and Seyfert 2 galaxies, we find a much broader range, extending up to  $\log(N_{\rm H}/{\rm cm^{-2}})$ = 23.8 (median column density:  $\log(N_{\rm H}/{\rm cm^{-2}})$ = 23.3).
         We find that for at least 10 \% of Seyfert 2 galaxies, one can detect broad components, so-called hidden broad lines, in the NIR, which then can be used to estimate the mass of the central black hole.
        The reason why broad lines are detected in the NIR and not in the optical is mainly the decreased obscuration at longer wavelengths. \cite{Lamperti2017} found that sources with hidden broad lines are often merger systems, so the optical broad emission component is most likely obscured by the host galaxy's dust rather than by the nuclear torus.  The \oiii\ to X-ray luminosity ratio is also found to be lower in merging BAT AGN systems \citep[e.g.,][]{Koss:2010:L125,Koss:2011:L42, Koss:2012:L22} and most of the late-stage, close nuclear ($<$3 kpc) mergers are found in optical Seyfert 2 systems \citep{Koss:2018:214a} rather than in broad-line AGN consistent with this claim.  Higher X-ray obscuration is also found to correlate with later merger stages \citep{Koss:2016:85,Koss:2016:L4, Ricci2017MNRAS} and has been predicted by theoretical studies \citep[e.g.,][]{Hopkins:2006:864,Blecha:2018:3056}.  

        If true, this indicates that hidden broad lines are not a confutation of the unification model, because the obscuration of the broad components is not due to the torus, but rather due to more extended host galaxy dust and gas.
        Indeed, one of the AGN counterparts with hidden broad lines we detect,  2MASX J042340.80+04080.17, shows a spiral companion indicating a possible merger event \citep{Goncalves1999}. The second example, ESO~383-18, shows dust winds and Compton-thin dust lanes, which could cause the optical broad lines to be obscured \citep{Ricci2010}.
       NGC 4941 is a Seyfert 2 galaxy and is marked as a galaxy without hidden broad lines in \cite{Yu2005}, even though we detect a broad Pa$\alpha$ emission component. This galaxy shows no signs of large-scale interactions. 
        
        Sources with hidden broad lines are also interesting to investigate in terms of potential differences between the optical and NIR broad-line properties, and how they affect the estimation of the mass of the black hole. In the case of Seyfert 1.9 galaxies, broad H$\alpha$ can be attenuated by dust leading to a low black hole mass.
        \autoref{fig:Dust_atten} presents the value of the mass estimation from velocity dispersion measurements and the broad Paschen line method. No structural offset can be seen and points are spread out equally on both sides of the 1:1 relation line and hence the Paschen lines appear to yield reliable estimates. 
        We compare the mass estimates from the various hydrogen recombination lines H$\alpha$, Pa$\beta$ and Pa$\alpha$ in \autoref{tab:mass_estimate_comp} and \autoref{fig:Dust_atten}. The mass estimation from H$\alpha$ emission is  lower by approximately 1 dex. 
        A similar result is found when comparing the broad H$\alpha$ estimates for black hole mass based on velocity dispersion measurements.
        
        To study whether the cause of the bias when using the broad H$\alpha$ to estimate the mass is dust extinction, we compare the ratio between the FWHMs of broad Pa$\alpha$ and H$\alpha$ with the hydrogen column density and Balmer decrement (\autoref{fig:Dust_atten_2}).
        Based on the limited number of cases, a trend can be seen between the FWHM ratio and the hydrogen column density, potentially indicating that the Paschen lines are broader than H$\alpha$ for higher column densities. This lets us conclude that obscuration indeed causes the H$\alpha$ line to be attenuated as the Paschen lines are less affected by reddening. Is the obscuration indeed due to dust? 
        To understand any potential trend, however, a larger sample of Seyfert 1.9 galaxies would be necessary. The reason for this is most likely that the Balmer decrement is more complex: \cite{Pottasch1960} noted that the Balmer line optical depth can also lead to larger Balmer decrements when the gas is optically thick, H$\alpha$ is scattered, and H$\beta$ is absorbed and reemitted as Pa$\alpha$ or H$\alpha$. As a consequence, H$\alpha$ emission gets stronger and the Balmer decrement increases.  This means that the use of the Balmer decrement as an indicator of reddening due to dust is potentially not valid.

        To conclude our analysis on the use of the Paschen lines, we find that they provide reliable estimates of the black hole mass  for Seyfert 1.9 and 2 galaxies assuming that velocity dispersion measurements of the black hole mass are robust.
        The use of broad H$\alpha$ for mass estimation is already established for Seyfert 1 -- 1.8 galaxies \citep{Mejia2016}. For Seyfert 1.9 galaxies, the use of H$\alpha$ for the mass estimation is shown to be less robust, as there is clearly an offset when comparing with measurements based on Paschen lines and stellar velocity dispersions.
        To fully quantify the bias when using H$\alpha$ in the case of Seyfert 1.9 galaxies, we need more sources. Our analysis relies on sources for which the column density is determined in order to find cases with hidden broad lines (see \autoref{fig:HBL}). With upcoming data releases from the BASS project, we will have more Seyfert 1.9 galaxies to work with.\\

        \subsection{Outlook for Studies Using JWST}
        {In this study, we provide the largest NIR spectroscopic census and legacy database for nearby AGN using the large collecting area of the VLT.  The AGN luminosities of our sample ($L_{\rm bol}\sim10^{43}-5\times 10^{45}$ \ergps) are similar to the luminosities of AGN at the epoch of  the peak of black hole growth at $z\sim1{-}2$ (e.g., \citealt{Aird2015}). Our spectra thus provide a useful high-resolution, high-S/N template for higher-redshift AGN ($\sim z=1{-}2$). With the advent of \textit{JWST}, unprecedentedly deep CL surveys will be possible. On board the satellite is the Near-infrared Spectrograph (NIRSPEC; \citealt{Dorner2016}), which is an NIR multi-object dispersive spectrograph. It operates in the $1{-}5\,\mu$m regime and can simultaneously observe more than 100 slits.  This large spectroscopic sample will have immense legacy value for NIRSPEC/\textit{JWST} in the full 1-5 \micron\ range (z=1--3, $\sim$0.3--2 \micron\ rest frame). While the spectral resolution is slightly lower than that of X-shooter spectra ($R\sim1000$), it is still sufficient for resolving CLs \citep[e.g. see][]{Lamperti2017}. 
        An exposure time of $10^5\,$s is expected to yield an ${\rm S/N}=3$ sensitivity of $\sim 2\times 10^{-19}\,{\rm erg\,cm^{-2}\,s^{-1}}$ at $2\,\mu$m. This is 1000 times more sensitive than our \SiVI\ observations ($\sim2\times 10^{-16}\,{\rm erg\,cm^{-2}\,s^{-1}}$). Translating this to 14-195 keV X-ray flux using the line ratio we find between \SiVI\ and the X-ray flux, the limit  corresponds to a flux of  $\sim5\times 10^{-15}\,{\rm erg\,cm^{-2}\,s^{-1}}$.
        This is ${\sim}1000$ more sensitive than the sensitivity limit of the 105\,month deep \textit{Swift-BAT} survey (the 105 month survey reaches >50\% completeness at that sensitivity; \citealt{Oh_2018}).
        Consequently, with NIRSPEC/\textit{JWST}, it is potentially possible to observe highly obscured (Compton-thick) AGN missed by X-ray surveys as well as much fainter sources. As X-ray confusion can be a problem for low-luminosity AGN, it may be possible to detect them in the NIR with the CLs discussed in this paper, besides other NIR (high-ionization) lines \citep{Satyapal2021}.
        We note, however, that with greater sensitivity AGN CL emission may be difficult to distinguish from other emission mechanisms such as shocks \citep{Rich2011}.  }


\section{Summary and conclusions}
In this work, we analyze 168 NIR spectra of nearby  ($z < 0.6$) hard X-ray selected AGN from BASS.
    First, we look at high-ionization lines in the NIR spectrum of these nearby AGN: 
      \begin{itemize}
        \item  We find CLs in more cases than found by previous studies. We find that 49/109 (45\%) Seyfert 2 and 35/58 (60\%)  Seyfert 1 -- 1.9 galaxies show at least one NIR high-ionization line. 
        
        \item 
        The correlation of \SiVI\ with the X-ray emission shows considerably less scatter (0.37 dex) than the correlation of the \OIII\ emission line (0.71 dex) with the X-ray;  however, its scatter of $\sigma \gtrapprox 0.4$ dex is still significant.
        
        \item The \SiVI, \SiX, \SVIII, \SIX, and \SXI\ emission line FWHMs and offsets all show dependence on the IP of the line. This is a clear indicator that the emission is coming from different locations within the CLR and cannot be explained by a homogeneous distribution of the ionized species.
        
        \item Studying the sources with hidden broad lines case by case, we find indications of galaxy-scale interactions and obscuration from extended dust lanes. The lack of broad optical emission line components can be explained by obscuration due to dust or gas in the environment of the host galaxy rather than by obscuration by the nuclear torus.
        
        \item  NIR hidden broad lines can be used to estimate the black hole mass. Mass estimations using the FWHM of Pa$\alpha$ and Pa$\beta$ are in accordance with estimations from velocity dispersion measurements. On the other hand, the H$\alpha$ width underestimates the mass in Seyfert 1.9 galaxies.
    \end{itemize}
    
    {This study provides a benchmark investigation of the use of CL emission as a tracer of AGN activity using the largest assembled NIR rest-frame sample to date. With next-generation NIR instruments, particularly JWST, deeper and more sensitive observation will be possible. As such, it will be possible to observe highly obscured (Compton-thick) AGN missed by X-ray surveys and much fainter sources, expanding our understanding of the AGN  population.}




\begin{acknowledgments} 
We acknowledge support from NASA through ADAP award NNH16CT03C and 80NSSC19K0749 (M.J.K.); the European Research Council (ERC) under the European Union’s Horizon 2020 research and innovation programme through grant agreement No.726384/Empire (J.S.d.B.);
the Israel Science Foundation through grant number 1849/19 (B.T.); 
the European Research Council (ERC) under the European Union's Horizon 2020 research and innovation program, through grant agreement number 950533 (B.T.); the Comunidad de Madrid through the Atracción de Talento Investigador Grant 2018-T1/TIC-11035 (I.L.);
the National Research Foundation of Korea award NRF-2020R1C1C1005462 (K.O.); the Japan Society for the Promotion of Science,  ID: 17321 (K.O.); the Jet Propulsion Laboratory, California Institute of Technology, under a contract with NASA (D.S.); Swiss National Science Foundation grants PP00P2\_163824 and PP00P2\_190092 and the ERC under the European Union’s Horizon 2020 research and innovation programme grant agreement No 864361 (S.C.); ANID grants CATA-Basal AFB-170002 (F.E.B., F.R., E.T.) and FB210003 (F.E.B., E.T.), FONDECYT Regular 1190818 (E.T., F.E.B.) and 1200495 (F.E.B., E.T.), Fondecyt Iniciacion 11190831 (C.R.), FONDECYT Postdoctorado 3180506 (F.R.), and Millennium Science Initiative ICN12\_009 (F.E.B.); the Conselho Nacional de Desenvolvimento Cient\'ifico e Tecnol\'ogico
(CNPq), CAPES and FAPERGS (R.R.);
 CNPq, through grant 312036/2019-1 (A.R.-A.); and the Ministry of Education, Science and Technological Development of the Republic of Serbia through contract No.
451-03-9/2021-14/200002 and the Science Fund of the Republic of Serbia, PROMIS 6060916, BOWIE (M.S.). This work was performed in part at the Aspen Center for Physics, which is supported by National Science Foundation grant PHY-1607611. 
\end{acknowledgments}

\facilities{ESO-VLT, Swift (BAT)}

\software{\texttt{astropy} \citep{Collaboration:2013:A33},  
          \texttt{Matplotlib} \citep{Hunter:2007:90}, 
          \texttt{NumPy} \citep{vanderWalt:2011:22},
          \texttt{Linmix} (https://github.com/jmeyers314/linmix),
          \texttt{PySpecKit} v0.1.20 \citep{Pyspeckit},
          ESO Reflex software v2.9.3 \citep{Freudling2013},
          \texttt{molecfit} v1.5.9 \citep{Kausch2015,Smette2015}.}



\bibliography{references,bib_add,DR2_bib} 



%
\renewcommand\thefigure{\thesection.\arabic{figure}}    
\setcounter{figure}{0} 

\renewcommand\thetable{\thesection.\arabic{figure}}    
\setcounter{table}{0} 

\appendix
\section{Observational Data}
\label{app:obsdat}
\begin{figure*}
    \centering
    \includegraphics[width=\textwidth]{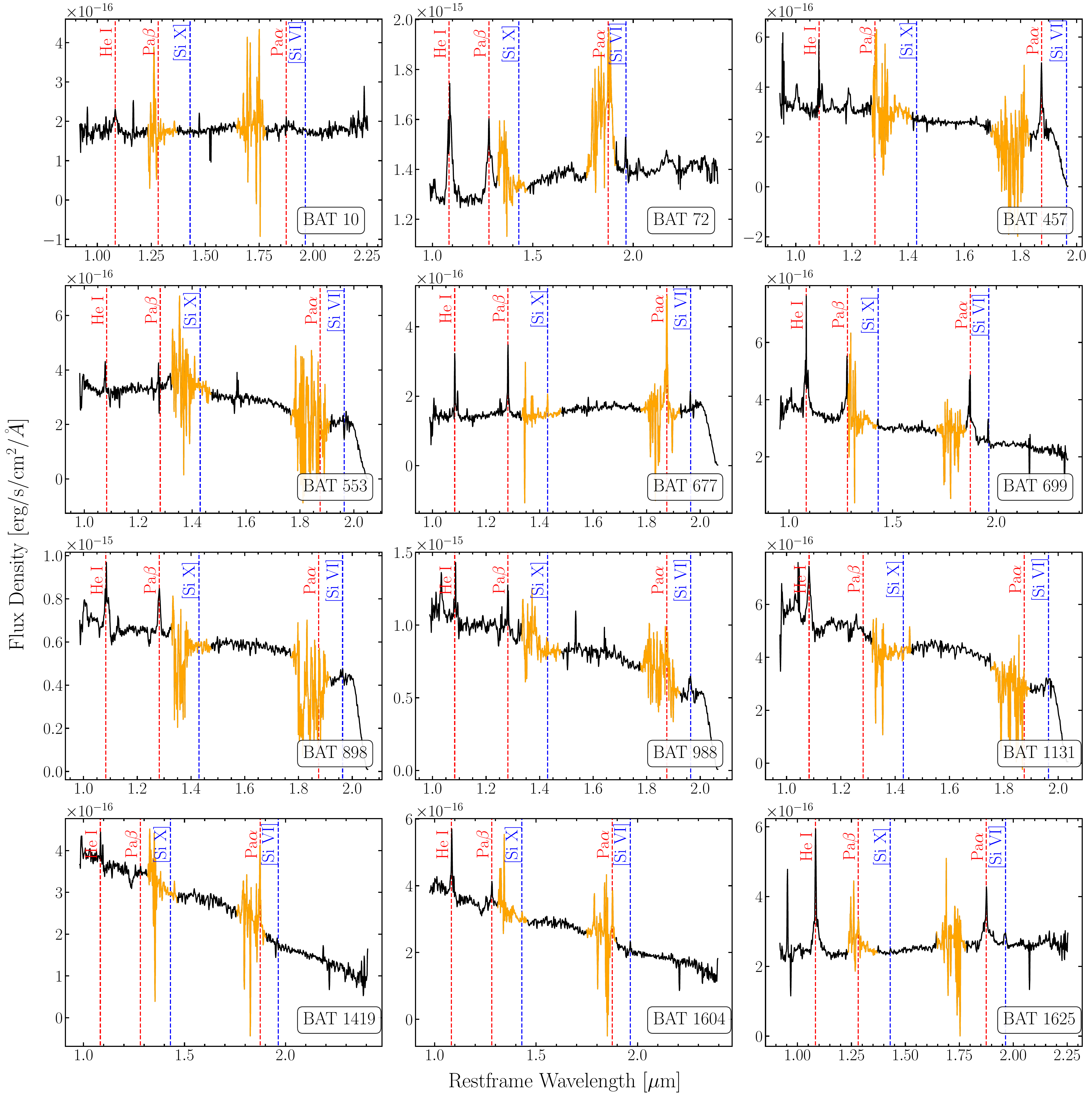}
    \caption{Compilation of a selection of 12 spectra from the sample investigated and presented in this study. The spectra have been corrected for telluric absorption. Regions that are potentially heavily affected by telluric absorption even after correction are indicated in orange.  }
    \label{fig:spectra_all}
\end{figure*}

\autoref{tb:summary_obs} lists the observations that are part of the sample studied in this work. This portion of the whole table is a guide to the reader. The complete table can be found in the online journal. The table gives information about the observational setup and further properties of the sources. {In \autoref{fig:spectra_all} we show the full observed NIR X-shooter arm observations for a set of 12 sources from our sample.}

\begin{table*}
    \caption{Summary of Observations Used in This Work.}
    \label{tb:summary_obs}
      
    \begin{center}
    
    \scriptsize
    
        \begin{tabular}{l l c c c c c c c c c c}%
    
        \hline
        \bfseries ID & \bfseries Counterpart & \bfseries Redshift & \bfseries Date & \bfseries Exp. Time & \bfseries Airmass & \bfseries Seeing& \bfseries $R$  & \bfseries Program & \bfseries slit &\bfseries slit &\bfseries Coverage\\
        &&&[dd.mm.yy]&[sec]&&[$\arcsec$]&&&[$\arcsec$]&[kpc]&\\
        (1)&&&&&&&&(2)&(3)&(3)&(4)\\
        \hline
        \begin{filecontents*}{Tables/sample_description_short.csv}
batid,Counterpart,z,slitsizeinch,slitkpc,date,coverage,airmass,exptime,Type,resol,seeing,ProgID
17,ESO112-6,0.029,0.9 / 4,0.48 / 2.13,01.02.17,lim.,1.93,480,2,5400,0.76,098.A-0635(B)
20,2MASX J00-04,0.213,0.9 / 4,3.37 / 14.98,02.12.16,lim.,1.12,960,2,5400,0.88,098.A-0635(B)
31,MCG -02-02-095,0.018,0.9 / 4,0.3 / 1.3,30.09.17,lim.,1.04,960,2,5400,0.77,099.A-0403(B)
32,ESP 39607,0.201,0.9 / 4,3.18 / 14.13,26.11.16,lim.,1.13,960,2,5400,0.98,098.A-0635(B)
37,2MASX J00-27,0.077,0.9 / 4,1.25 / 5.56,03.12.16,lim.,1.06,960,2,5400,1.26,098.A-0635(B)
50,ESO 243- G 026,0.019,0.9 / 4,0.31 / 1.38,24.08.18,full,1.16,480,2,5400,0.82,0101.A-0765(A)
52,HE 0103-3447,0.057,0.9 / 4,0.94 / 4.18,28.01.17,lim.,1.55,480,1,5400,1.08,098.A-0635(B)
53,2MASX J01+06,0.041,0.9 / 4,0.68 / 3.02,01.10.17,lim.,1.27,960,2,5400,0.69,099.A-0403(B)
57,3C 033,0.059,0.9 / 4,0.97 / 4.31,30.09.17,lim.,1.37,960,2,5400,0.67,099.A-0403(B)
65,2MASX J01-12,0.142,0.9 / 4,2.29 / 10.18,19.07.17,lim.,1.05,1920,1.9,5400,0.99,099.A-0403(B)
\end{filecontents*}
\csvreader[head to column names]{Tables/sample_description_short.csv}{}
        {\\ \batid & \Counterpart & \z &\date & \exptime& \airmass &\seeing&\resol  & \ProgID & \slitsizeinch&\slitkpc&\coverage}\\
        \hline\\
        \end{tabular}
        \end{center}
        \footnotesize
        
        \noindent \textbf{Note.} (This table is available in its entirety in a machine-readable form in the online journal. A part is shown here as guidance for the reader regarding its content.) \\(1) ID number in \textit{Swift}--BAT 105 month survey. (2) The observational program ID. The IDs 098.A-0635, 0 99.A-0403, 0101.A-0765, 0102.A-0433, and 0103.A-0521 are BASS project observation runs. (3) Slit width / slit length (4) Wavelength coverage setup for the NIR arm. Full coverage goes from 0.994 to 2.479\,$\mu$m. The limited coverage ranges from 0.994 to 2.101\,$\mu$m.
    
    \end{table*}

\section{Emission Line Table}
\label{sec:em_lines}
    In \autoref{tab:emission lines} the emission lines in the NIR range of 0.9 -- 2\,$\mu$m\ are shown based on the fitting routine. The lines are sorted by increasing wavelength and spectral region. In \autoref{tab:coronal lines} the CLs that are part of this study are listed, including their IP and critical density values.

\setcounter{table}{0}
\begin{table}
    \centering
    \footnotesize
    \caption[List of Emission Lines]{Emission lines arranged by wavelength and spectral region, adopted from \cite{Lamperti2017}.}
    \label{tab:emission lines}
    \begin{tabular}{l c c}
        \hline
         \textbf{Line} & \textbf{Wavelength} & \textbf{Region}  \\ 
         & $[$$\mu$m$]$& \\ \hline \hline
         
         $[$S \textsc{iii}$]$ & 0.9531 & \multirow{2}{*}{Pa$\epsilon$}\\
         Pa$\epsilon$ & 0.9546 & \\
         & & \\
         $[$C \textsc{i}$]$ & 0.9827 & \multirow{3}{*}{$[$S \textsc{viii}$]$}\\
         $[$C \textsc{i} $]$ & 0.9853 & \\
         $[$S \textsc{viii}$]$ & 0.9915 & \\
         & & \\
         Pa$\delta$ & 1.0049& \multirow{10}{*}{Pa$\gamma$} \\
         He II $\lambda$4686 & 1.0126 & \\
         $[$Fe \textsc{vi}$]$ & 1.0109 & \\
         $[$S \textsc{ii}$]$ & 1.0290 & \\
         $[$S \textsc{ii}$]$ & 1.0320 & \\
         $[$S \textsc{ii}$]$ & 1.0336 & \\
         $[$S \textsc{ii}$]$ & 1.0370 & \\
         $[$Fe \textsc{xiii}$]$ & 1.0747 & \\
         He \textsc{i} & 1.0830 & \\
         Pa$\gamma$ & 1.0938 & \\
         & & \\
         He \textsc{ii} $\lambda$4686 & 1.1620 & \multirow{9}{*}{Pa$\beta$}\\
         $[$P \textsc{ii}$]$ & 1.1886 & \\
         $[$S \textsc{ix}$]$ & 1.2520 & \\
         $[$Fe \textsc{ii}$]$ & 1.2570 & \\
         $[$Fe \textsc{ii}$]$ & 1.2788 & \\
         Pa$\beta$ & 1.2818 & \\
         $[$Fe \textsc{ii}$]$ & 1.2950 & \\
         O \textsc{i} & 1.3169& \\
         $[$Fe \textsc{ii}$]$ & 1.3201 & \\
         & & \\
         $[$Si \textsc{x}$]$ & 1.4300 & $[$Si \textsc{x}$]$ \\
         & & \\
         $[$Fe \textsc{ii}$]$ & 1.6436 & \multirow{2}{*}{ $[$Fe \textsc{ii}$]$}\\
          $[$Fe \textsc{ii}$]$ & 1.6807 & \\
         & & \\
         H$_2$ 1-0S(5) & 1.8345& \multirow{8}{*}{Pa$\alpha$}\\
         He \textsc{i} & 1.8635 & \\
         Pa$\alpha$ & 1.8751& \\
         $[$S \textsc{xi}$]$ & 1.9196 & \\
         $[$Si \textsc{xi}$]$ & 1.9320 & \\
         Br $\delta$ & 1.9446 & \\
         H$_2$ & 1.9564 & \\
         $[$Si \textsc{vi}$]$ & 1.9641 & \\
         
    \end{tabular}

    \end{table}

\begin{table*}
    \centering
    \small
    \caption{NIR CLs Arranged by Wavelength, Adopted from \cite{Lamperti2017} and \cite{Rodriguez2011}}
    \label{tab:coronal lines}
    \begin{tabular}{l c c c}
        \hline
         \textbf{CL} & \textbf{Wavelength} & \textbf{IP} & \textbf{Critical Density}  \\ 
         & $[$$\mu$m$]$& [eV] & [cm$^{-3}$]\\ \hline \hline

         $[$S \textsc{viii}$]$ & 0.9915 & 280.9 &$4.0\times10^{10}$ \\
         $[$Fe \textsc{xiii}$]$ & 1.0747 & 330.8 & $6.3\times10^8$ \\
         $[$S \textsc{ix}$]$ & 1.2520 & 328.2 & $2.5\times10^9$ \\
         $[$Si \textsc{x}$]$ & 1.4300 & 351.1 & $6.3\times10^8$\\
         $[$S \textsc{xi}$]$ & 1.9196 & 447.1 & $3.2\times10^8$ \\
         $[$Si \textsc{xi}$]$ & 1.9320 & 401.4 & $1.1\times10^8$\\
         $[$Si \textsc{vi}$]$ & 1.9641 & 166.8 & $6.3\times10^8$\\
         
    \end{tabular}
    
\end{table*}

\section{Measured Data}
\label{sec:measdata}
\autoref{tb:summary_measurements_all} includes the flux measurements from the spectral fits done in this work. The table is available in its entirety in the online journal. \autoref{tb:summary_measurements_all} includes the flux values of the Pa$\epsilon$ spectral range; the other spectral ranges can be found in the online journal.

\begin{table}[h]
    \caption{Flux measurements for All Lines.  }
    \label{tb:summary_measurements_all}
    \begin{tabular}{l c c c c c c c}\hline
        Line&Position & Flux & FWHM &S/N &Error Position & Error Flux &Error FWHM\\
        & [$\mu$m] & [erg/s/cm$^2$] & [km/s] & &[nm] & [erg/s/cm$^2$] & [km/s]\\
        (a) &  & & (b)\\\hline \hline
        SIII&0.953&4.23e-15&167&15.90&0.0030&2.232e-16&5\\
        SIII\_blue&0.9529&2.05e-15&350&3.68&0.023&2.57e-16&1\\
        Pa\_Epsilon&0.9545&7.15e-16&277&1.62&0.021&1.17e-16&32\\
        SIII\_broad&0.9531&-1.71e-15&--&--&--&--&--\\
        Pa\_Epsilon\_broad&0.9546&-1.74e-15&--&--&--&--&--\\
        SVIII&0.9915&-2.32e-16&--&--&--&--&--\\
        CIa&0.9853&-2.30e-16&--&--&--&--&--\\
        CIb&0.9827&-2.08e-16&--&--&--&--&--\\
        SVIII\_broad&0.9915&-5.45e-15&--&--&--&--&--\\
        CIa\_broad&0.9853&-5.42e-15&--&--&--&--&--\\
        CIb\_broad&0.9827&-4.8e-15&--&--&--&--&--\\
        Pa\_Gamma&1.0939&1.26e-15&254&7.71&0.033&2.43e-16&33\\
        Pa\_Delta&1.003&9.48e-16&313&5.32&0.0492&7.646e-17&33\\
        HeII&1.0123&1.085e-15&264&7.18&0.027&5.34e-17&13\\
        SiII\_a&1.029&-1.93e-16&--&--&--&--&--\\
        SiII\_b&1.032&-1.70e-16&--&--&--&--&--\\
        SiII\_c&1.0336&-1.62e-16&--&--&--&--&--\\
        SiII\_d&1.037&-1.59e-16&--&--&--&--&--\\
        He I&1.0830&3.65e-15&387&15.41&0.015&2.57e-16&15\\
        Fe XIII&1.0748&4.53e-16&464&1.61&0.10&6.67e-17&75\\
        Fe VI&1.0108&-3.02e-16&--&--&--&--&--\\
        Pa\_Gamma\_broad&1.0934&1.93e-15&1703&1.83&0.450&3.11e-16&184\\
        He I broad& 1.0839 & 3.99e-15 & 1981 &3.29 &0.398&5.32e-16&894\\
        ...& \\ \hline
    \end{tabular}
    \centering
    {\\{\bf Notes.} This example shows the measured lines for BAT ID 677. Negative values indicate 2$\sigma$ upper limits. The position corresponds to the expected rest-frame wavelength in the case of a nondetection. (a) Line name as found in the data table; (b) the S/N based on the amplitude of the line.}
\end{table}

\section{Spectral Fits}
\label{sec:fits}
As an example, \autoref{fig:fitting_plts} shows the different spectral regions where the emission lines have been fitted. The spectrum taken as an example is of the source 2MASX J214805.31-535941.3 (BAT ID 1604). For each object, we show the spectra (in black), the components fitted for the emission lines (in blue),
the overall best-fit model (in red), and the residuals (below).

In \autoref{fig:fitting_plts_2} source ESO 103-035 (BAT ID 988) is shown, where the fitting routine is more difficult to apply due to irregular line shapes and heavy telluric absorption. In the Pa$\alpha$ spectral range, a spline fit is applied to estimate the continuum level.

\setcounter{figure}{0}   
\begin{figure*}
    \centering
    \includegraphics[width=\textwidth]{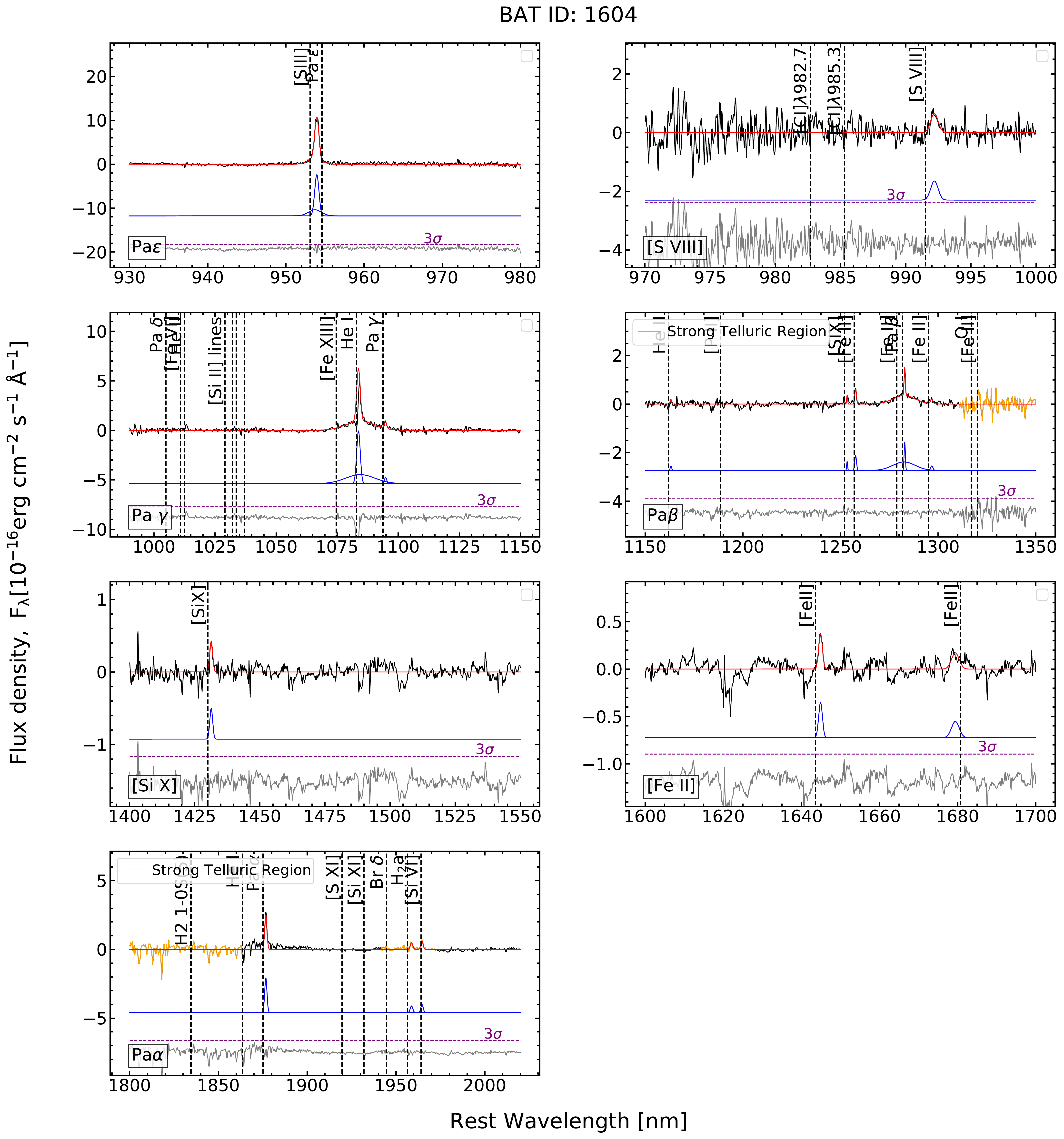}
    \caption{Compilation of the various fitting regions separated into the individual spectral regions. The spectrum of 2MASX J214805.31-535941.3 (BAT ID 1604) is shown. For this source, the emission lines are fitted well. The vertical dashed lines indicate the location of the nonsystemic corrected emission line.}
    \label{fig:fitting_plts}
\end{figure*}
\begin{figure*}
    \centering
      \includegraphics[width=\textwidth]{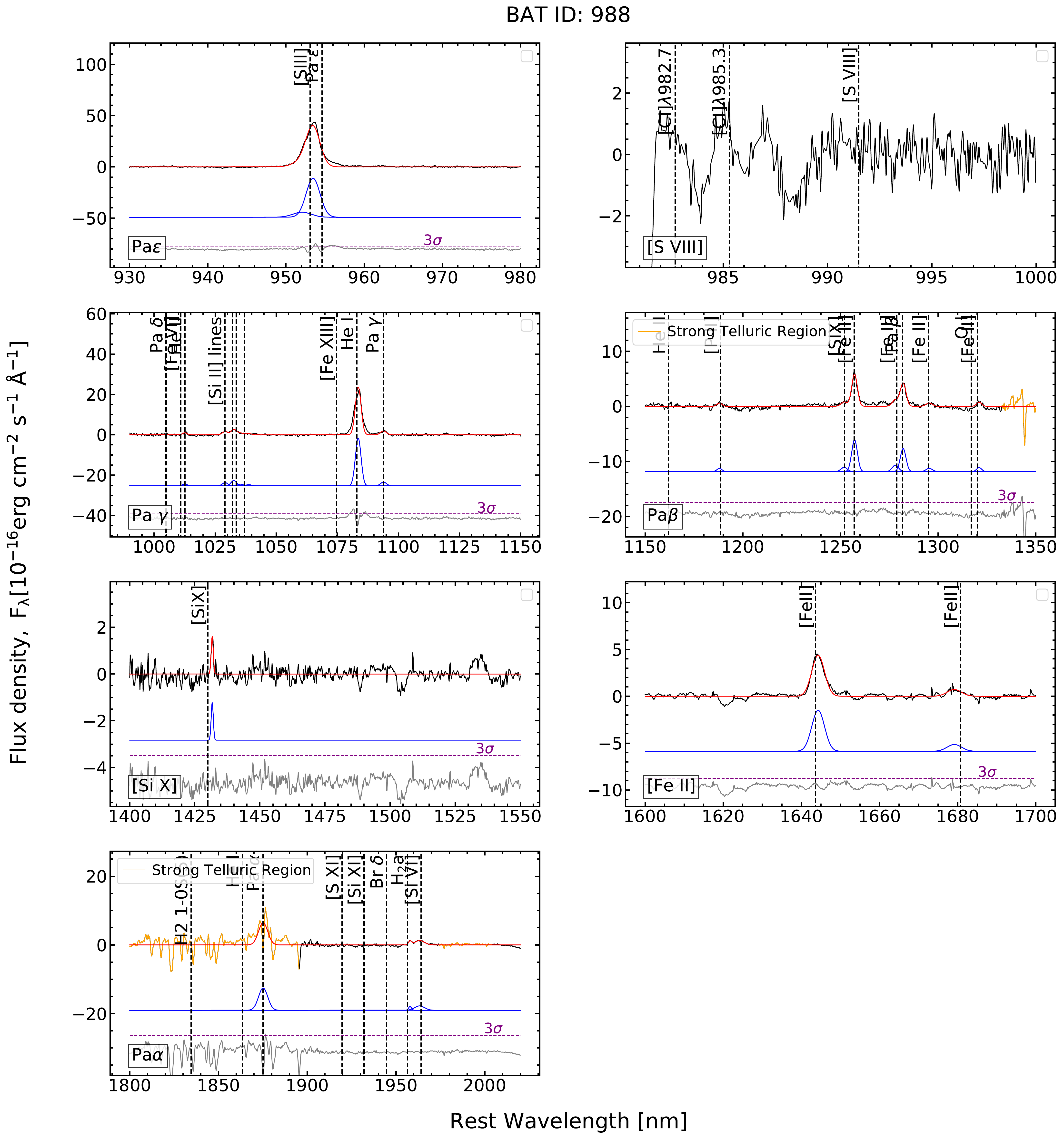}
    \caption{Compilation of the various fitting regions separated into the individual spectral regions. The spectrum of ESO 103-035 (BAT ID 988) is shown. For this source, the S/N is lower and for the Pa$\alpha$ spectral region, spline fitting for continuum estimation is applied. The vertical dashed lines indicate the location of the nonsystemic corrected emission line.}
    \label{fig:fitting_plts_2}
\end{figure*}

\section{CL and Black Hole Mass}
\label{App:bhmass}
Further comparisons between the black hole mass and the CL emission strength (in this case [Si \textsc{vi}] and [Si \textsc{x}]) are shown in

\setcounter{figure}{0}   
\autoref{fig:M_bh_CL}.
    \begin{figure*}
            \centering
            \includegraphics[width =0.9 \textwidth]{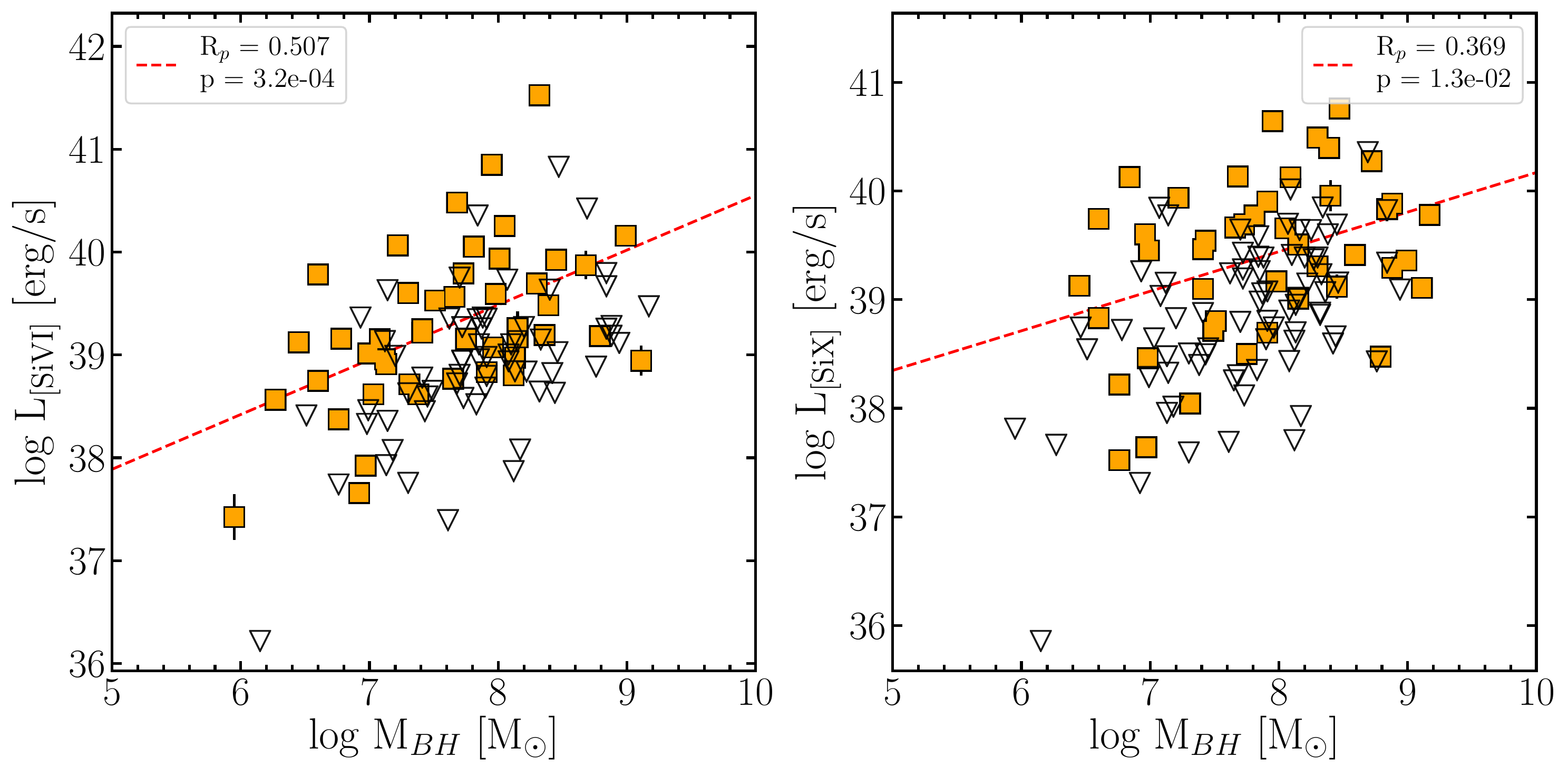}
             \caption{Correlation between the CL luminosities and the mass of the central black hole. Only moderate correlations are found. $R_{\rm pear}$ is the Pearson correlation and $p$ the Pearson $p$-value.}
            \label{fig:M_bh_CL}
        \end{figure*}

\section{Hidden Broad Lines}
\label{sec:hbl}
\autoref{fig:HBL_example} presents an example, of LEDA 157443 (BAT ID 597), showing hidden broad lines. Parts of the spectrum are shown. The H$\beta$ and H$\alpha$ emission lines show no clear sign of broad components. The Pa$\alpha$ emission line, on the other hand, shows a clear broad component.
\setcounter{figure}{0}
\begin{figure*}
            \centering
            \includegraphics[width=\textwidth]{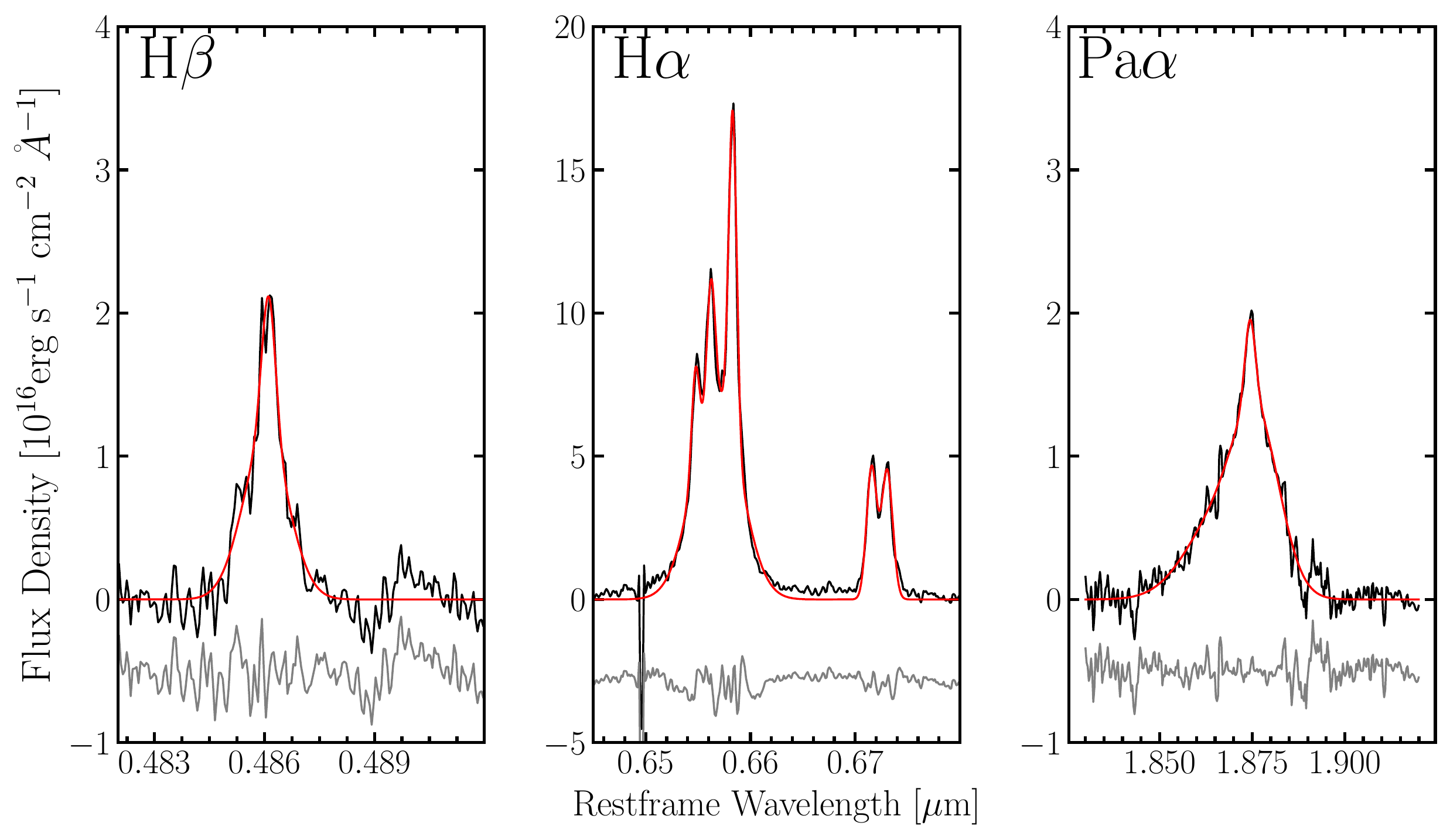}
            \caption{Example of hidden broad lines: the Pa$\alpha$ emission line in the NIR regime clearly shows a broad component, while the H$\beta$ and H$\alpha$ emission lines in the optical regime do not show clear broad components. The source shown is LEDA 157443 (BAT ID 597).  }
            \label{fig:HBL_example}
            
\end{figure*}

\section{Continuum Fitting}
\label{app:spline}
{For most of the spectral ranges, we fit the continuum using a fourth-order polynomial. In certain ranges, due to an irregular shape, we use a spline fit to correct the continuum. In total, we have applied a spline fit in 123 regions for 88/168 sources (as a reminder, we have separated each spectrum into seven spectral regions; see \autoref{tab:spec_reg}). \autoref{tab:spline_fit_list} lists the spectral regions where a spline fit is used. \autoref{fig:fit_comp} shows a comparison of the two continuum fit methods in the Pa$\alpha$ region in source BAT 1138.  }

\setcounter{table}{0}
\begin{table}
        \centering
        \caption{List of spectral regions where spline fit is applied (indicated by "x").}
        \label{tab:spline_fit_list}
        \begin{tabular}{c c c c c c c c}
             \textbf{BAT ID} & Pa$\epsilon$ &
             $[$S \textsc{viii}$]$&
             Pa$\gamma$& 
             Pa$\beta$&
             $[$Si \textsc{x}$]$&
             $[$Fe \textsc{xiii}$]$&
             Pa$\alpha$
             \\ \hline
             
             10&\\
             17&&&&&&&x\\
             20&\\
             31&&&&&&&x\\
             32&\\
             37&&&&&&&x\\
             50&\\
             52&&&&&x&&\\
             53&&&x&&&&x\\
             57&&&&x&&&x\\
             ...
        \end{tabular}
        
    \end{table}

\setcounter{figure}{0}
\begin{figure}
    \centering
    \includegraphics[width = \textwidth]{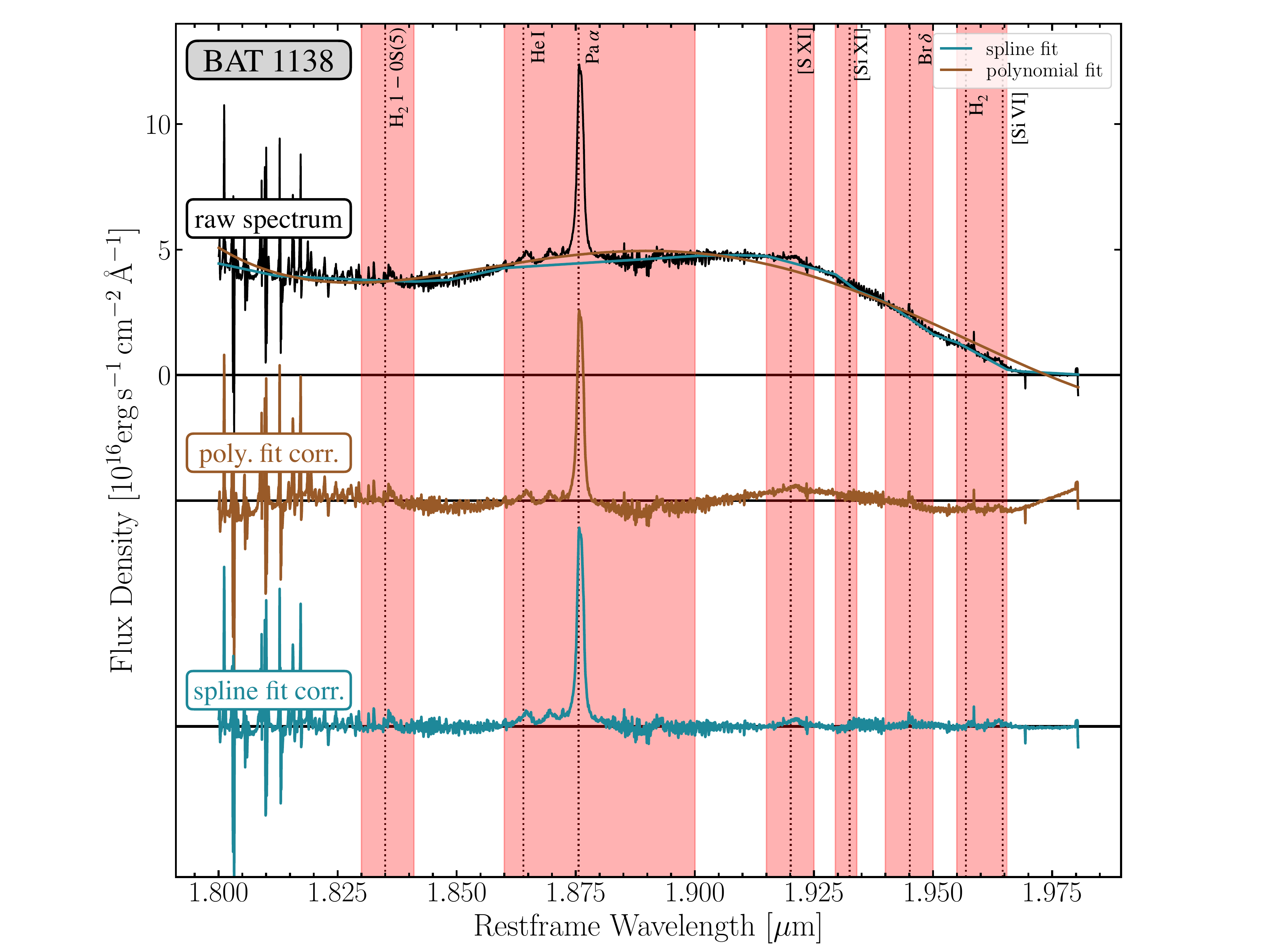}
    \caption{Comparison of the two continuum fitting methods. The black spectrum in the upper part shows the raw spectrum. The spline fit is indicated in blue and the fourth-order polynomial fit in brown. The fourth-order polynomial fit corrected spectrum is shown in the center (in brown) and the spline fit corrected spectrum is shown at the bottom (in blue). The red shaded regions are excluded from the continuum fit due to intrinsic emission.}
    \label{fig:fit_comp}
\end{figure}

\bibliographystyle{aasjournal}

\end{document}